\documentclass[10pt,aps,fleqn,superscriptaddress,notitlepage,nofootinbib,preprintnumbers,nobalancelastpage]{revtex4-1}
\pdfoutput=1
\usepackage{amsmath,amssymb,graphicx,xspace,subfigure}
\usepackage{todonotes}
\usepackage{bigints}

\newcommand{\Pythia}{P\protect\scalebox{0.8}{YTHIA}\xspace}
\newcommand{\Sherpa}{S\protect\scalebox{0.8}{HERPA}\xspace}

\newcommand{\Dire}{D\protect\scalebox{0.8}{IRE}\xspace}
\newcommand{\eps}{\varepsilon}
\newcommand{\mc}[1]{\mathcal{#1}}
\newcommand{\mr}[1]{\mathrm{#1}}

\usepackage{hyperref}
\hypersetup{
  pdfauthor={Falko Dulat, Stefan Hoeche, Stefan Prestel},
  pdftitle={Leading-color fully differential two-loop soft corrections to QCD dipole showers},
  pdfkeywords={QCD, Resummation, Parton Shower}
}
\begin{document}
\preprint{SLAC-PUB-17252}
\preprint{FERMILAB-PUB-18-147-T}
\preprint{MCNET-18-09}
\title{Leading-color fully differential two-loop soft corrections to QCD dipole showers}
\author{Falko~Dulat}
\affiliation{SLAC National Accelerator Laboratory,
  Menlo Park, CA, 94025, USA}
\author{Stefan~H{\"o}che}
\affiliation{SLAC National Accelerator Laboratory,
  Menlo Park, CA, 94025, USA}
\author{Stefan~Prestel}
\affiliation{Fermi National Accelerator Laboratory,
  Batavia, IL, 60510-0500, USA}
\begin{abstract}
  We compute the next-to-leading order corrections to soft-gluon radiation
  differentially in the one-emission phase space. We show that their
  contribution to the evolution of color dipoles can be obtained in a
  modified subtraction scheme, such that both one- and two-emission terms are
  amenable to Monte-Carlo integration. The two-loop cusp anomalous dimension
  is recovered naturally upon integration over the full phase space.
  We present two independent implementations of the new algorithm in the
  two event generators Pythia and Sherpa, and we compare the resulting
  fully differential simulation to the CMW scheme.
\end{abstract}
\maketitle

\section{Introduction}
Experiments at high-energy particle colliders have been integral to unraveling
the structure of our universe and have confirmed the validity of the Standard Model
of particle physics at an unprecedented accuracy. Going beyond the current level
of precision and possibly revealing new fundamental particles and forces
will require ever more detailed experimental analyses and theoretical calculations.
Monte-Carlo simulations by means of event generators play a vital role in this context,
as they link experiment and theory through the detailed description of fully exclusive
final states~\cite{Webber:1986mc,Buckley:2011ms}. They are required to describe the
dynamics of a large number of hadrons originating from QCD Bremsstrahlung,
which is modeled in the simulation through so-called parton showers. Modern parton showers
are typically based on a unified description of QCD radiative effects in a dipole picture,
which encompasses both the leading-order spin-averaged collinear radiation pattern,
and the leading-order color-averaged soft radiation pattern. The predictions generated
by such algorithms accurately describe many experimental measurements.
A notable exception to the success of the parton-shower method arises from its limited
phase-space coverage. This problem is alleviated by the matching and merging techniques
that allow to correct parton showers to any known fixed-order result at limited
final-state multiplicity, and that have been in the focus of interest of the theoretical
particle physics community in the past decade~\cite{Alwall:2007fs,Nason:2012pr}.

Currently the most pressing problem in the context of parton-shower simulations is the
lack of options to assess the intrinsic uncertainty of the method itself. The precision
of fixed-order perturbative QCD calculations is conventionally quantified by varying the renormalization and factorization
scales, and the dependence on these scales is reduced at higher orders in the perturbative
expansion if the perturbative series converges. No such technique is currently available
for parton showers, essentially because parton showers at higher precision do not yet
exist or their practical implementation is incomplete. First steps towards the construction
of next-to-leading order (NLO) parton showers have been made in~\cite{
  Kato:1986sg,Kato:1988ii,Kato:1990as,Kato:1991fs,
  Giele:2011cb,Hartgring:2013jma,Li:2016yez,Nagy:2017ggp,Hoche:2017iem,Hoche:2017hno},
but no method has yet been presented
that is capable of simulating fully exclusive final states at hadron colliders.
At the same time, it should be expected that a difference exists between parton showers
and analytical resummation. While it should be reduced at higher perturbative precision,
it cannot be completely eliminated due to differences in the treatment of momentum
and probability conservation~\cite{Hoeche:2017jsi}.

In this publication we address one of the most important aspects of next-to-leading
order parton showers, namely the simulation of the higher-order corrections to soft gluon
radiation, and we show how to implement these corrections in a fully differential form
in practice. In integrated form, they lead to the well-known two-loop cusp anomalous
dimension~\cite{Kodaira:1981nh,Davies:1984hs,Davies:1984sp,Catani:1988vd}, which is
included in improved leading-order parton showers by means of redefining the strong
coupling. This is known as the CMW method~\cite{Catani:1990rr}. At the differential level,
the corrections to soft gluon radiation induce spin correlations and sub-leading color corrections
that are not included in leading-order parton showers. As part of the extension to the
next-to-leading order, we adapt the algorithm in~\cite{Hoche:2015sya} to include these effects.
Moreover, the construction of a local modified subtraction procedure as anticipated 
in~\cite{Hoche:2017iem} mandates the computation of the two-loop cusp anomalous
dimension as an endpoint contribution corresponding to the iterated soft times collinear limit.
The resulting algorithm will be a key ingredient in the construction of a fully differential
technique for matching parton showers to next-to-next-to leading order calculations.

This paper is organized as follows: Section~\ref{sec:analytic} present an analytic
calculation of the local K-factor due to NLO corrections to soft-gluon radiation.
Based on this calculation, Sec.~\ref{sec:four_dimensions} introduces the modified subtraction
method and presents our approach to implementing the required changes in the dipole-like
parton shower. Section~\ref{sec:results} presents a numerical validation of the new
Monte-Carlo techniques and an assessment of the effect of the fully differential
simulation compared to the CMW method. A summary is given in Sec.~\ref{sec:conclusions}.

\section{Analytic computation of double-soft corrections}
\label{sec:analytic}
We employ the formalism for the construction of parton showers at next-to-leading
order accuracy originally proposed in~\cite{Hoche:2017iem}. This technique is based
on a modified subtraction method combined with a new algorithm for mapping
$n$-particle on-shell momentum configurations to $n+2$-particle on-shell
momentum configurations and the replacement of explicit symmetry factors
by appropriate light-cone momentum fractions that can be identified as
``tags'' for evolving partons. The extension of this method to soft evolution
at next-to-leading order requires the removal of overlap between the explicitly
included higher-order corrections in the CMW scheme~\cite{Catani:1990rr} and the 
potentially included triple-collinear splitting functions~\cite{Hoche:2017iem}.
In this section we will first derive analytic results for the double-soft corrections
at next-to-leading order. We define the kinematics in Sec.~\ref{sec:kinematics},
present the individual corrections in Sec.~\ref{sec:nlo_contributions} and collect 
the results in Sec.~\ref{sec:nlo_corrections}. Based on this calculation, 
Sec.~\ref{sec:four_dimensions} introduces a modified subtraction technique
and addresses the overlap removal.

\begin{figure}[t]
  \subfigure[]{\includegraphics[scale=0.33]{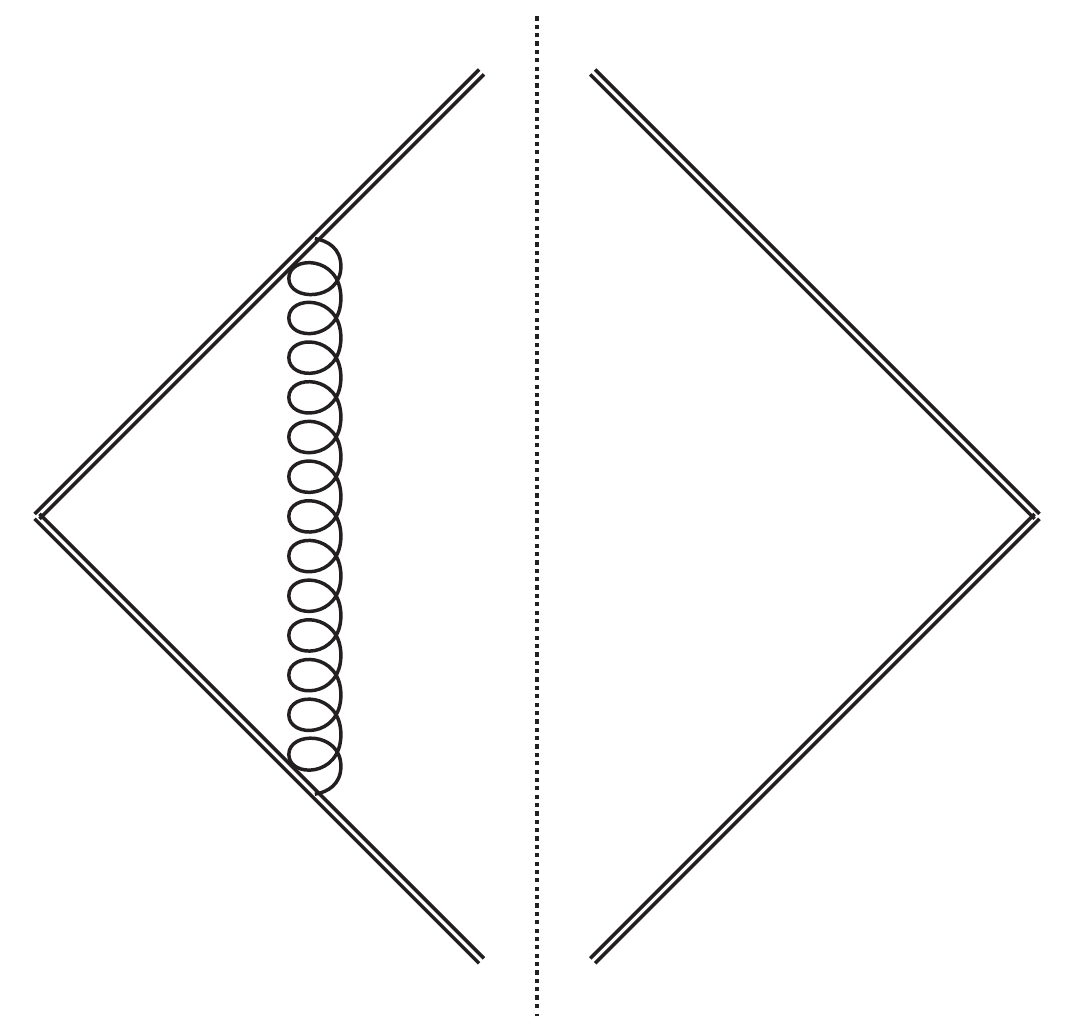}\label{fig:lo_virt}}
  \subfigure[]{\includegraphics[scale=0.33]{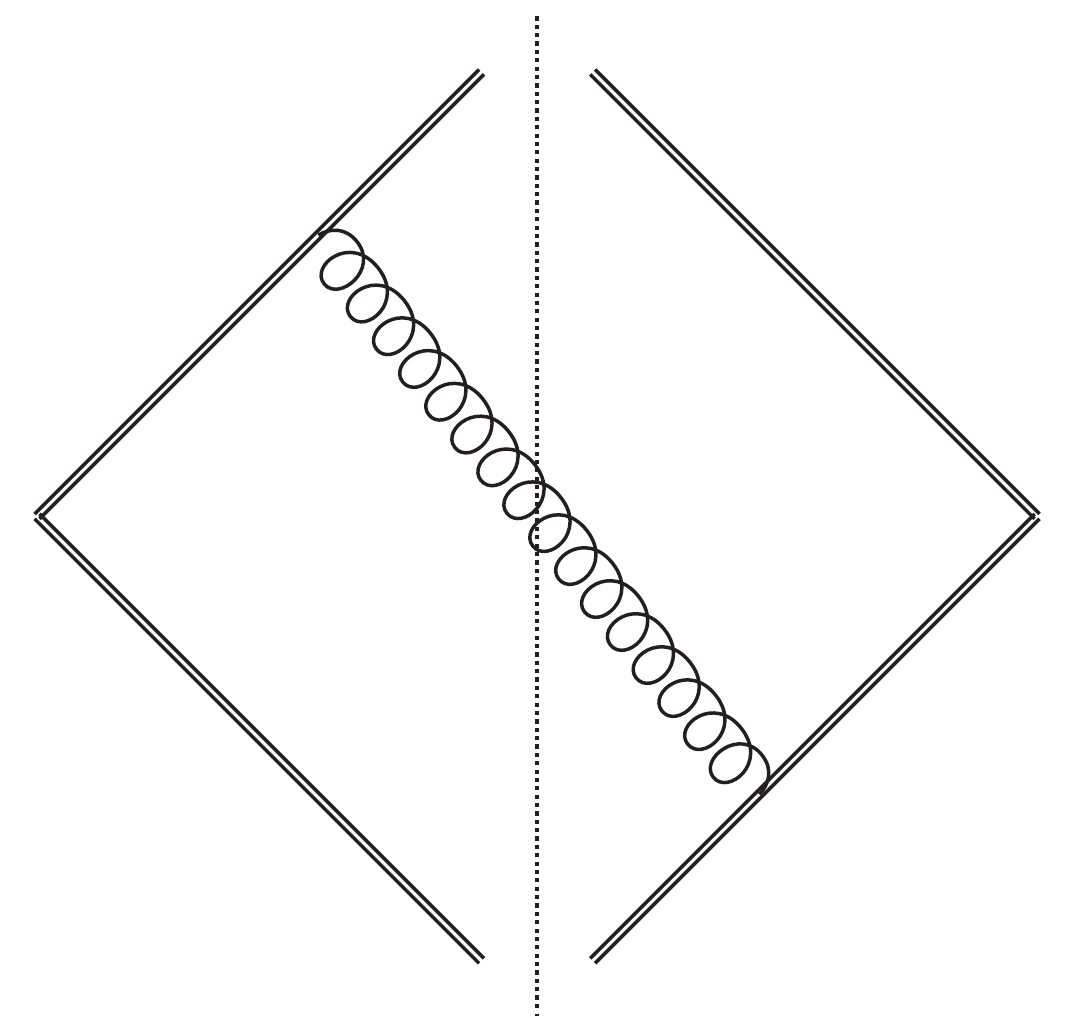}\label{fig:lo_real}}
  \caption{Leading order contributions to dipole-shower evolution in the soft limit.
    The double solid lines represent hard (identified) partons i.e.\ Wilson lines.
    \label{fig:lo}}
\end{figure}\noindent
The leading-order contributions to the soft function, which described the interaction
between two hard jets of collinear particles through soft gluon
exchange~\cite{Sterman:1986aj,Collins:1988ig,Collins:1989gx}
are shown in Fig.~\ref{fig:lo}. The double solid lines represent the hard legs,
and the dashed line indicates the cut. The virtual correction is given by a scaleless
integral and vanishes in dimensional regularization~\cite{Monni:2011gb}. The diagram
in Fig.~\ref{fig:lo_real} and its mirror conjugate generate the eikonal factor
\begin{equation}\label{eq:lo_eik_fc}
  {\bf S}_{ij}^{(0)}(q)=-{\bf T}_i{\bf T}_j\,\mc{S}_{ij}^{(0)}(q)\;,
  \qquad\text{where}\qquad
  \mc{S}_{ij}^{(0)}(q)=g_s^2\mu^{2\eps}\,\frac{p_ip_j}{2\,(p_iq)(q\,p_j)}
  =g_s^2\mu^{2\eps}\,\frac{s_{ij}}{s_{iq}s_{jq}}\;.
\end{equation}
Here and in the following we will label the Wilson lines by $i$ and $j$,
while the soft momenta will be denoted by $1$ and possibly $2$. We also refer
to the combined soft momentum as $q$, where $q=p_1$ and $q=p_1+p_2$ in one-
and two-emission configurations, respectively. We restrict our analysis to the
improved leading-color approximation typically used in parton-shower simulations.
In processes with $n$ possibly color-connected partons, the eikonal term,
Eq.~\eqref{eq:lo_eik_fc}, is first partial-fractioned~\cite{Catani:1996vz},
and subsequently the color-insertion operator ${\bf T}_i{\bf T}_j$ is
approximated by assuming independence of the kinematics. This leads
to the replacement
\begin{equation}\label{eq:leading_color}
  \sum_{\substack{i=1\\j=i+1}}^{n}{\bf S}_{ij}^{(0)}(q)=
  -\sum_{\substack{i,j=1\\j\neq i}}^n{\bf T}_i{\bf T}_j\,\mc{D}_{i,j}^{(0)}(q)\quad\to\quad
  \sum_{\substack{i,j=1\\j\neq i}}^n\frac{C_i}{n}\,\mc{D}_{i,j}^{(0)}(q)\;,
  \qquad\text{where}\qquad
  \mc{D}_{i,j}^{(0)}(q)=g_s^2\mu^{2\eps}\,\frac{1}{s_{iq}}\frac{s_{ij}}{s_{iq}+s_{jq}}\;.
\end{equation}
As the partial fraction $\mc{D}_{i,j}(q)$ can be matched to the collinear limit
unambiguously, the corresponding color Casimir operator, $C_i$, should indeed be
associated with the emission in the soft-collinear limit. This approximation
proves to be very accurate in practice. We therefore postpone the exact treatment
of the color insertion operators to future work and perform our analysis based on
$\mc{S}_{ij}^{(0)}(q)$. We also point out that including the full next-to-leading
order corrections to Eq.~\eqref{eq:leading_color} requires that the first sub-leading
color correction be implemented in the parton shower if the two-loop cusp
anomalous dimension is to be recovered in the fully differential calculation.
These terms are related to color factors of the form $C_F-C_A/2$, where the
first contribution is absorbed into the exponentiated leading-order soft result, and the
second term becomes part of the genuine two-loop result~\cite{Cornwall:1975ty,Frenkel:1976bj}.
This will be discussed in detail in Sec.~\ref{sec:four_dimensions},
and related numerical comparisons will be made in Sec.~\ref{sec:results}.

The virtual corrections to the single emission have been computed in~\cite{Bern:1999ry,Catani:2000pi}. 
They are given by
\begin{equation}\label{eq:virt_soft}
  \mc{S}_{ij}^{\rm(virt)}(q)=-C_A\frac{g_s^4}{8\pi^2}
  \frac{(4\pi\mu^4)^\eps}{\eps^2}\frac{\Gamma^4(1-\eps)\Gamma^3(1+\eps)}{
  \Gamma^2(1-2\eps)\Gamma(1+2\eps)}\left(\frac{s_{ij}}{s_{iq}s_{jq}}\right)^{1+\eps}\;.
\end{equation}
\begin{figure}[t]
  \subfigure[]{\includegraphics[scale=0.33]{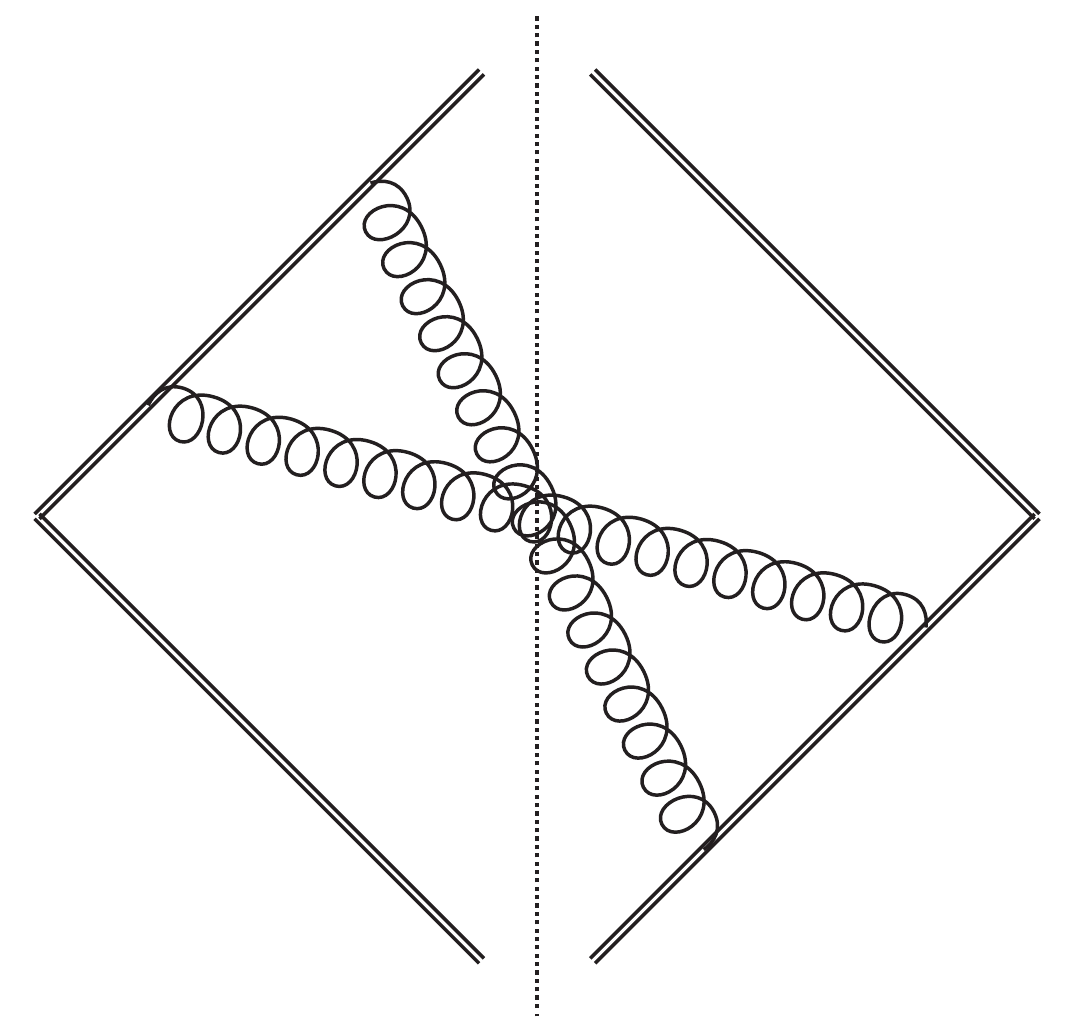}\label{fig:nlo_box1}}
  \subfigure[]{\includegraphics[scale=0.33]{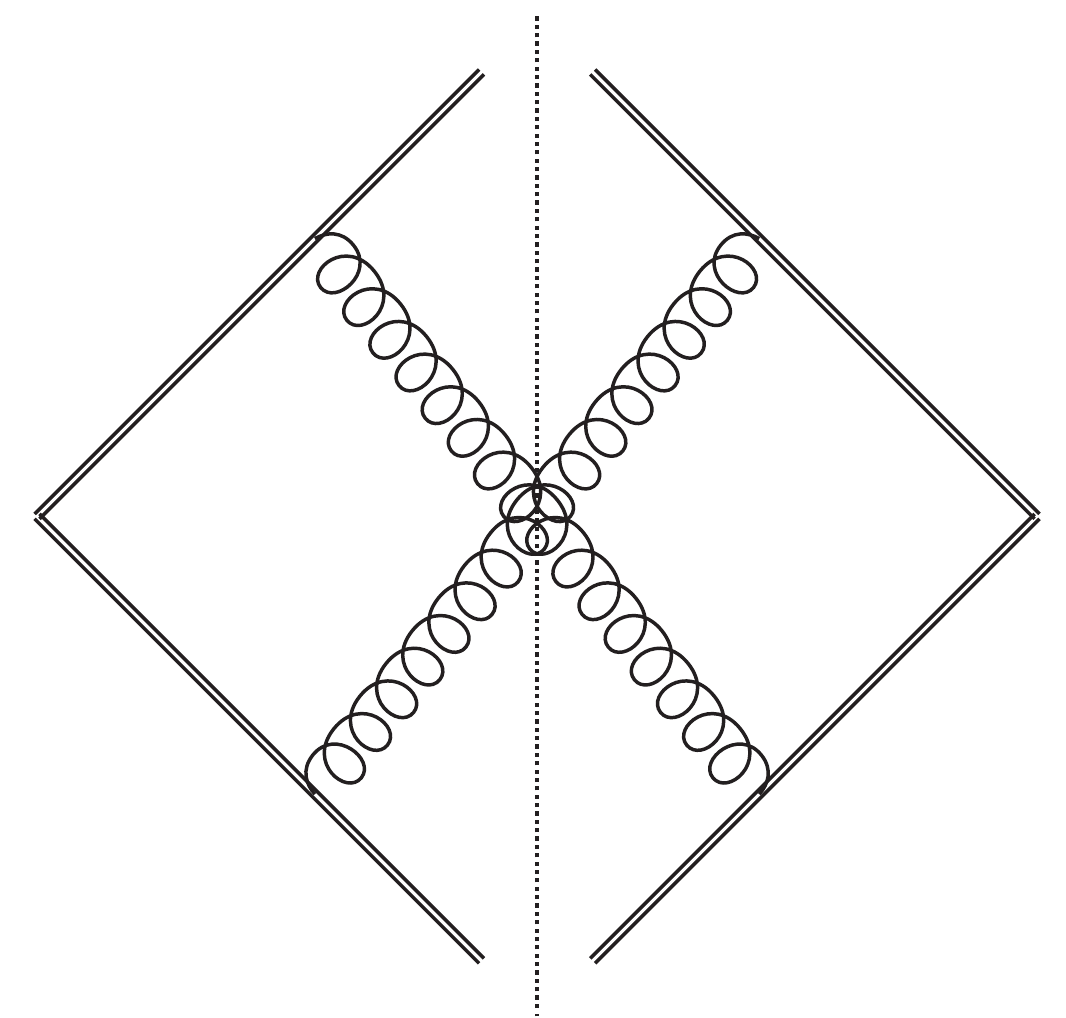}\label{fig:nlo_box2}}
  \subfigure[]{\includegraphics[scale=0.33]{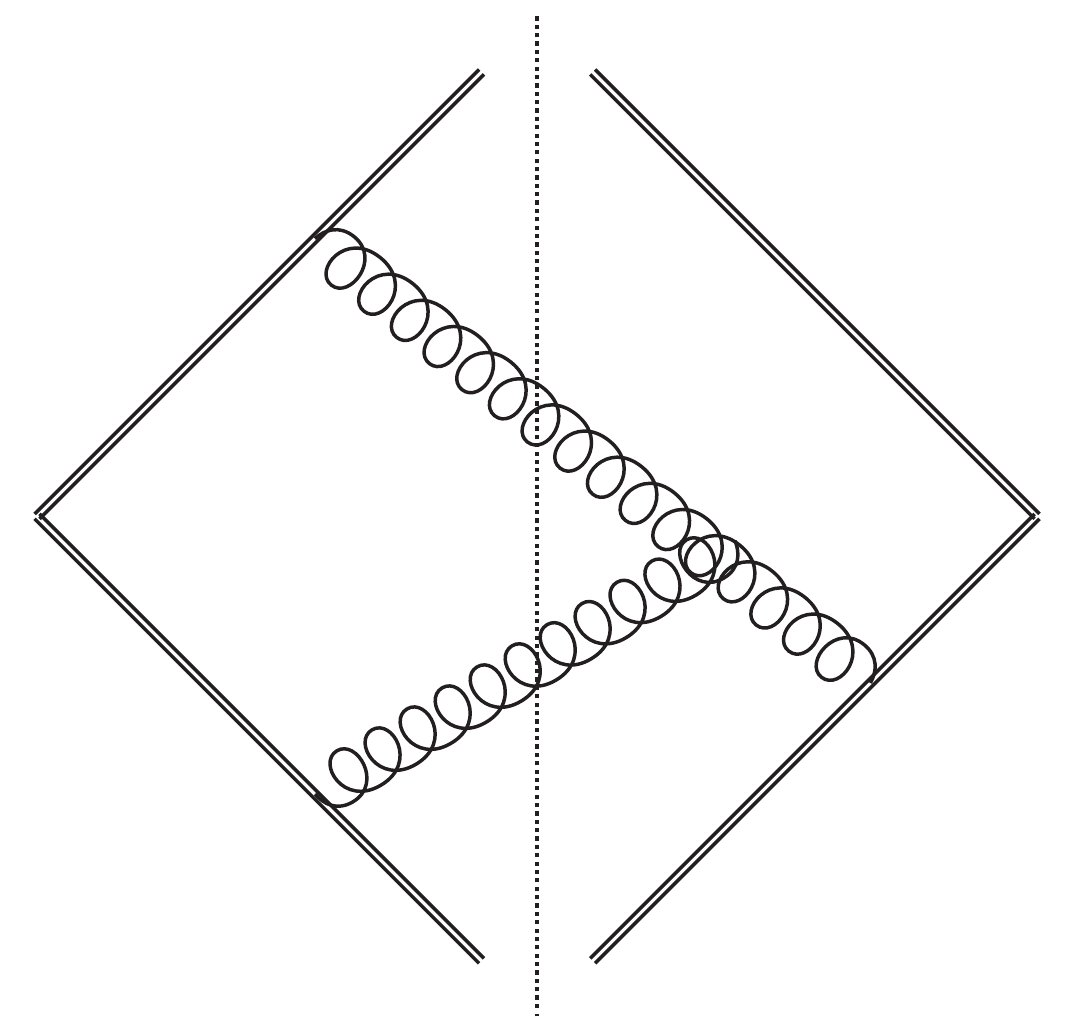}\label{fig:nlo_tgc1}}
  \subfigure[]{\includegraphics[scale=0.33]{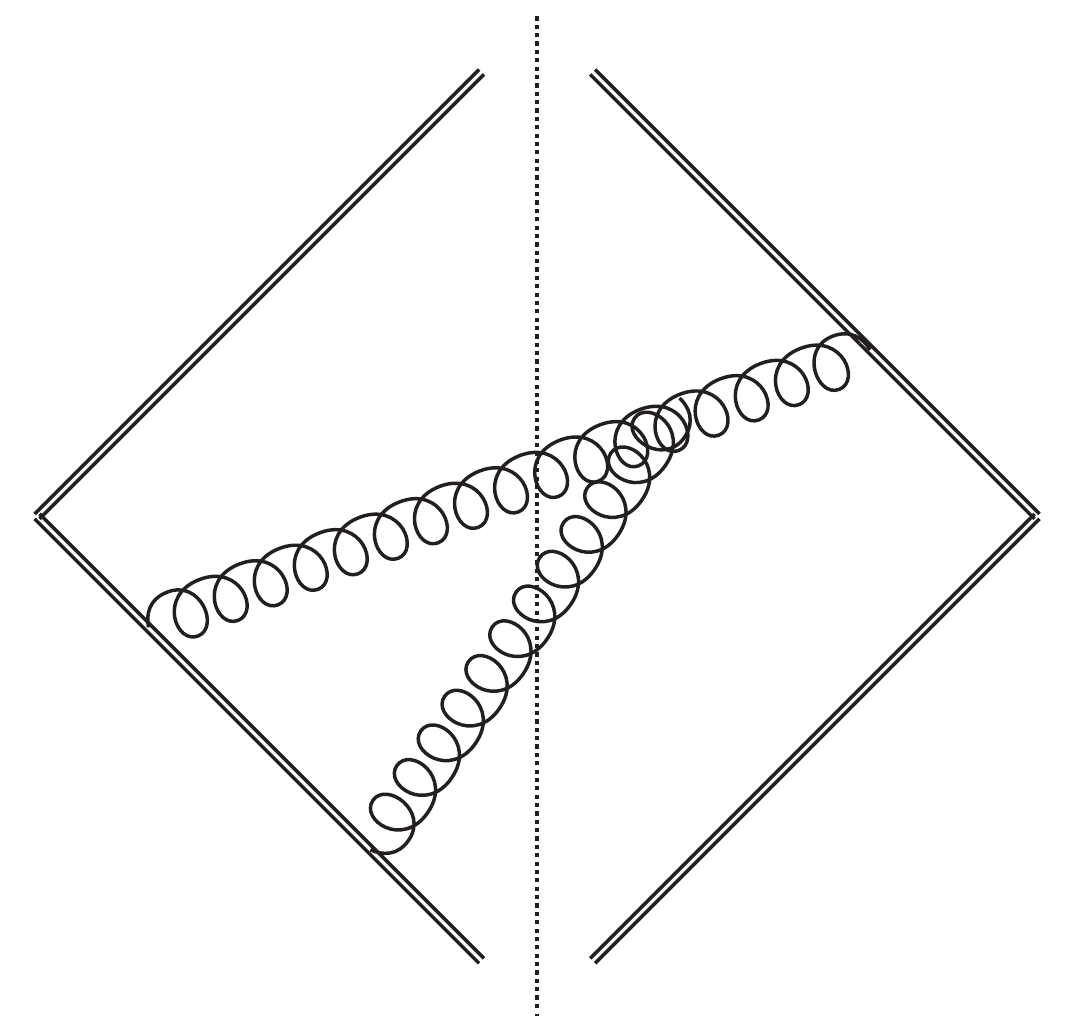}\label{fig:nlo_tgc2}}
  \subfigure[]{\includegraphics[scale=0.33]{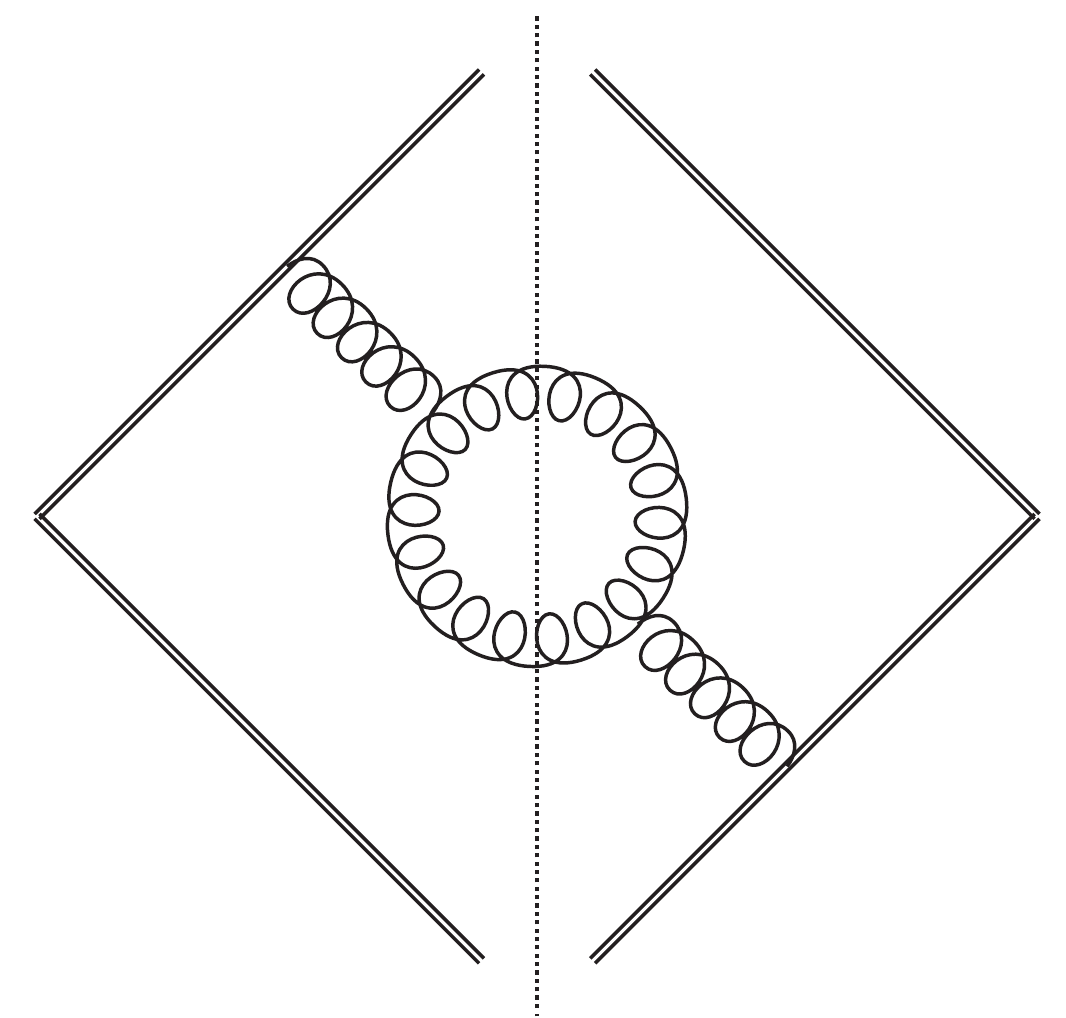}\label{fig:nlo_vpsg}}
  \subfigure[]{\includegraphics[scale=0.33]{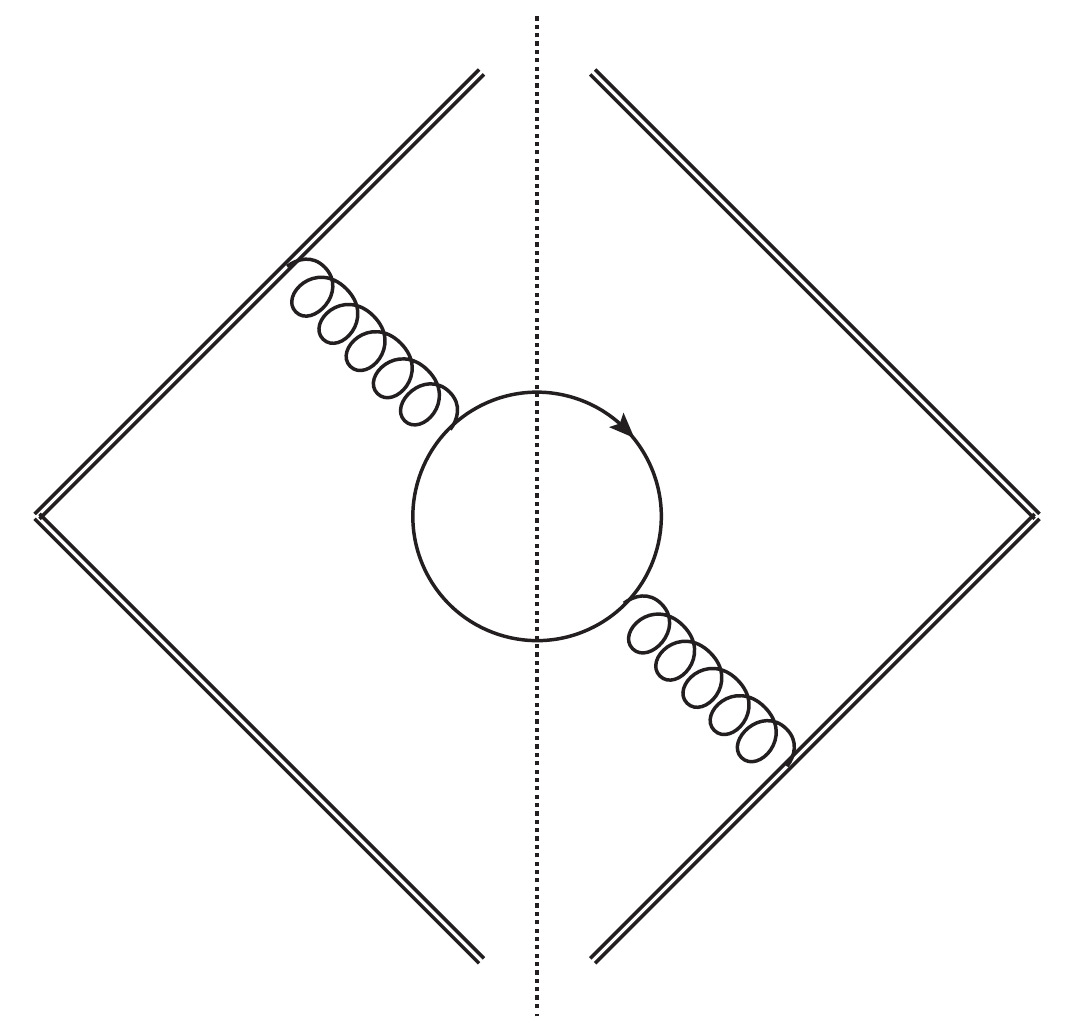}\label{fig:nlo_vpsq}}
  \subfigure[]{\includegraphics[scale=0.33]{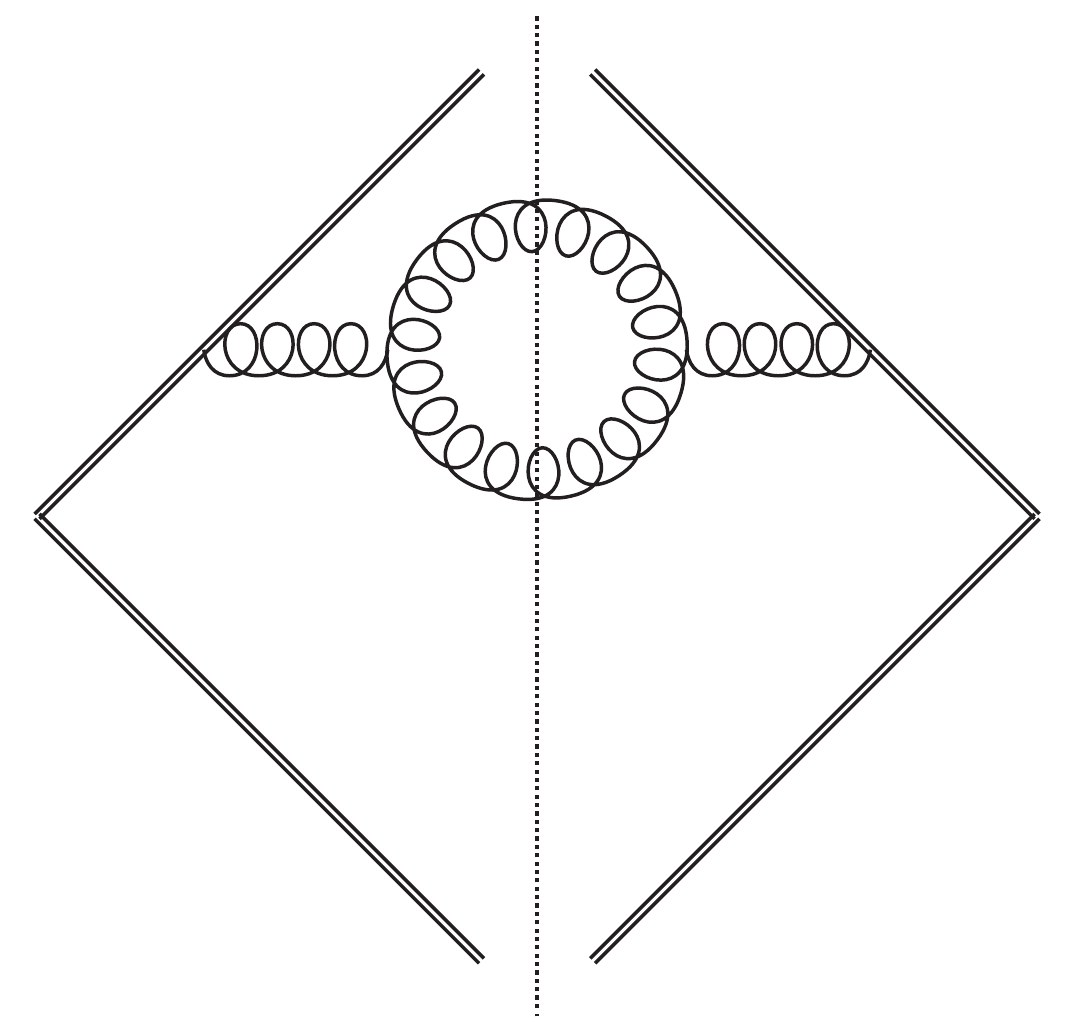}\label{fig:nlo_vpcg}}
  \subfigure[]{\includegraphics[scale=0.33]{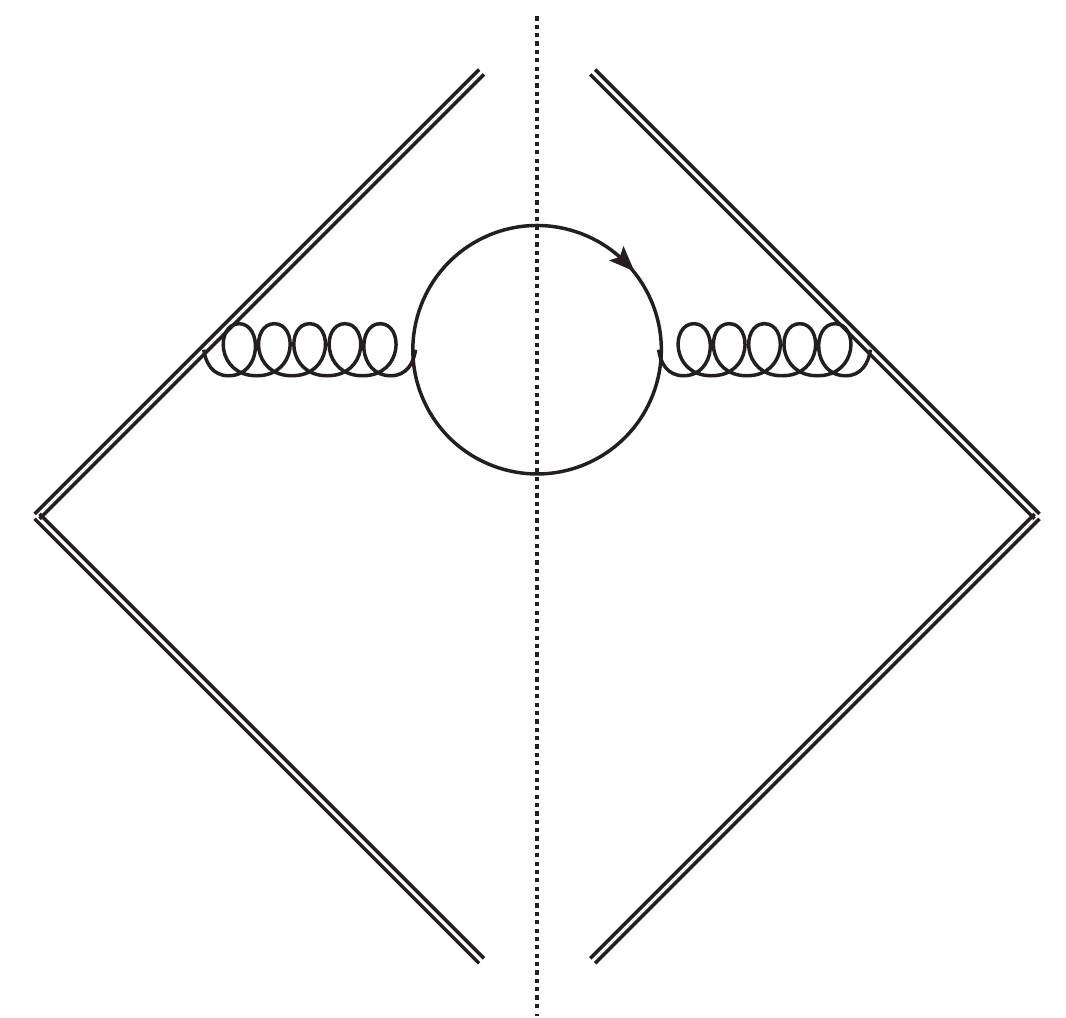}\label{fig:nlo_vpcq}}
  \caption{Next-to-leading order real-emission contributions to dipole-shower evolution
    in the soft limit. The double solid lines represent hard (identified) partons i.e.\
    Wilson lines.
    \label{fig:nlo}}
\end{figure}\noindent
The diagrams contributing to the gluonic real-emission corrections are schematically
displayed in Fig.~\ref{fig:nlo_box1}-\subref{fig:nlo_vpsg}, while the quark contribution
is shown in Fig.~\ref{fig:nlo_vpsq}. The vacuum polarization diagrams with gluons have
corresponding ghost diagrams, and all terms also occur in the mirror symmetric configuration.
Their sum is given by the soft insertion operators computed in~\cite{Catani:1999ss}
\begin{equation}\label{eq:real_soft}
  \begin{split}
    \mc{S}_{ij}^{(q\bar{q})}(1,2)=&\;T_R\;\frac{s_{i1}s_{j2}+s_{i2}s_{j1}-s_{12}s_{ij}}{
      s_{12}^2(s_{i1}+s_{i2})(s_{j1}+s_{j2})}\\
    \mc{S}_{ij}^{(gg)}(1,2)=&\;C_A\;\frac{(1-\eps)[s_{i1}s_{j2}+s_{i2}s_{j1}]-2s_{12}s_{ij}}{
      s_{12}^2(s_{i1}+s_{i2})(s_{j1}+s_{j2})}+\mc{S}_{ij}^{\rm(s.o.)}(1,2)\,\frac{C_A}{2}
    \left(1+\frac{s_{i1}s_{j1}+s_{i2}s_{j2}}{(s_{i1}+s_{i2})(s_{j1}+s_{j2})}\right)\;.
  \end{split}
\end{equation}
In the limit of strongly ordered soft emissions, $\mc{S}_{ij}^{(gg)}(1,2)$
reduces to $C_A\,\mc{S}_{ij}^{\rm(s.o.)}(1,2)$, where
\begin{equation}\label{eq:soft_so}
  \mc{S}_{ij}^{\rm(s.o.)}(1,2)=
  \frac{s_{ij}}{s_{i1}s_{12}s_{j2}}+\frac{s_{ij}}{s_{j1}s_{12}s_{i2}}
  -\frac{s_{ij}^2}{s_{i1}s_{j1}s_{i2}s_{j2}}\;.
\end{equation}
The full real-emission corrections are obtained by adding the cut vacuum polarization diagrams
displayed in Fig.~\ref{fig:nlo_vpcg} and~\subref{fig:nlo_vpcq}, as well as the corresponding terms
with the gluons attached to the other Wilson line. They are given by~\cite{Monni:2011gb}
\begin{equation}\label{eq:real_collinear}
  \begin{split}
    \mc{C}_{ij}^{(q\bar{q})}(1,2)=&\;-\frac{T_R}{s_{12}^2}
    \left(\frac{s_{i1}s_{i2}}{(s_{i1}+s_{i2})^2}+\frac{s_{j1}s_{j2}}{(s_{j1}+s_{j2})^2}\right)\\
    \mc{C}_{ij}^{(gg)}(1,2)=&\;-(1-\eps)\,\frac{C_A}{s_{12}^2}
    \left(\frac{s_{i1}s_{i2}}{(s_{i1}+s_{i2})^2}+\frac{s_{j1}s_{j2}}{(s_{j1}+s_{j2})^2}\right)\\
  \end{split}
\end{equation}
To simplify the integration, we define the soft remainder as well as
two collinear coefficients
\begin{equation}\label{eq:soft_helper_funcs}
  \begin{split}
    \mc{S}_{ij}^{\rm(rem)}(1,2)=&\;\mc{S}_{ij}^{\rm(s.o.)}(1,2)\,
    \frac{s_{i1}s_{j2}+s_{i2}s_{j1}}{(s_{i1}+s_{i2})(s_{j1}+s_{j2})}\\
    \mc{S}_{ij,B}^{\rm(coll)}(1,2)=&\;\frac{s_{ij}}{(s_{i1}+s_{i2})(s_{j1}+s_{j2})}
    \frac{1}{s_{12}}\\
    \mc{S}_{ij,A}^{\rm(coll)}(1,2)=&\;\mc{S}_{ij,B}^{\rm(coll)}(1,2)\,
    4\,z_1z_2\cos^2\phi_{12}^{\,ij}\,
    \qquad\text{where}\qquad
    4\,z_1z_2\cos^2\phi_{12,ij}=\frac{(s_{i1}s_{j2}-s_{i2}s_{j1})^2}{
      s_{12}s_{ij}(s_{i1}+s_{i2})(s_{j1}+s_{j2})}\;.
  \end{split}
\end{equation}
The precise meaning of $z$ and $\phi$ will be discussed in Sec.~\ref{sec:subtraction}.
In terms of the above functions we can write
\begin{equation}\label{eq:real_soft_2}
  \begin{split}
    \mc{S}_{ij}^{(q\bar{q})}(1,2)+\mc{C}_{ij}^{(q\bar{q})}(1,2)
    =&\;T_R\left(\mc{S}_{ij,A}^{\rm(coll)}(1,2)-\mc{S}_{ij,B}^{\rm(coll)}(1,2)\right)\\
    \mc{S}_{ij}^{(gg)}(1,2)+\mc{C}_{ij}^{(gg)}(1,2)=&\;
    C_A\left((1-\eps)\,\mc{S}_{ij,A}^{\rm(coll)}(1,2)-2\,\mc{S}_{ij,B}^{\rm(coll)}(1,2)
    +\mc{S}_{ij}^{\rm(s.o.)}(1,2)-\frac{1}{2}\,\mc{S}_{ij}^{\rm(rem)}(1,2)\right)\;.
  \end{split}
\end{equation}

\subsection{Kinematics}
\label{sec:kinematics}
We perform the calculation in a scheme that is applicable to both initial-
and final-state evolution. We parametrize the final-state momenta using two
light-like momenta $l$ and $n$ as
\begin{equation}\label{eq:def_sudakov_decomposition}
  p^\mu=\alpha_p\,l^\mu+\beta_p\,n^\mu+p^\mu_T
  \qquad\text{where}\qquad
  \alpha_p=\frac{pn}{ln}\;,\qquad
  \beta_p=\frac{pl}{ln}\;.
\end{equation}
The component along $l$ is denoted as $p^+$ and the component along $n$ as $p^-$.
The reference momenta for the Sudakov decomposition are defined in terms of
rescaled hard momenta,
\begin{equation}\label{eq:def_lc_momenta}
  l^\mu=\frac{p_i^\mu}{\sqrt{1-\alpha_q-\beta_q-q^2/Q^2}}
  \qquad\text{and}\qquad
  n^\mu=\frac{p_j^\mu}{\sqrt{1-\alpha_q-\beta_q-q^2/Q^2}}\;,
\end{equation}
where $q=p_1$ in configurations with one, and $q=p_1+p_2$ in configurations with
two soft gluons, and where $Q^2=(p_i+p_j+q)^2$. This implies in particular that
$2ln=Q^2$, irrespective of the number of gluons in the final state, and that
$0<\alpha,\beta<1$ for any of the final-state momenta. We parametrize the integrations
over the soft momenta $p_1$ and $p_2$ as follows~\cite{Monni:2011gb}
\begin{equation}
  d^Dp=\frac{1}{2}\,dp^+dp^-d^{D-2}p_T
  =\frac{Q^2}{2}\,d\alpha_p d\beta_p\,d^{D-2}p_T\;.
\end{equation}
The transverse momentum integrals can be written as
\begin{equation}\label{eq:pt_int_1p}
  \int d^{D-2}p_{T1}\,\delta^+(p_1^2)=
  \Omega(2-2\eps)\,Q^{-2\eps}\big(\alpha_1\beta_1\big)^{-\eps}\;,
\end{equation}
and
\begin{equation}\label{eq:pt_int_2p}
  \int d^{D-2}p_{T1} d^{D-2}p_{T2}\,\delta^+(p_1^2)\delta^+(p_2^2)=
  \Omega(2-2\eps)\,Q^{-4\eps}\big(\alpha_1\beta_1\,\alpha_2\beta_2\big)^{-\eps}\,
  \Omega(1-2\eps)\int_0^\pi(\sin^2\phi)^{-\eps}d\phi\;,
\end{equation}
where $\Omega(n)=2\pi^{n/2}/\Gamma(n/2)$ and where we have used the relation
$p^2=Q^2\alpha_p\beta_p-\mr{p}_T^2$ to perform the integrals over the magnitudes
of the transverse momenta. The remaining angular integral has to be carried out
differently for different powers of the invariant 
$s_{12} = Q^2(\alpha_1\beta_2+\alpha_2\beta_1-2\sqrt{\alpha_1\beta_1\,\alpha_2\beta_2}\cos\phi_{12})$
that appears in the expressions of the soft current.

To parametrize the measurement as well as the mapping from four- to three-particle
topologies, we introduce the observables
\begin{equation}
  \begin{split}
    \mc{F}_{ij}(1)=&\;\delta(\alpha-2np_1/Q^2)\,\delta(\beta-2lp_1/Q^2)\\
    \mc{F}_{ij}(1,2)=&\;\delta(\alpha-2np_{12}/Q^2)\,\delta(\beta-2lp_{12}/Q^2)\;.
  \end{split}
\end{equation}

\subsection{Contributions at leading and next-to-leading order}
\label{sec:nlo_contributions}
The leading order momentum space soft function is given by the integral
of Eq.~\eqref{eq:lo_eik_fc}
\begin{equation}
  \begin{split}
    S_{ij}^{\rm(1)}(q)=&\;\int \frac{d^Dp_1}{(2\pi)^{D-1}}\,\delta^+(p_1^2)\,
    \mc{S}_{ij}^{(0)}(1)\,\mc{F}_{ij}(1)\\
    =&\;\frac{\Omega(2-2\eps)}{(2\pi)^{3-2\eps}}\,\frac{Q^2}{2}\,
    (Q^2\alpha\beta)^{-\eps}\,g_s^2\mu^{2\eps}\,\frac{1}{Q^2\alpha\beta}
    =\frac{\alpha_s^0(4\pi)^\eps}{2\pi\,\Gamma(1-\eps)}\,
    \Big(\frac{\mu}{\kappa}\Big)^{2\eps}\,\frac{Q^2}{\kappa^2}\;.
  \end{split}
\end{equation}
To simplify the notation we have defined $\kappa^2=Q^2\alpha\beta$.
Next we replace the bare coupling, $\alpha_s^0$, by the renormalized one
in the $\overline{\rm MS}$ scheme,
\begin{equation}\label{eq:alphas_msbar}
  \alpha_s^0=\alpha_s(\mu)\,\frac{e^{\eps\gamma_E}}{(4\pi)^\eps}
  \left(1-\frac{1}{\eps}\frac{\alpha_s(\mu)}{2\pi}\beta_0+\mc{O}(\alpha_s^2)\right)\;,
  \qquad\text{where}\qquad
  \beta_0=\frac{11}{6}\,C_A-\frac{2}{3}\,T_R n_f\;.
\end{equation}
Thus the leading-order soft function in the dipole shower scheme reads
\begin{equation}\label{eq:lo_soft}
  S_{ij}^{\rm(0)}(q)=\frac{\bar{\alpha}_s}{2\pi}\,\frac{Q^2}{\kappa^2}\;,
  \qquad\text{where}\qquad
  \bar{\alpha}_s=\alpha_s(\mu)\,\frac{e^{\eps\gamma_E}}{\Gamma(1-\eps)}\,
  \Big(\frac{\mu}{\kappa}\Big)^{2\eps}\;.
\end{equation}
Similarly, the contribution from the virtual corrections, Eq.~\eqref{eq:virt_soft},
to the next-to-leading order soft dipole evolution is given by
\begin{equation}\label{eq:nlo_virt}
  \begin{split}
    S_{ij}^{\rm(virt)}(q)=&\;\int \frac{d^Dp_1}{(2\pi)^{D-1}}\,\delta^+(p_1^2)\,
    \mc{S}_{ij}^{\rm(virt)}(1)\,\mc{F}_{ij}(1)\\
    =&\;-\frac{\Omega(2-2\eps)}{(2\pi)^{3-2\eps}}\,\frac{Q^2}{2}\,
    (Q^2\alpha\beta)^{-\eps}\,C_A\frac{g_s^4}{8\pi^2}
    \frac{(4\pi\mu^4)^\eps}{\eps^2}\frac{\Gamma^4(1-\eps)\Gamma^3(1+\eps)}{
      \Gamma^2(1-2\eps)\Gamma(1+2\eps)}\Big(\frac{1}{Q^2\alpha\beta}\Big)^{1+\eps}\\
    =&\;-C_A\,\frac{\bar{\alpha}_s^2}{(2\pi)^2}\,\frac{1}{\eps^2}\,
    \frac{\Gamma^5(1-\eps)\Gamma^3(1+\eps)}{\Gamma^2(1-2\eps)\Gamma(1+2\eps)}\,
    \frac{Q^2}{\kappa^2}\;.
  \end{split}
\end{equation}
The calculation of the real-emission contributions is straightforward but tedious.
We discuss the details in App.~\ref{sec:real_corrections}. The contribution
from the strong ordering approximation, Eq.~\eqref{eq:soft_so}, reads
\begin{equation}\label{eq:soft_so_int}
  \begin{split}
    S_{ij}^{\rm(so)}(q)=&\;
    \frac{\bar{\alpha}_s^2}{(2\pi)^2}\,\frac{Q^2}{\kappa^2}
    \left(\frac{1}{\eps^2}-\frac{2}{3}\,\pi^2-14\,\eps\,\zeta_3+\mc{O}(\eps^2)\right)\;,\\
  \end{split}
\end{equation}
The contributions from the soft remainder and the collinear terms,
Eq.~\eqref{eq:soft_helper_funcs}, are given by
\begin{equation}\label{eq:soft_rem_coll}
  \begin{split}
    S_{ij}^{\rm(rem)}(q)
    =&\;\frac{\bar{\alpha}_s^2}{(2\pi)^2}\,\frac{Q^2}{\kappa^2}
    \left(-\frac{2}{\eps}-4-\frac{\pi^2}{3}
    +\eps\left(\frac{2}{3}\,\pi^2-8-10\,\zeta_3\right)+\mc{O}(\eps^2)\right)\;,\\
    S_{ij,gg}^{\rm(coll)}(q)
    =&\;\frac{\bar{\alpha}_s^2}{(2\pi)^2}\,\frac{Q^2}{\kappa^2}
    \left(\frac{5}{6\eps}+\frac{31}{18}
    +\eps\left(\frac{94}{27}-\frac{5}{18}\pi^2\right)+\mc{O}(\eps^2)\right)\;,\\
    S_{ij,q\bar{q}}^{\rm(coll)}(q)
    =&\;\frac{\bar{\alpha}_s^2}{(2\pi)^2}\,\frac{Q^2}{\kappa^2}
    \left(\frac{2}{3\eps}+\frac{10}{9}
    +\eps\left(\frac{56}{27}-\frac{2}{9}\pi^2\right)+\mc{O}(\eps^2)\right)\;.
  \end{split}
\end{equation}
In Sec.~\ref{sec:four_dimensions} we will devise a modified subtraction method
that allows to compute the coefficients of the above functions in four dimensions.
The results obtained here are used as a cross-check on the new technique.

\subsection{Complete next-to-leading order corrections}
\label{sec:nlo_corrections}
The complete Born-local next-to-leading order corrections to the soft function
in the dipole approach are given by the sum of Eqs.~\eqref{eq:nlo_virt},
\eqref{eq:soft_so_int} and~\eqref{eq:soft_rem_coll}.
The coupling renormalization, Eq.~\eqref{eq:alphas_msbar}, contributes
an additional counterterm
\begin{equation}\label{eq:alphas_ren}
  \begin{split}
    S_{ij}^{\rm(ren)}(q)=&\;-\frac{\alpha_s^2(\mu)}{(2\pi)^2}\,
    \frac{e^{\eps\gamma_E}}{\Gamma(1-\eps)}\Big(\frac{\mu}{\kappa}\Big)^{2\eps}\frac{Q^2}{\kappa^2}
    \frac{\beta_0}{\eps}\;.
  \end{split}
\end{equation}
We finally obtain the fully differential two-loop momentum space soft function
\begin{equation}\label{eq:nlo}
  \begin{split}
    S_{ij}^{\rm(2)}(q)=&\;S_{ij}^{\rm(virt)}(q)+S_{ij}^{\rm(ren)}(q)+
    C_A\left(S_{ij}^{\rm(s.o.)}(q)-\frac{S_{ij}^{\rm(rem)}(q)}{2}
      +S_{ij,gg}^{\rm(coll)}(q)\right)-T_R\,n_f\;S_{ij,qq}^{\rm(coll)}(q)\\
    =&\;\frac{\alpha_s^2(\mu)}{(2\pi)^2}\,\frac{Q^2}{\kappa^2}\,
    \frac{e^{2\eps\gamma_E}}{\Gamma(1-\eps)^2}\Big(\frac{\mu}{\kappa}\Big)^{4\eps}
    \left[\,\beta_0\left(\frac{1}{\eps}\left(1-\frac{\Gamma(1-\eps)}{e^{\eps\gamma_E}}
      \Big(\frac{\kappa}{\mu}\Big)^{2\eps}\right)-\eps\,\frac{\pi^2}{6}\right)
      +\Gamma_{\rm cusp}^{(2)}+2\,\eps\,\Gamma_{\rm soft}^{(2)}+\mc{O}(\eps^2)\,\right]\\
  \end{split}
\end{equation}
Note that Eq.~\eqref{eq:nlo} only depends on $\alpha$ and $\beta$ through $\kappa^2$,
which is a consequence of rescaling invariance in the soft limit~\cite{Gardi:2009qi,Becher:2009cu}.
The constant $\Gamma_{\rm cusp}^{(2)}$ is the well known two-loop cusp anomalous
dimension~\cite{Kodaira:1981nh,Davies:1984hs,Davies:1984sp,Catani:1988vd}
\begin{equation}\label{eq:two_loop_cusp}
  \Gamma_{\rm cusp}^{(2)}=\left(\frac{67}{18}-\frac{\pi^2}{6}\right)C_A-\frac{10}{9}\,T_R\,n_f\;,
\end{equation}
and the constant $\Gamma_{\rm soft}^{(2)}$ is the two-loop soft anomalous dimension
computed in~\cite{Belitsky:1998tc,Li:2011zp},
\begin{equation}\label{eq:two_loop_soft}
  \Gamma_{\rm soft}^{(2)}=\left(\frac{101}{27}-\frac{11}{72}\,\pi^2-\frac{7}{2}\,\zeta_3\right)C_A
  -\left(\frac{28}{27}-\frac{\pi^2}{18}\right)\,T_R\,n_f\;.
\end{equation}
Using Eq.~\eqref{eq:plus_expansion} to expand Eqs.~\eqref{eq:lo_soft} and~\eqref{eq:nlo}
about the poles in the light-cone momenta $q^+$ and $q^-$, defined according to
Eq.~\eqref{eq:def_sudakov_decomposition}, we obtain
\begin{equation}\label{eq:lo_expanded}
  \begin{split}
    S_{ij}^{\rm(1)}(q)=&\;\frac{\alpha_s(\mu)}{2\pi}\,Q^2
    \Bigg[\,L_{0,0}\left(\frac{1}{\eps^2}-\frac{\pi^2}{12}\right)
      +\frac{L_{0,1}}{\eps}+(L_{0,2}+L_{1,1})+\mc{O}(\eps)\,\Bigg]\;,
  \end{split}
\end{equation}
and
\begin{equation}\label{eq:nlo_expanded}
  \begin{split}
    S_{ij}^{\rm(2)}(q)=&\;\frac{\alpha_s^2(\mu)}{(2\pi)^2}\,Q^2
    \Bigg[\,\frac{L_{0,0}}{2}\left(-\frac{3\beta_0}{2\eps^3}
      +\frac{\Gamma_{\rm cusp}^{(2)}}{2\eps^2}+\frac{\Gamma_{\rm soft}^{(2)}}{\eps}
      -\frac{\pi^2}{12}\,\Gamma_{\rm cusp}^{(2)}+\frac{\zeta_3}{3}\beta_0\right)\\
      &+\frac{L_{0,1}}{2}\left(-\frac{\beta_0}{\eps^2}+\frac{\Gamma_{\rm cusp}^{(2)}}{\eps}
      +2\Gamma_{\rm soft}^{(2)}-\frac{\pi^2}{6}\beta_0\right)
      +\Big(L_{0,2}+L_{1,1}\Big)\Gamma_{\rm cusp}^{(2)}
      +\Big(L_{0,3}+L_{1,2}\Big)\beta_0+\mc{O}(\eps)\,\Bigg]\;.
  \end{split}
\end{equation}
In this context we have defined the functions
\begin{equation}\label{eq:def_log_expansion}
  \begin{split}
    L_{0,0}=&\;\delta(q_+)\delta(q_-)\;,\\
    L_{0,n}=&\;\frac{(-1)^n}{\mu}\left[\frac{\ln^{n-1}(q_+/\mu)}{q_+/\mu}\right]_+\delta(q_-)
    +\frac{(-1)^n}{\mu}\left[\frac{\ln^{n-1}(q_-/\mu)}{q_-/\mu}\right]_+\delta(q_+)\;,\\
    L_{n,m}=&\;\frac{(-1)^{n+m}}{1+\delta_{nm}}\bigg(\,
    \frac{1}{\mu^2}\left[\frac{\ln^{n-1}(q_+/\mu)}{q_+/\mu}\right]_+
    \left[\frac{\ln^{m-1}(q_-/\mu)}{q_-/\mu}\right]_+
    +\frac{1}{\mu^2}\left[\frac{\ln^{n-1}(q_-/\mu)}{q_-/\mu}\right]_+
    \left[\frac{\ln^{m-1}(q_+/\mu)}{q_+/\mu}\right]_+\bigg)\;.\\
  \end{split}
\end{equation}
Note that only the two terms proportional to $L_{1,n}$ in Eq.~\eqref{eq:nlo_expanded}
contribute to the differential radiation pattern as $\kappa>0$. They correspond to a
next-to-leading order K-factor modifying the soft eikonal, such that the soft-gluon
emission probability becomes
\begin{equation}\label{eq:nlo_kappa}
  S_{ij}^{\rm(2)}(q)\big|_{\kappa>0}=\frac{\alpha_s^2(\mu)}{(2\pi)^2}\,\frac{Q^2}{\kappa^2}
  \left[\,\beta_0\ln\frac{\mu^2}{\kappa^2}+\Gamma_{\rm cusp}^{(2)}+\mc{O}(\eps)\,\right]\;.
\end{equation}
In the CMW scale scheme~\cite{Catani:1990rr} the $\Gamma_{\rm cusp}^{(2)}$ contribution
is absorbed into the definition of the strong coupling as
\begin{equation}\label{eq:cmw}
  \alpha_s(\mu)\to\alpha_s(\mu)\left(1+\frac{\alpha_s(\mu)}{2\pi}\,\Gamma^{(2)}\right)\;.
\end{equation}
Upon setting $\mu_R=\kappa$ we can further eliminate the explicit $\beta_0$ term
in Eq.~\eqref{eq:nlo_kappa}~\cite{Amati:1980ch}. In this scheme, which is commonly
used in parton showers and dipole showers~\cite{Buckley:2011ms}, the Monte-Carlo simulation
correctly accounts for the effects of next-to-leading order soft QCD corrections
at the inclusive level, i.e.\ integrated over all real-emission configurations.
This approximation is valid in principle only for finite $\kappa$, whereas in the
double-soft limit additional corrections arise from the $L_{0,0}$ and $L_{0,n}$
terms in Eq.~\eqref{eq:nlo_expanded}. However, we will detail in the following
that the net effect of implementing two-loop soft corrections fully differentially
in the parton shower phase space indeed reduces to generating Eq.~\eqref{eq:nlo_kappa}
at the inclusive level, thereby confirming the findings of~\cite{Catani:1990rr}.
The connection to analytic soft-gluon resummation is established in App.~\ref{sec:resummation}.

\section{Implementation of the calculation in four dimensions}
\label{sec:four_dimensions}
A general scheme to implement higher-order corrections in parton showers in the form 
of a modified local subtraction method was suggested in~\cite{Hoche:2017iem}.
Here we proceed to work out the details of the method in the double soft limit.
Regarding the divergence structure of the full double real corrections, this is 
one of the most demanding regions due to the overlap between various singular
configurations, and it can be viewed as a part of the complete solution which
will include the simulation of higher-order corrections also in all
triple-collinear limits.

\subsection{Modified subtraction method}
\label{sec:ps_correspondence}
Our technique is based on the modified subtraction method discussed in~\cite{Frixione:2002ik}.
We identify the parton-shower splitting kernels with generalized factorization terms
in the $\overline{\rm MS}$ scheme. These terms can be computed by expanding the differential
cross section for a particular final state of interest in terms of plus distributions
corresponding to light-cone singularities along the directions of the fast partons.
Schematically, for a process with no infrared divergences at the leading order,
we can use the next-to-leading order factorization formula~\cite{Catani:1996vz} for
real-emission corrections
\begin{equation}\label{eq:cs_real}
  d\sigma_{n+1}=d\Phi^{(n)}\sum_{i<j,k}\mc{D}_{ij,k}\big(\Phi^{(n)}\big)\;,
\end{equation}
where
\begin{equation}\label{eq:cs_dipole}
  \mc{D}_{ij,k}\big(\Phi^{(n)}\big)=
  d\Phi^{(+1)}_{ij,k}\,\big|M_{n}^{ij,k}\big(\Phi^{(n)},\Phi^{(+1)}_{ij,k}\big)\big|^2\;
  \frac{\alpha_s}{2\pi}\,\frac{1}{s_{ij}}\,\hat{V}_{ij,k}\big(\Phi^{(+1)}_{ij,k}\big)\;.
\end{equation}
In this context, $|M_{n}^{ij,k}|^2$ are the color-correlated Born matrix elements for
the $n$-particle final state, and $d\Phi^{(n)}$ is the corresponding differential
phase-space element. The $\hat{V}_{ij,k}$ are the dipole insertion operators defined
in~\cite{Catani:1996vz}. They reduce to $-{\bf T}_{ij}{\bf T}_k\,\mc{S}_{ik}^{(0)}(j)$
in the soft limit, cf.\ Eq.~\eqref{eq:leading_color}. The corresponding one-emission
differential phase-space element is given by $d\Phi^{(+1)}_{ij,k}$. 
The Monte-Carlo integration of NLO real-emission corrections in four dimensions can now
be performed by subtracting Eq.~\eqref{eq:cs_real} from the real-emission corrections 
and computing only the remainder, while the subtraction terms $\mc{D}_{ij,k}$ are 
usually integrated over $d\Phi^{(+1)}_{ij,k}$ analytically to extract the poles in $\eps$.
We will instead perform these integrals in a Monte-Carlo fashion. We first parametrize
the emission phase space in the collinear limit $s_{ij}\to 0$ in terms of the virtuality
$t=s_{ij}$ and the light-cone momentum fraction $z=s_{ik}/(s_{ik}+s_{jk})$ for final state
radiation and $z=1-s_{jk}/s_{ik}$ for initial-state radiation
\begin{equation}\label{eq:coll_phasespace_ddim}
  \begin{split}
    d\Phi^{\rm(+1),F/I}_{ij,k}=\Omega(1-2\eps)\,dt\,dz\;t^{-\eps}(1-z)^{-\eps}z^{\mp\eps}(\sin^2\phi_i)^{-\eps}\;.
  \end{split}
\end{equation}
Note the sign of the exponent of the $z^{\mp\eps}$ term, which is negative for emissions
from final-state particles and positive for initial-state radiation. The integrand
in Eq.~\eqref{eq:cs_dipole} can now be expanded in in powers of the dimensional regularization
 parameter, $\eps$, using the relation
\begin{equation}\label{eq:plus_expansion}
  \frac{1}{v^{1+\eps}}=-\frac{1}{\eps}\,\delta(v)+\sum_{i=0}^\infty\frac{\eps^n}{n!}\left(\frac{\ln^n v}{v}\right)_+\;,
\end{equation}
which is applied to both the $t$ and the $z$ integral. The $1/\eps$ poles generated
in this manner will cancel against the virtual corrections and renormalization terms.
This produces a non-locality of the finite remainder which is corrected by the resummation,
as the first-order expansion of the parton-shower generates the complementary distribution
of the real-emission corrections in phase space~\cite{Frixione:2002ik}.
In order to compute the finite remainder, we simply need to compute the $\mc{O}(\eps^0)$
terms of Eq.~\eqref{eq:plus_expansion} applied to Eq.~\eqref{eq:cs_dipole}. This can be done
fully differentially in the remaining phase-space variables, however we need to take into
account that the underlying $n$-particle phase space and matrix element have an $\eps$
dependence that contributes finite terms when combined with the poles from real and
virtual corrections. This technique was used in~\cite{Hoche:2017iem} to obtain the matching
coefficients for the flavor-changing splitting functions. In the following we will describe
how it is implemented in the context of the two-loop soft corrections.

\subsection{Separation of iterated double-collinear endpoints}
\label{sec:ep_separation}
\begin{figure}[t]
  \begin{center}
  \subfigure[]{
    \includegraphics[scale=0.5]{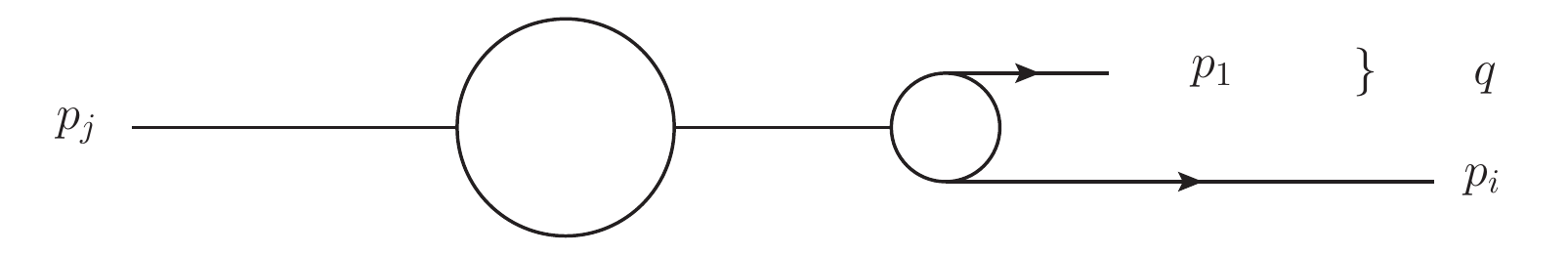}
    \label{fig:ep_separation_lo}}\hskip 5mm
  \subfigure[]{
    \includegraphics[scale=0.5]{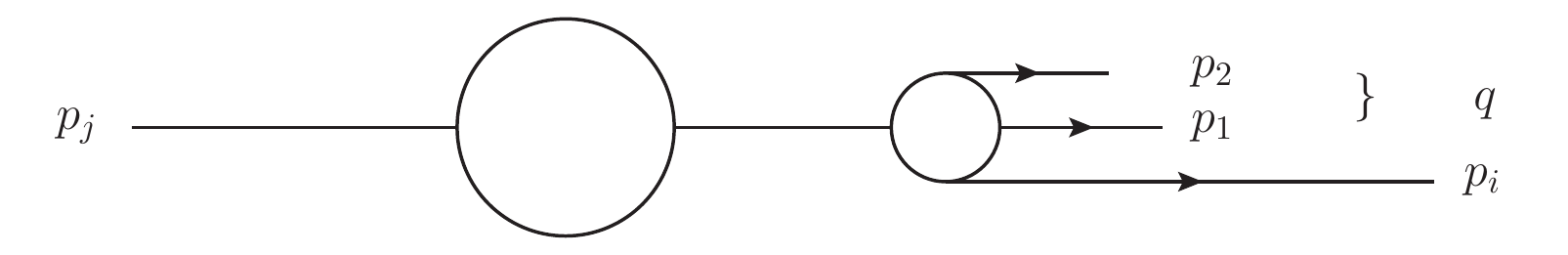}
    \label{fig:ep_separation_nlo}}
  \end{center}
  \caption{Illustration of the kinematical configurations corresponding to the
    endpoint contributions $L_{0,n}$ in Eq.~\eqref{eq:lo_nlo_expanded} in one-particle
    (left) and two-particle (right) emissions. Note in particular that all partons
    in the two-particle configuration are forced to be collinear to the same Wilson line,
    cf.\ the explanation in Sec.~\ref{sec:ep_separation}.
    \label{fig:ep_separation}}
\end{figure}
First, we must account for the fact that there is no equivalent of $L_{n,m}$
in the parton shower. The factorized plus distributions are instead replaced by a
double-plus distribution and two related endpoint terms. We define the double plus
distribution by its action on a test function
\begin{equation}\label{eq:double_plus}
  \big[f(x,y)\big]_{++}\,g(x,y)=f(x,y)\Big(g(x,y)-g(0,0)\Big)\;.
\end{equation}
Using this relation, we can write Eqs.~\eqref{eq:lo_expanded} and~\eqref{eq:nlo_expanded} as
\begin{equation}\label{eq:lo_nlo_expanded}
  \begin{split}
    S_{ij}^{\rm(1)}(q)=&\;\frac{\alpha_s(\mu)}{2\pi}\,Q^2
    \Bigg[\,\frac{L_{0,0}}{\eps^2}+\frac{L_{0,1}}{\eps}-\frac{\pi^2}{12}\,L_{0,0}
      +\Big(L_{0,2}-L_{0,1}\Big)+\tilde{L}_{1,1}
      +\mc{O}(\eps)\,\Bigg]\;,\\
    S_{ij}^{\rm(2)}(q)=&\;\frac{\alpha_s^2(\mu)}{(2\pi)^2}\,Q^2
    \Bigg[\,\frac{L_{0,0}}{2}\left(-\frac{3\beta_0}{2\eps^3}
      +\frac{\Gamma_{\rm cusp}^{(2)}}{2\eps^2}+\frac{\Gamma_{\rm soft}^{(2)}}{\eps}\right)
      +\frac{L_{0,1}}{2}\left(-\frac{\beta_0}{\eps^2}+\frac{\Gamma_{\rm cusp}^{(2)}}{\eps}\right)
      +\frac{L_{0,0}}{2}\left(-\frac{\pi^2}{12}\,\Gamma_{\rm cusp}^{(2)}+\frac{\zeta_3}{3}\beta_0\right)\\
      &+L_{0,1}\Gamma_{\rm soft}^{(2)}+\left(L_{0,3}-L_{0,2}-\frac{\pi^2}{12}\,L_{0,1}\right)\beta_0
      +\Big(L_{0,2}-L_{0,1}\Big)\,\Gamma_{\rm cusp}^{(2)}
      +\tilde{L}_{1,2}\,\beta_0+\tilde{L}_{1,1}\Gamma_{\rm cusp}^{(2)}+\mc{O}(\eps)\,\Bigg]\;,
  \end{split}
\end{equation}
where
\begin{equation}\label{eq:def_log_expansion_2}
  \begin{split}
    \tilde{L}_{n,m}=&\;\frac{(-1)^{n+m}}{1+\delta_{nm}}\,
    \frac{1}{\mu^2}\left[\frac{\mu^2}{q_+q_-}
      \left(\ln^{n-1}\frac{q_+}{\mu}\ln^{m-1}\frac{q_-}{\mu}
      +\ln^{m-1}\frac{q_+}{\mu}\ln^{n-1}\frac{q_-}{\mu}\right)\right]_{++}\;.
  \end{split}
\end{equation}
Note that because $L_{0,n}$ is located at either $q_-=0$ or $q_+=0$,
the corresponding terms are not included in a standard parton shower.
In order to add these contributions we will need to implement endpoint
terms in the (iterated) double collinear limit. The relevant kinematical
configurations are depicted in Fig.~\ref{fig:ep_separation}. They can
be explained as follows: Suppose that a single soft momentum $q$ is emitted
off the two light-like partons $i$ and $j$ as in Fig~\ref{fig:ep_separation_lo}.
As $q_-=0$ or $q_+=0$, Eqs.~\eqref{eq:def_sudakov_decomposition}
and~\eqref{eq:def_lc_momenta} imply $q\,||\,p_i$ or $q\,||\,p_j$,
hence all terms $L_{0,n}$ are related to single soft gluon radiation
in the collinear limit. In the case of double-soft radiation of momenta
$p_1$ and $p_2$, depicted in Fig.~\ref{fig:ep_separation_nlo},
the situation is similar, but slightly more involved. Because the
radiated partons are both in the final state, $s_{i1}$ and $s_{i2}$
must have the same sign. The limit $q_-=0$ then implies that
$s_{i1}+s_{i2}=0$, which can only be fulfilled if $s_{i1}=s_{i2}=0$.
Therefore, $p_i||\,p_1$ and $p_i||\,p_2$, which leads to $p_1||\,p_2$,
such that $s_{12}=0$ and $s_{i12}=0$. The conclusion is that all
$L_{0,n}$ terms correspond to the regions where soft emissions
are collinear to one of the Wilson lines. If there are two emissions,
they must be collinear to the same Wilson line.
The change in color flow generated by the soft radiation then reduces
the phase space available to subsequent gluon radiation by a factor
proportional to the light-cone momentum fraction of the gluon that is
color-adjacent to the anti-collinear Wilson line. Let us assume that
the corresponding dipole is spanned by $p_j$ and $p_1$, then the phase
space for subsequent gluon radiation is $\alpha_1Q^2$. As we are
interested in the soft gluon limit, $\alpha_1\to 0$, the remaining
phase space is typically close to zero. Any further QCD radiation
from the $j1$-dipole will be suppressed by $\alpha_1$, and radiation
from the remaining dipoles cannot occur because $s_{i1}=s_{i2}=s_{12}=0$.
It follows that the effect of the collinear configurations
corresponding to $L_{0,n}$ is to generate a radiator of reduced
invariant mass, oriented along the light-cone directions of the
original Wilson lines. The phenomenologically relevant branching probability
for such configurations cannot be determined in the double soft limit alone,
but requires in addition the computation of endpoint contributions
in the triple-collinear limit. We therefore postpone the discussion
of these terms to a forthcoming publication.

\subsection{Differential subtraction terms}
\label{sec:subtraction}
We will now derive the modified subtraction terms needed for implementing
the two-loop soft corrections in the parton-shower. Due to the non-abelian
exponentiation theorem~\cite{Cornwall:1975ty,Frenkel:1976bj} it is sufficient
to consider the gluon splitting functions and include the complete soft eikonal
instead of the partial-fractioned term, Eq.~\eqref{eq:leading_color}.
However, we must include sub-leading color configurations, corresponding
to double soft-gluon radiation off the hard Wilson lines in order to account
for coherence effects. The subtraction terms related to the functions
$\tilde{L}_{n,m}$ in Eq.~\eqref{eq:lo_nlo_expanded} can be defined as
\begin{equation}\label{eq:real_ps}
  \begin{split}
    \mc{S}_{ij}^{(q\bar{q})\rm ct}(1,2)=&\;T_R\left(\mc{S}_{ij}^{\rm(coll),2}(1,2)
      -2\,\mc{S}_{ij}^{\rm(coll),1}(1,2)\right)\;,\\
      \mc{S}_{ij}^{(gg)\rm ct}(1,2)=&\;C_A\left(\mc{S}_{ij}^{\rm(s.o.)}(1,2)
      -\frac{1}{2}\,\mc{S}_{ij}^{\rm(rem)}(1,2)-2\,\mc{S}_{ij,B}^{\rm(coll)}(1,2)
      +(1-\eps)\,\mc{S}_{ij}^{\rm(coll),1}(1,2)\right)\;.
  \end{split}
\end{equation}
where $\mc{S}_{ij}^{\rm(s.o.)}$, $\mc{S}_{ij}^{\rm(rem)}$ and
$\mc{S}_{ij}^{\rm(coll)}$ are given by Eqs.~\eqref{eq:soft_so}
and~\eqref{eq:soft_helper_funcs}, respectively.
In the collinear limit, Eqs.~\eqref{eq:real_ps} reduce to
\begin{equation}\label{eq:real_ps_coll}
  \mc{S}_{ij}^{\rm\, ct}(1,2)=g_{\mu\rho}g_{\nu\sigma}\,
  J_{ij}^{\mu}(1,2)J_{ij}^{*\nu}(1,2)\,
  \frac{P^{\rho\sigma}(z_1)}{s_{12}}\;,
\end{equation}
where $z_1$ is the light-cone momentum fraction of $p_1$ in the direction of
$p_1+p_2-s_{12}/(s_{n1}+s_{n2})n$, with $n$ an auxiliary light-like vector
satisfying $(p_1+p_2)n\neq 0$. The spin-dependent DGLAP splitting functions,
$P^{\mu\nu}(z)$, are given by
\begin{equation}\label{eq:dglap_corr}
  \begin{split}
    P_{gq}^{\mu\nu}(z)=&\;T_R\left(-g^{\mu\nu}
    +4\,z(1-z)\,\frac{k_\perp^\mu k_\perp^\nu}{k_\perp^2}\right)\;,\\
    P_{gg}^{\mu\nu}(z)=&\;2C_A\left(-g^{\mu\nu}\left(\frac{z}{1-z}+\frac{1-z}{z}\right)
    -2\,(1-\eps)z(1-z)\,\frac{k_\perp^\mu k_\perp^\nu}{k_\perp^2}\right)\;.
  \end{split}
\end{equation}
The soft gluon current $J_{ij}^\mu$ is given by the standard expression in the eikonal limit.
\begin{equation}\label{eq:eikonal_current_1}
  J_{ij}^{\mu}(q)=\frac{p_i^\mu}{2p_iq}-\frac{p_j^\mu}{2p_jq}\;.
\end{equation}
Note the minus sign in this expression, which arises from color conservation along
the hard Wilson line, i.e.\ ${\bf T}_i=-{\bf T}_j$. In processes with a non-trivial
color structure this condition holds only at leading color.
We rewrite Eq.~\eqref{eq:eikonal_current_1} such that the transversality of the current
becomes manifest:
\begin{equation}\label{eq:eikonal_current_2}
  J_{ij}^{\mu}(q)=\sqrt{\frac{p_ip_j}{2\,(p_iq)(p_jq)}}\;j_{ij,\perp}^\mu(q)\;,
  \qquad\text{where}\qquad
  j_{ij,\perp}^{\mu}(q)=\frac{(p_jq)\,p_i^\mu-(p_iq)\,p_j^\mu}{
    \sqrt{2\,(p_ip_j)(p_iq)(p_jq)}}\;.
\end{equation}
The transverse momentum in Eq.~\eqref{eq:dglap_corr} can be parametrized as
$k_\perp^\mu=j_{12,\perp}^{\mu}(n)$.
We can now prove that $\phi_{12}^{ij}$ defined in Eq.~\eqref{eq:soft_helper_funcs}
is indeed an azimuthal angle, as $\cos\phi_{12}^{ij}=k_\perp\,j_{ij,\perp}(p_1+p_2)$,
and we can replace $z_1z_2\to s_{n1}s_{n2}/(s_{n1}+s_{n2})^2$. In order to obtain
the correct differential radiation pattern in the leading-order simulation,
we implement $2\cos^2\phi_{12}^{ij}$ as a correction factor applied to the
purely collinear parts of the $g\to q\bar{q}$ and $g\to gg$ splitting functions,
see Sec.~\ref{sec:mc} for details.

The pure soft terms of Eq.~\eqref{eq:real_ps} can be rewritten as
\begin{equation}\label{eq:psct_soft_sum}
  \mc{S}_{ij}^{\rm(sct)}(1,2)=
  \frac{1}{2}\left(\mc{S}_{ij,A}^{\rm(sct)}(1,2)+
  \mc{S}_{ij}^{\rm(s.o.)}(1,2)\right)\;.
\end{equation}
where
\begin{equation}\label{eq:psct_soft}
  \mc{S}_{ij,A}^{\rm(sct)}(1,2)=
  \mc{S}_{ij}^{\rm(s.o.)}(1,2)-\mc{S}_{ij}^{\rm(rem)}(1,2)=
  \frac{s_{ij}}{(s_{i1}+s_{i2})(s_{j1}+s_{j2})}\left(
  \frac{s_{i2}}{s_{i1}s_{12}}+\frac{s_{j2}}{s_{j1}s_{12}}
  -\frac{s_{ij}}{s_{i1}s_{j1}}\right)+\Big(1\leftrightarrow 2\Big)\;.
\end{equation}
The first contribution in Eq.~\eqref{eq:psct_soft} can be interpreted as the
eikonal expression for emission of the combined soft-gluon cluster ${12}$
from the hard Wilson lines $i$ and $j$, and the subsequent radiation
of gluon $2$ off the leading-color dipoles spanned by $i1$, $j1$ or the
sub-leading color dipole spanned by $ij$. The second term describes
the same situation with the two gluons interchanged. The last term is
a negative contribution arising from the dipole spanned by $i$ and $j$.
This contribution is sub-leading in the global $1/N_c$ expansion, but it
contributes at leading color in the double-soft limit and must therefore
be included in the parton-shower simulation as the first correction
to leading-color evolution. Partial fractioning Eq.~\eqref{eq:psct_soft}
following the approach in~\cite{Catani:1996vz}, we find
\begin{equation}\label{eq:psct_soft_dec}
  \mc{S}_{ij,A}^{\rm(sct)}(1,2)=
  \mc{S}_{i,j,A}^{\rm(sct)}(1,2)
  +\Big(1\leftrightarrow 2\Big)
  +\Big(i\leftrightarrow j\Big)
  +\Big(\begin{array}{c}1\leftrightarrow 2\\
    i\leftrightarrow j\end{array}\Big)\;,
\end{equation}
where
\begin{equation}\label{eq:psct_soft_tc}
  \begin{split}
    \mc{S}_{i,j,A}^{\rm(sct)}(1,2)=&\;
    \frac{s_{ij}}{(s_{i1}+s_{i2})(s_{j1}+s_{j2})}\;
    \left[\,\frac{1}{s_{12}}\frac{s_{i2}}{s_{i1}+s_{12}}
      +\frac{1}{s_{i1}}\left(\frac{s_{i2}}{s_{i1}+s_{12}}-\frac{s_{ij}}{s_{i1}+s_{j1}}\right)
      \,\right]\;.
  \end{split}
\end{equation}
Equation~\eqref{eq:psct_soft_tc} can be interpreted as the soft enhanced part
of the dipole shower splitting function in the limit where partons $i$, 1 and 2
become triple-collinear, with parton $j$ defining the anti-collinear direction.
Note that in the $i1$-collinear limit, Eq.~\eqref{eq:psct_soft_tc} develops
an integrable singularity that vanishes upon azimuthal integration.
This problem will be discussed in Sec.~\ref{sec:mc}.
The only remaining two-particle singularity is approached as partons $1$ and $2$
become collinear.

The integrals of Eq.~\eqref{eq:psct_soft} have been computed in
Eqs.~\eqref{eq:soft_so_int} and~\eqref{eq:soft_rem}. They combine to give
\begin{equation}\label{eq:pscoll_sum}
  \begin{split}
    S_{ij,A}^{\rm(sct)}(q)
    =\frac{\bar{\alpha}_s^2}{(2\pi)^2}\,\frac{Q^2}{\kappa^2}
    \bigg[\,&\left(\frac{11}{6\eps}+\frac{67}{18}-\frac{\pi^2}{3}
      +\eps\left(\frac{202}{27}-\frac{11}{18}\pi^2-4\zeta_3\right)\right)C_A\\
      &\;-\left(\frac{2}{3\eps}+\frac{10}{9}
      +\eps\left(\frac{56}{27}-\frac{2}{9}\pi^2\right)\right)T_R n_f
      +\mc{O}(\eps^2)\,\bigg]\;.
  \end{split}
\end{equation}
Upon defining approximate virtual corrections as
\begin{equation}\label{eq:psvirt}
  \begin{split}
    \frac{\bar{\alpha}_s^2}{(2\pi)^2}\,\frac{Q^2}{\kappa^2}
    \;C_A\left(-\frac{1}{\eps^2}+\frac{\pi^2}{6}-3\,\eps\zeta_3\right)\;,
  \end{split}
\end{equation}
we would readily obtain the desired result, Eq.~\eqref{eq:nlo}, at $\mc{O}(\eps)$.
We have verified that the corresponding subtracted real-emission contribution
could be computed directly in four dimensions and cross-checked the finite
term against the difference between Eqs.~\eqref{eq:nlo} and~\eqref{eq:pscoll_sum}.
Nevertheless, $\mc{S}_{ij,A}^{\rm(sct)}$ is not a suitable local subtraction term
for Monte-Carlo simulation, because the difference to the full real-emission
correction contains integrable singularities. In the following section, we will
therefore devise a technique to simulate the complete soft subtraction term,
Eq.~\eqref{eq:psct_soft_sum}, by reweighting the leading-order parton shower.

\subsection{Monte Carlo implementation details}
\label{sec:mc}
We employ the techniques described in~\cite{Hoche:2015sya,Hoche:2017iem}
to generate the final-state momenta, and we evaluate the splitting functions
directly in terms of the kinematic invariants $s_{nm}$ with $n,m\in\{1,2,i,j\}$.
The kinematics mapping in $2\to 3$ branchings is based on~\cite{Catani:1996vz,Catani:2002hc}
and is summarized in App.~A of~\cite{Hoche:2015sya}. The kinematics mapping
and (D-dimensional) phase-space factorization in $2\to 4$ splittings was
derived in~\cite{Hoche:2017iem}, App.~A. Note that in both cases we use the
Lorentz invariant and numerically stable technique of~\cite{Hoche:2017iem}
to construct the transverse components of the momenta.

In order to simulate Eq.~\eqref{eq:psct_soft_tc} in the parton shower, we must
correct the leading-order soft radiation pattern. First we need to account
for the fact that the eikonal generated by the leading-order parton shower
is not identical to $s_{ij}/((s_{i1}+s_{i2})(s_{j1}+s_{j2}))$ if the soft-gluon
emission is followed by a subsequent branching of any of the emerging momenta.
In the transition $(\widetilde{\imath 12},\tilde{\jmath})\to(\tilde{\imath},\widetilde{12},j)$
followed by $(\widetilde{12},\tilde{\imath})\to (1,2,i)$, we obtain instead
the following probability for the emission of the final soft cluster $\widetilde{12}$
\begin{equation}\label{eq:eikonal_from_iterated_ps_12}
  \begin{split}
    \frac{\tilde{p}_ip_j}{2(\tilde{p}_i\tilde{p}_{12})(\tilde{p}_{12}p_j)}
    =&\;\bigg[\,p_ip_j\,\frac{p_ip_1+p_ip_2\pm p_1p_2}{p_ip_1+p_ip_2}\bigg]
    \left[2\bigg(p_ip_1+p_ip_2\pm p_1p_2\bigg)
      \bigg(p_jp_1+p_jp_2-\frac{p_1p_2}{p_ip_1+p_ip_2}\;p_ip_j\bigg)\right]^{-1}\\
    =&\;\frac{s_{ij}}{(s_{i1}+s_{i2})(s_{j1}+s_{j2})-s_{ij}s_{12}}\;.
  \end{split}
\end{equation}
The similarity of the kinematics mapping in final-state splittings
with a final- and initial-state spectator~\cite{Catani:1996vz,Hoche:2015sya}
implies that Eq.~\eqref{eq:eikonal_from_iterated_ps_12} holds for both final-
and initial-state Wilson lines, $i$ (corresponding to the $\pm$ sign).
Note that the term proportional to $s_{12}$ in the denominator cannot be
neglected in the double-soft limit. We can correct the mismatch between
Eq.~\eqref{eq:eikonal_from_iterated_ps_12} and the target distribution
$s_{ij}/((s_{i1}+s_{i2})(s_{j1}+s_{j2}))$ in Eq.~\eqref{eq:psct_soft}
by applying a reweighting factor in the branching of the soft gluon
$(\widetilde{12},\tilde{\imath})\to (1,2,i)$
\begin{equation}\label{eq:psct_soft_tc_12_weight}
  w_{ij}^{12}=1-\frac{s_{ij}s_{12}}{(s_{i1}+s_{i2})(s_{j1}+s_{j2})}\;.
\end{equation}
In the transition $(\widetilde{\imath 12},\tilde{\jmath})\to(\widetilde{\imath 1},\tilde{2},j)$
followed by $(\widetilde{\imath 1},\tilde{2})\to (i,1,2)$, with $i$ in the final state,
we obtain the following probability for the emission of the final soft cluster
$\widetilde{12}$
\begin{equation}\label{eq:eikonal_from_iterated_ps_i1}
  \begin{split}
    \frac{\tilde{p}_{i1}p_j}{2(\tilde{p}_{i1}\tilde{p}_2)(\tilde{p}_2p_j)}
    =&\;\bigg[\,p_ip_j+p_1p_j-\frac{p_ip_1}{p_ip_2+p_1p_2}\;p_2p_j\bigg]
    \left[2\bigg(p_ip_2+p_1p_2+p_ip_1\bigg)
      \bigg(\frac{p_ip_2+p_1p_2+p_ip_1}{p_ip_2+p_1p_2}\;p_2p_j\bigg)\right]^{-1}\\
    =&\;\frac{(s_{ij}+s_{j1})(s_{i2}+s_{12})-s_{i1}s_{j2}}{(s_{i1}+s_{i2}+s_{12})^2s_{j2}}\;.
  \end{split}
\end{equation}
If the radiator $\widetilde{\imath 1}$ is in the initial state, we obtain instead
Eq.~\eqref{eq:eikonal_from_iterated_ps_12} with $\pm\to -$. The weight factor arising
from Eq.~\eqref{eq:eikonal_from_iterated_ps_i1} generates pseudo-singularities in the
parton shower phase space, which is undesirable in a Monte-Carlo simulation.
We will therefore choose to implement a different strategy in the leading order
parton shower. The kinematics in the soft enhanced part of the $i1$-collinear emission
will be chosen according to the identified particle prescription of~\cite{Catani:1996vz}.
Note that due to our definition of the evolution and splitting variable in final-state
evolution with final-state spectator~\cite{Hoche:2015sya}, the Jacobian factor related
to this modification is unity. Eventually, all kinematical correction factors
are then given by Eq.~\eqref{eq:psct_soft_tc_12_weight}.

Upon including the phase-space correction factors, the collinear terms in the gluon splitting
functions implementing the spin correlations present in Eq.~\eqref{eq:real_ps_coll} read
\begin{equation}\label{eq:psct_soft_tcsfs}
  \begin{split}
    P_{gg,ij}^{\rm(coll)}(1,2)=&\;2C_A\,z(1-z)\,2\,w_{ij}^{12}\,\cos^2\phi_{12}^{ij}\;,\\
    P_{gq,ij}^{\rm(coll)}(1,2)=&\;-T_R\,2\,z(1-z)\,2\,w_{ij}^{12}\,\cos^2\phi_{12}^{ij}\;.
  \end{split}
\end{equation}
The remaining phase-space effects leading to $\mc{S}_{i,j,A}^{\rm(sct)}$
are taken into account by multiplying the $1$-soft parts of the $i1$- and
$12$-collinear splitting functions by $w_{ij}^{12}$.
Finally, we need to account for the additional strongly ordered term in
Eq.~\eqref{eq:psct_soft_sum}. This is achieved by means of the identities
\begin{equation}\label{eq:psct_softso_identity}
  \begin{split}
  \frac{s_{ij}}{s_{i1}s_{12}s_{j2}}=&\;
  \frac{s_{i2}}{s_{i1}s_{12}}\left[\frac{s_{ij}}{s_{i2}s_{j2}+s_{i1}s_{j1}}\right]
  +\frac{s_{j1}}{s_{j2}s_{12}}\left[\frac{s_{ij}}{s_{i2}s_{j2}+s_{i1}s_{j1}}\right]\;,\\
  \frac{s_{ij}^2}{s_{i1}s_{j1}s_{i2}s_{j2}}=&\;
  \frac{s_{ij}}{s_{i1}s_{j1}}\left[\frac{s_{ij}}{s_{i1}s_{j1}+s_{i2}s_{j2}}\right]
  +\frac{s_{ij}}{s_{i2}s_{j2}}\,\left[\frac{s_{ij}}{s_{i1}s_{j1}+s_{i2}s_{j2}}\right]\;.
  \end{split}
\end{equation}
Using Eq.~\eqref{eq:eikonal_from_iterated_ps_12} we can then write
\begin{equation}\label{eq:psct_soft_tc_full}
  \begin{split}
    \mc{S}_{i,j}^{\rm(sct)}(1,2)=&\;
    \frac{s_{ij}}{(s_{i1}+s_{i2})(s_{j1}+s_{j2})-s_{ij}s_{12}}
    \left[\,\frac{1}{s_{12}}\frac{s_{i2}}{s_{i1}+s_{12}}
      +\frac{1}{s_{i1}}\left(\frac{s_{i2}}{s_{i1}+s_{12}}
      -\frac{s_{ij}}{s_{i1}+s_{j1}}\right)\,\right]
    \frac{w_{ij}^{12}+\bar{w}_{ij}^{12}}{2}\;,
  \end{split}
\end{equation}
where we have defined the weight factor
\begin{equation}\label{eq:psct_soft_tc_so_weight}
  \bar{w}_{ij}^{12}=\frac{(s_{i1}+s_{i2})(s_{j1}+s_{j2})-s_{ij}s_{12}}{s_{i1}s_{j1}+s_{i2}s_{j2}}\;.
\end{equation}
Note that the negative contribution in Eq.~\eqref{eq:psct_soft_tc_full}
does not have a parton-shower correspondence. At the same time, we have
so far omitted the squared leading-order contribution arising from
Eq.~\eqref{eq:lo_eik_fc}. We can correct for both mismatches by adding
a subleading color contribution to the $i1$-collinear terms of the
splitting function of the Wilson lines. This term reads
\begin{equation}\label{eq:psct_soft_slc_1}
  P_{ij,A}^{\rm(slc)}(1,2)=\frac{2\,s_{ij}}{s_{i1}+s_{j1}}\,
  \frac{w_{ij}^{12}+\bar{w}_{ij}^{12}}{2}
  \Big(\bar{C}_{ij}-C_A\Big)\;
  \qquad\text{where}\qquad
  \bar{C}_{ij}=\left\{\begin{array}{cc}
  2C_F&\text{if $i$ \& $j$ quarks}\\
  C_A&\text{else}\end{array}\right.\;.
\end{equation}
The weight factor of the $\bar{C}_{ij}$ term in Eq.~\eqref{eq:psct_soft_slc_1}
was derived by considering its diagrammatic representation, which arises from
the Abelian parts of Figs.~\ref{fig:nlo_box1} and~\subref{fig:nlo_box2}~\cite{Monni:2011gb}.
We may also consider the result of the integration in Sec.~\ref{sec:analytic}
and its Fourier transform in impact parameter space, cf.\ \ App.~\ref{sec:resummation}.
In fact, Eqs.~\eqref{eq:soft_so_3} and~\eqref{eq:soft_rem_3} generate the 
exact same result up to $\mc{O}(1)$ as the square of the leading-order term,
Eq.~\eqref{eq:lo_b}, hence proving that Eq.~\eqref{eq:psct_soft_slc_1} is a
valid form in the double-soft region that will allow us to reproduce the 
squared leading-order term at the integrated level. In our numerical implementation
we include Eq.~\eqref{eq:psct_soft_slc_1} in the $i1$-collinear sector with spectator
$2$. This means that we mis-identify in principle the related evolution variable,
which should be $\kappa^2$ in the notation of~\cite{Hoche:2015sya}, and hence 
proportional to $s_{i1}s_{1j}$ instead of $s_{i1}s_{12}$. We correct for this effect
by reweighting with a ratio of strong couplings, taken at the current vs.\ the correct
evolution variable, and by setting Eq.~\eqref{eq:psct_soft_slc_1} to zero as the
evolution variable falls below the parton-shower cutoff. 
A second sub-leading color contribution is given by the difference
\begin{equation}\label{eq:psct_soft_slc_2}
  P_{ij,B}^{\rm(slc)}(1,2)=\frac{2\,s_{i2}}{s_{i1}+s_{12}}\,
  \frac{w_{ij}^{12}+\bar{w}_{ij}^{12}}{2}\,\Big(C_A-\bar{C}_{ij}\Big)\;.
\end{equation}
It accounts for the fact that the second soft emission off the Wilson lines
occurs with the color charge $C_A$ due to the interference with a color octet.
We can now define the combined sub-leading color contribution to the
parton-shower evolution as
\begin{equation}\label{eq:psct_soft_slc}
  P_{ij}^{\rm(slc)}(1,2)=
  P_{ij,A}^{\rm(slc)}(1,2)+P_{ij,B}^{\rm(slc)}(1,2)=
  \Big(C_A-\bar{C}_{ij}\Big)
  \left(\frac{2\,s_{i2}}{s_{i1}+s_{12}}
  -\frac{2\,s_{ij}}{s_{i1}+s_{j1}}\right)\,
  \frac{w_{ij}^{12}+\bar{w}_{ij}^{12}}{2}\;.
\end{equation}
Note that $P_{ij}^{\rm(slc)}$ vanishes in the $i1$-collinear limit,
such that the correct color factor is recovered in collinear evolution.
The remaining parts of the improved leading-order, fully differential
splitting functions related to the $12$-collinear, $1$-soft final-state
singularities are given by the leading-color expressions
\begin{equation}\label{eq:final_ps_kernels}
  \begin{split}
    (P_{qq})_2^k(1,i)=&\;C_F\left(\frac{2\,s_{i2}}{s_{i1}+s_{12}}
    \frac{w_{2k}^{1i}+\bar{w}_{2k}^{1i}}{2}\right)\;,\\
    (P_{gg})_{ij}(1,2)=&\;
    C_A\,\left(\frac{2\,s_{i2}}{s_{i1}+s_{12}}\frac{w_{ij}^{12}+\bar{w}_{ij}^{12}}{2}
    +w_{ij}^{12}\left(-1+z(1-z)\,2\cos^2\phi_{12}^{ij}\right)\right)\;,\\
    (P_{gq})_{ij}(1,2)=&\;T_R\,w_{ij}^{12}\left(1-4z(1-z)\cos^2\phi_{12}^{ij}\right)\;.
  \end{split}
\end{equation}
Note that we omitted the collinear parts of the splitting functions related to the
Wilson lines $i$ and $j$, as these are unchanged by the double-soft corrections.
The notation is such that the subscripts indicate the partons which are color-adjacent
to the splitting products, while the superscript indicates one that is color-adjacent
to the adjacent parton. In particular, the color connection in $(P_{qq})_{2}^{k}(1,i)$
would be $2\leftrightarrow 1\leftrightarrow i \leftrightarrow k$.
Note that the ordering of the arguments and lower indices is important. This is apparent
in the $12$-collinear limit, where the gluon-to-gluon kernel receives a second contribution,
$(P_{gg})_{ji}^{\,l}(2,1)$, which is related to a different evolution variable in the
dipole shower approach because it corresponds to the $12$-collinear, $2$-soft 
singularity~\cite{Hoche:2015sya}. The leading-order $g\to gg$ splitting function
in the collinear limit is recovered only upon adding $(P_{gg})_{ij}^k(1,2)$ and
$(P_{gg})_{ji}^{\,l}(2,1)$. The pure collinear term in $(P_{qq})_i^k(1,2)$ could
in principle be modified by the weight, Eq.~\eqref{eq:psct_soft_tc_12_weight}, but there
is no indication, based on the double-soft limit, as to whether this would constitute an
improvement of the parton shower or not. We postpone the analysis of this term
to a future publication.

We emphasize that, despite the reweighting of the leading-order parton shower to the
full real-emission pattern of the double-soft limit, a hard correction remains to be
computed using the techniques of~\cite{Hoche:2017iem}. This correction arises
because the leading-order parton shower does not fill the complete two-emission
phase space, see for example~\cite{Fischer:2017yja}. The correction is given by
\begin{equation}\label{eq:subtracted_real}
  \begin{split}
    &\tilde{\mc{S}}_{ij}^{(gg)}(1,2)=
    \frac{2\,s_{ij}}{(s_{i1}+s_{i2})(s_{j1}+s_{j2})-s_{ij}s_{12}}\,
    \left[\,(P_{gg})_{ij}(1,2)+(P_{ii})_{ij}(1,2)+P_{ii}^{\rm(slc)}(1,2)\,\right]\\
    &\qquad\times\left[1-
      \Theta\left(\frac{(s_{i1}+s_{i2})(s_{j1}+s_{j2})-s_{ij}s_{12}}{S_{ij,12}}
      -\frac{s_{i1}s_{12}}{S_{i,12}}\right)\right]
    \Theta\left(\frac{(s_{i1}+s_{i2})(s_{j1}+s_{j2})-s_{ij}s_{12}}{S_{ij,12}}-t_c\right)\\
    &\;\qquad\qquad+\Big(1\leftrightarrow 2\Big)+\Big(i\leftrightarrow j\Big)
    +\Big(\begin{array}{c}1\leftrightarrow 2\\
      i\leftrightarrow j\end{array}\Big)\;,\\
    &\tilde{\mc{S}}_{ij}^{(qq)}(1,2)=
    \frac{2\,s_{ij}}{(s_{i1}+s_{i2})(s_{j1}+s_{j2})-s_{ij}s_{12}}\;
    (P_{gq})_{ij}(1,2)\\
    &\qquad\times\left[1-
      \Theta\left(\frac{(s_{i1}+s_{i2})(s_{j1}+s_{j2})-s_{ij}s_{12}}{S_{ij,12}}
      -\frac{s_{i1}s_{12}}{S_{i,12}}\right)\right]
    \Theta\left(\frac{(s_{i1}+s_{i2})(s_{j1}+s_{j2})-s_{ij}s_{12}}{S_{ij,12}}-t_c\right)\\
    &\;\qquad\qquad+\Big(1\leftrightarrow 2\Big)+\Big(i\leftrightarrow j\Big)
    +\Big(\begin{array}{c}1\leftrightarrow 2\\
      i\leftrightarrow j\end{array}\Big)\;,
  \end{split}
\end{equation}
where $t_c$ is the infrared cutoff of the parton shower.
The two terms in the $\Theta$-function correspond to the ordering variables in the first
and second emission, respectively. To simplify the notation we have defined $S_{i,12}$,
which is given as $s_{i12}$ in the case of final-state Wilson lines and $2p_i(p_1+p_2)$ 
in the case of initial-state Wilson lines. Correspondingly, $S_{ij,q}$ is given by
$Q^2$ for two final-state Wilson lines, $2p_ip_j$ for two initial-state Wilson lines,
and $2p_i(p_j+q)$ if $i$ is in the initial, and $j$ is in the final state~\cite{Hoche:2015sya}.

Following the discussion in Sec.~\ref{sec:ps_correspondence}, the $\mc{O}(1)$
remainder in Eqs.~\eqref{eq:lo_nlo_expanded} is implemented as an endpoint
contribution at $s_{12}=0$ and $s_{i1}=0$, $s_{i2}=0$, $s_{j1}=0$, $s_{j2}=0$
for all $\kappa>0$. This allows us to simulate the radiation pattern fully differentially
at the next-to-leading order. The related endpoint terms are given by
\begin{equation}\label{eq:cusp_diff_endpoints}
  \begin{split}
    \tilde{\mc{S}}_{gq,ij}^{\rm(cusp)}(1,2)=&\;\delta(s_{12})\,
    \frac{2\,s_{ij}}{s_{i12}s_{j12}}\,T_R
    \Big(2z(1-z)+\big(1-2z(1-z)\big)\ln(z(1-z))\Big)\;,\\
    \tilde{\mc{S}}_{gg,ij}^{\rm(cusp)}(1,2)=&\;\delta(s_{12})\,
    \frac{2\,s_{ij}}{s_{i12}s_{j12}}\,2C_A\left(\frac{\ln z}{1-z}+\frac{\ln(1-z)}{z}
    +\big(-2+z(1-z)\big)\ln(z(1-z))\right)\;,\\
    \tilde{\mc{S}}_{wl,ij}^{\rm(cusp)}(1,2)=&\;-\delta(s_{i1})\,
    \frac{1}{2}\,
    \frac{C_A}{2}\,\frac{2\,s_{ij}}{s_{i12}s_{j12}}\,
    \left(\frac{\ln z_i}{1-z_i}+\frac{\ln (1-z_i)}{z_i}\right)
    +\Big(1\leftrightarrow 2\Big)
    +\Big(i\leftrightarrow j\Big)
    +\Big(\begin{array}{c}1\leftrightarrow 2\\
      i\leftrightarrow j\end{array}\Big)\;.
  \end{split}
\end{equation}
The factor $1/2$ in $\tilde{\mc{S}}_{gq,ij}^{\rm(cusp)}$ removes the double counting of
soft-collinear regions when swapping the role of $i$ and $j$. It would in principle
be desirable to work with partial fractions of the eikonals $s_{ij}/(s_{i1}s_{j1})$ and
$s_{ij}/(s_{i2}s_{j2})$. However, these partial fractions cannot be defined
unambiguously in the exact limits $s_{i1}\to 0$, $s_{j1}\to 0$, $s_{i2}\to 0$, $s_{j2}\to 0$.
One possible solution would be to introduce an additional rapidity regulator,
similar to~\cite{Curci:1980uw} or~\cite{Chiu:2011qc,Li:2016axz}.
We leave the investigation of this possibility to future work.
We implement the contributions proportional to the beta function as a
double endpoint which contributes an additional term to the soft enhanced
parts of the leading order splitting functions.~\footnote{We could in principle
  implement the terms proportional to $\beta_0$ in the same manner as
  Eq.~\eqref{eq:cusp_diff_endpoints} by splitting them into real and virtual
  contributions, corresponding to uncanceled infrared and ultraviolet singularities.
  When $\mu\approx\kappa$, the impact on the Monte-Carlo predictions will be minor,
  and we will therefore leave the investigation of this possibility to future work.}
\begin{equation}
  \begin{split}
    \tilde{\mc{S}}_{gg,ij}^{\rm(coll)}(q)=&\;\delta(q^2)\,
    \frac{2\,s_{ij}}{s_{iq}s_{jq}}\,\ln\frac{\mu^2s_{ij}}{s_{iq}s_{jq}}\;\beta_0\;.
  \end{split}
\end{equation}

\section{Numerical results}
\label{sec:results}
In this section we present numerical cross-checks of our algorithm, and we compare the
magnitude of the corrections generated by the double-soft splitting functions to the
leading-order parton shower result in the CMW scheme~\cite{Catani:1990rr}.
We restrict the analysis to pure final-state evolution, but we stress that the
formulae relevant to initial-state evolution have also been presented in Sec.~\ref{sec:mc}.
We have implemented our algorithm into the \Dire parton showers, which implies
two entirely independent realizations within the general purpose event generation
frameworks \Pythia~\cite{Sjostrand:1985xi,Sjostrand:2014zea} and
\Sherpa~\cite{Gleisberg:2003xi,Gleisberg:2008ta} that are cross-checked point by point
and in the full simulation at high statistical precision.
We use the strong coupling according to the CT10nlo PDF set~\cite{Lai:2010vv}.
The process under investigation is $e^+e^-\to$hadrons at LEP I energy (91.2~GeV).
We choose to exemplify the effects of the double-soft corrections using the
$k_T$ jet rates $y_{23}$ and $y_{34}$ in the Durham algorithm~\cite{Catani:1991hj}
and the angle $\alpha_{34}$ between the two softest jets~\cite{Abreu:1990ce}.

\begin{figure}[t]
  \subfigure{
    \begin{minipage}{0.31\textwidth}
      \begin{center}
        \includegraphics[scale=0.5]{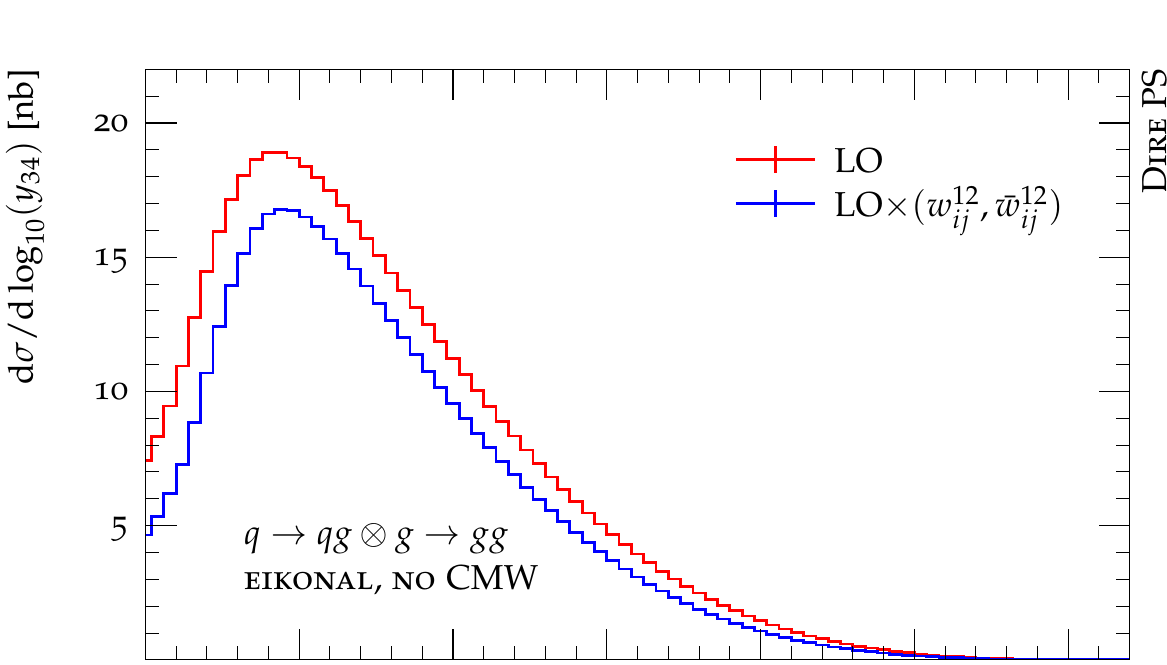}\\[-0.5mm]
        \includegraphics[scale=0.5]{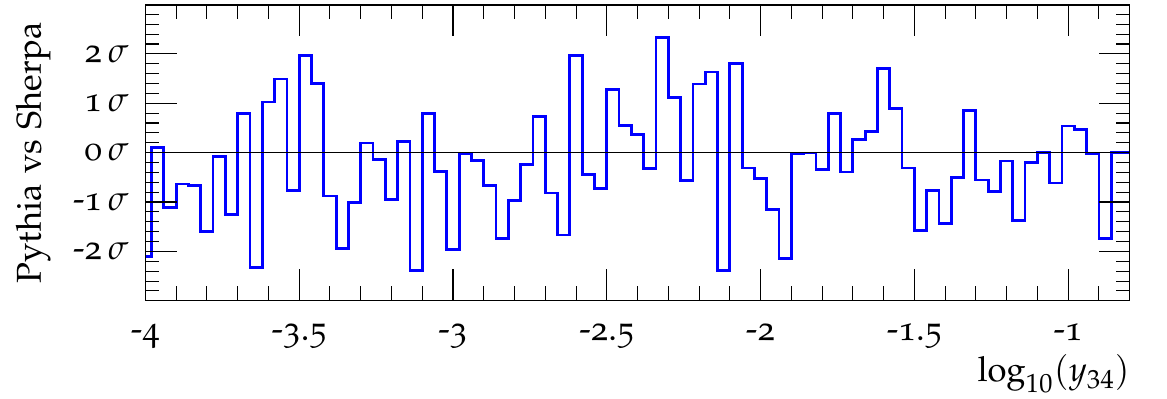}\\[2mm]
        \includegraphics[scale=0.5]{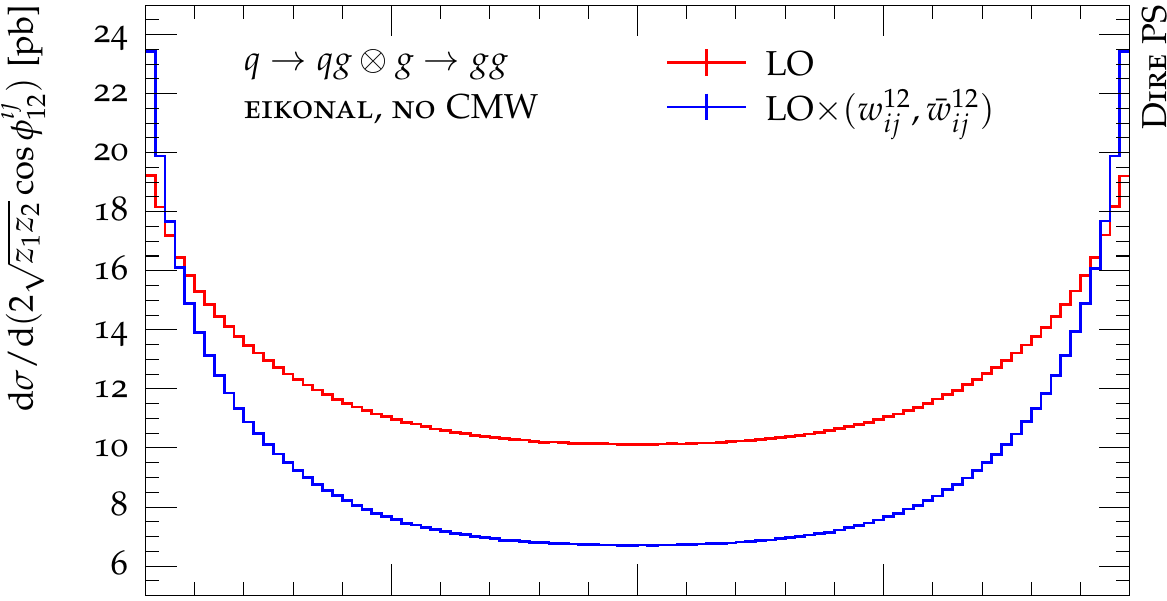}\\[-0.5mm]
        \includegraphics[scale=0.5]{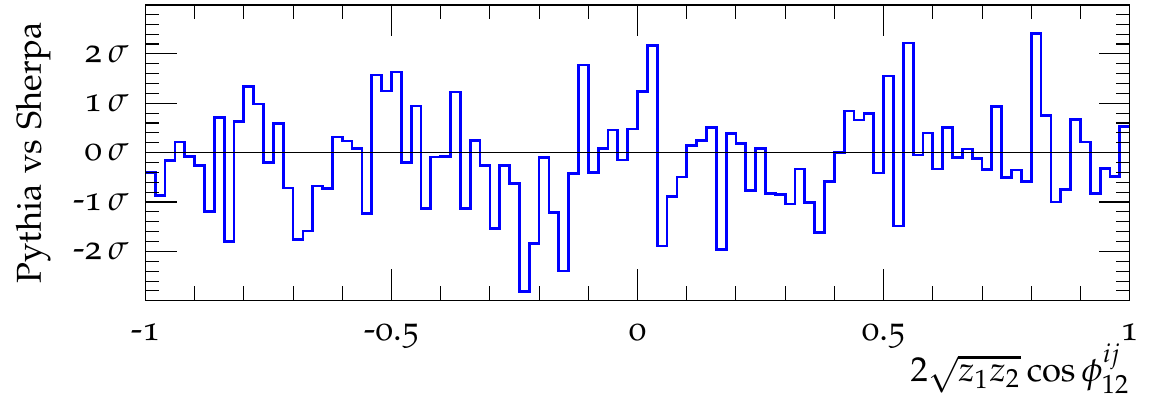}
      \end{center}
    \end{minipage}
    \label{fig:psw_validation_gg}}\hfill
  \subfigure{
    \begin{minipage}{0.31\textwidth}
      \begin{center}
        \includegraphics[scale=0.5]{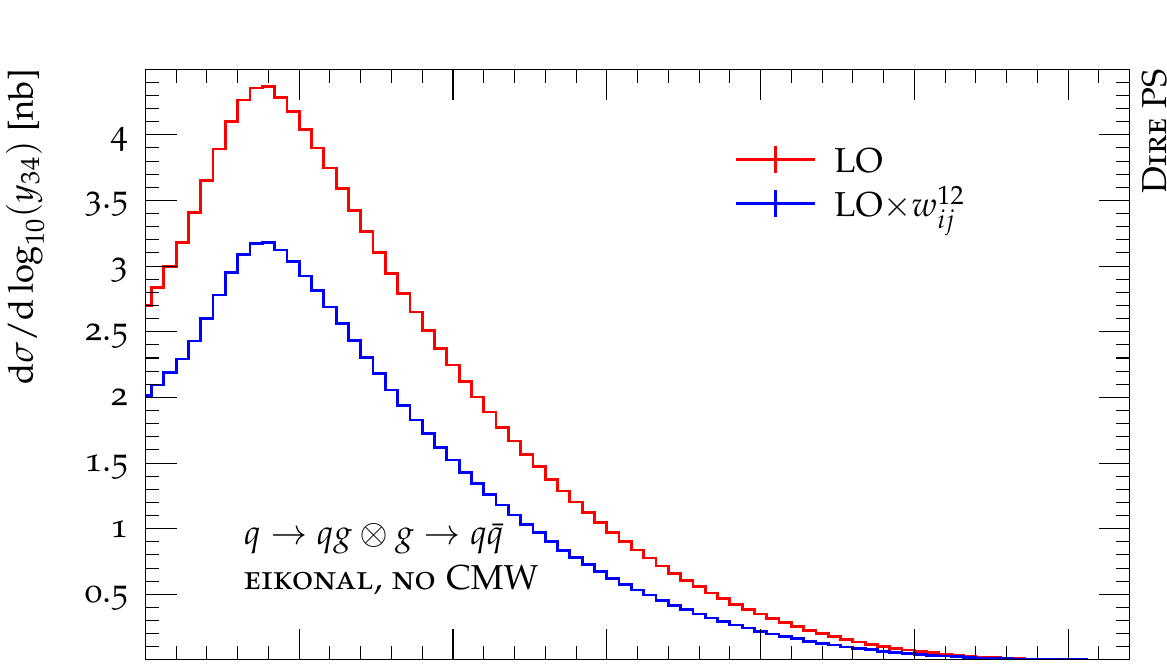}\\[-0.5mm]
        \includegraphics[scale=0.5]{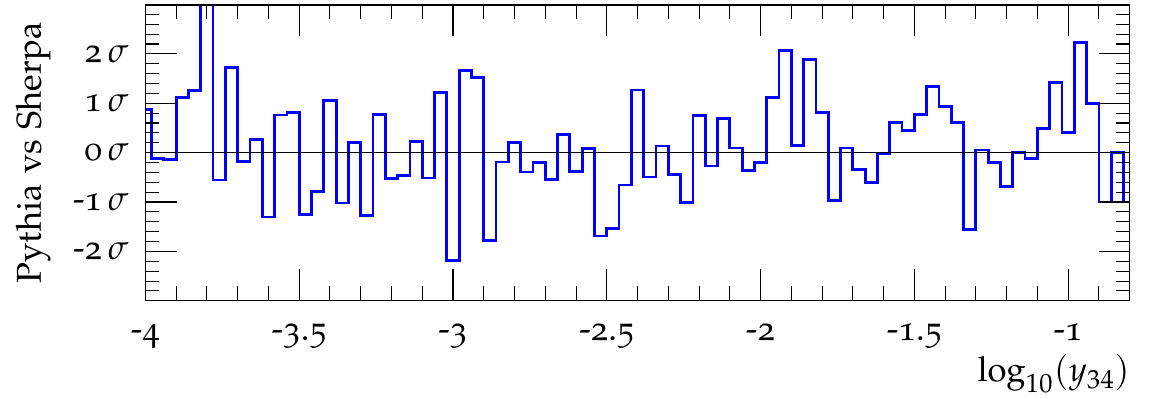}\\[2mm]
        \includegraphics[scale=0.5]{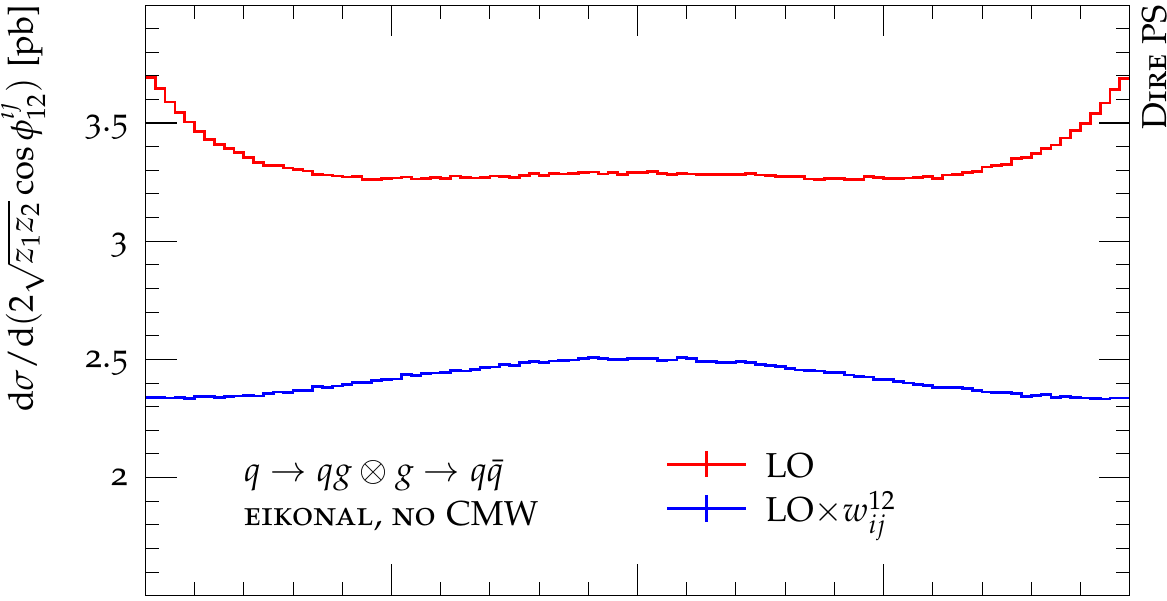}\\[-0.5mm]
        \includegraphics[scale=0.5]{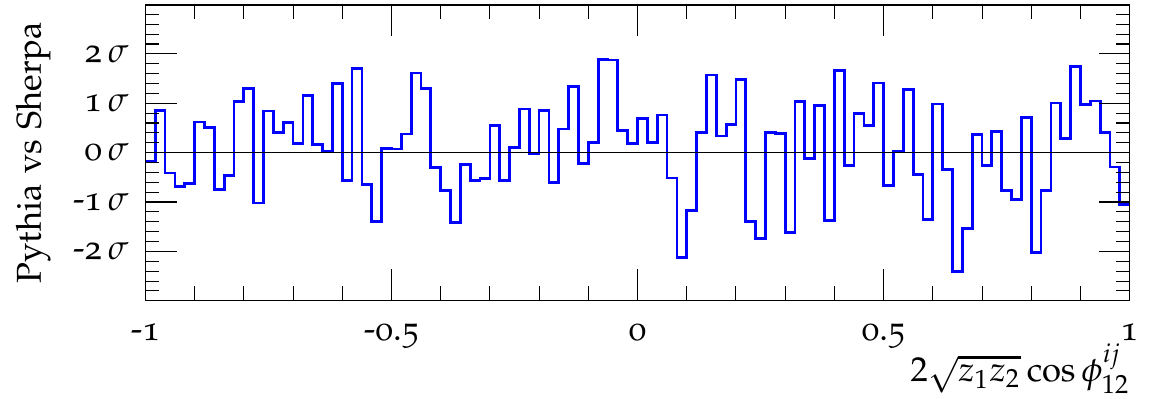}
      \end{center}
    \end{minipage}
    \label{fig:psw_validation_gq}}\hfill
  \subfigure{
    \begin{minipage}{0.31\textwidth}
      \begin{center}
        \includegraphics[scale=0.5]{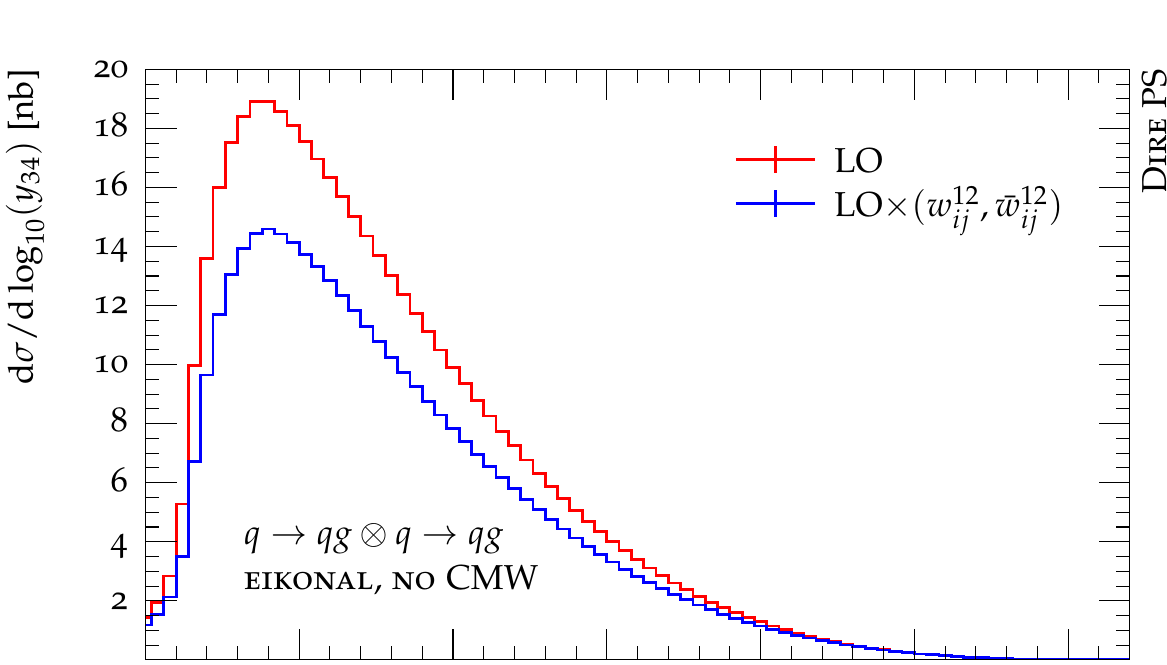}\\[-0.5mm]
        \includegraphics[scale=0.5]{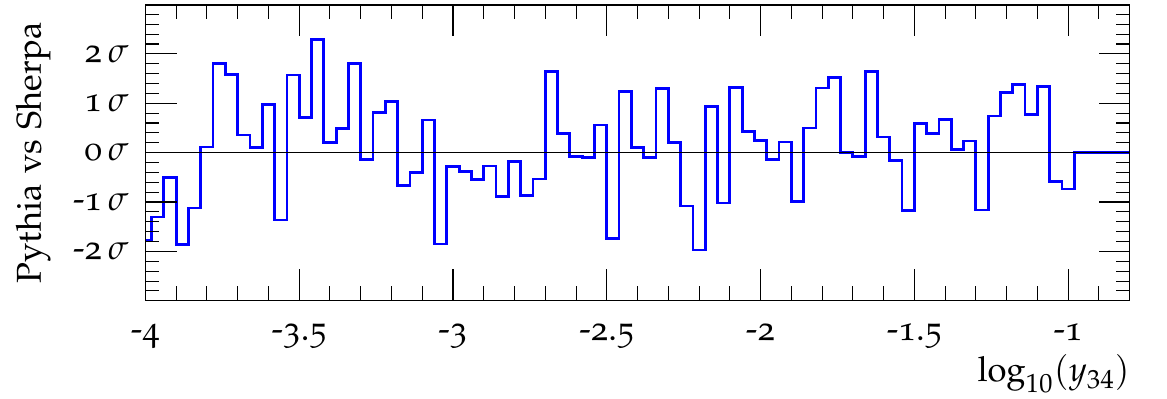}\\[2mm]
        \includegraphics[scale=0.5]{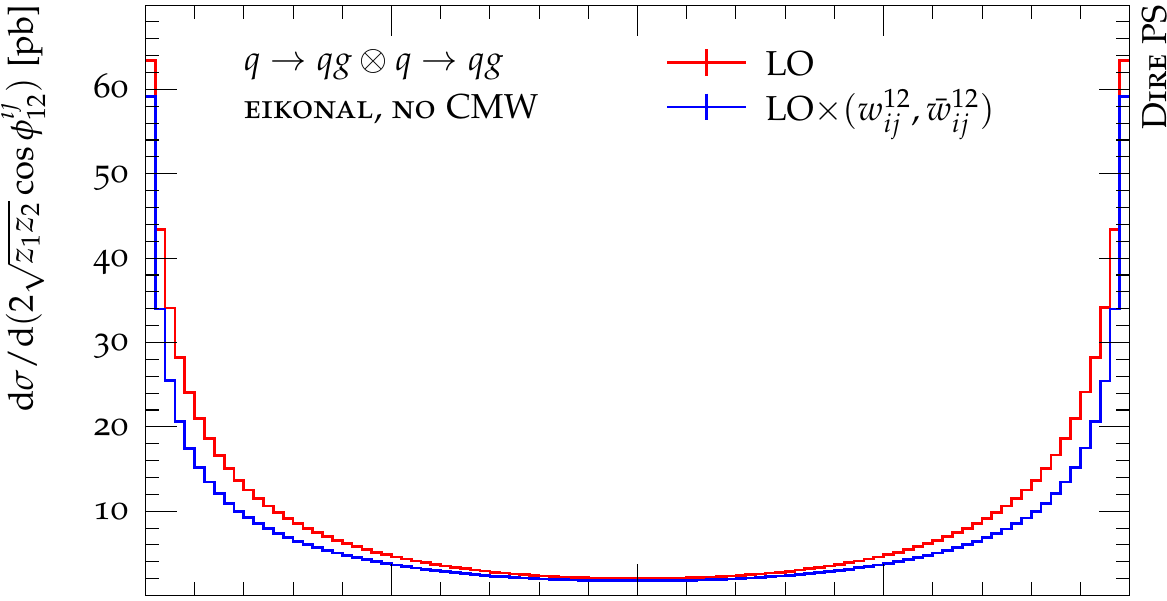}\\[-0.5mm]
        \includegraphics[scale=0.5]{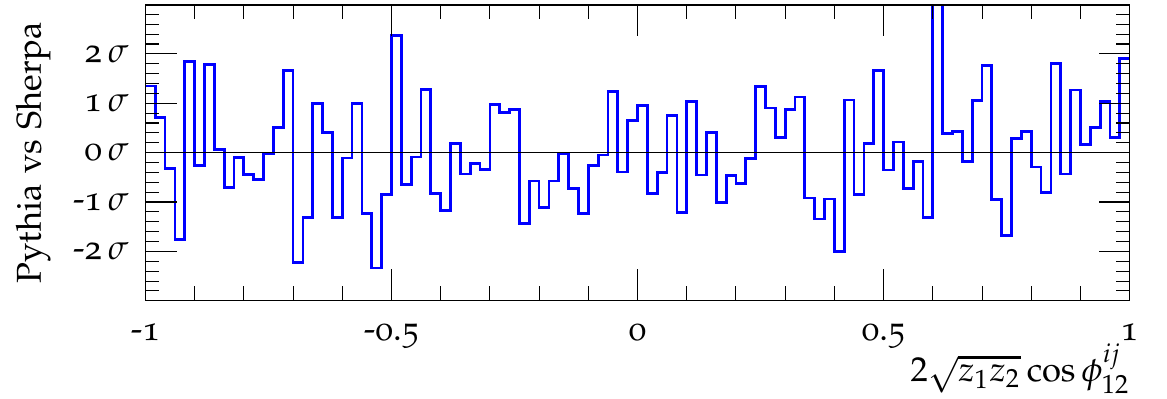}
      \end{center}
    \end{minipage}
    \label{fig:psw_validation_qq}}
  \caption{The effect of the phase-space weights $w_{ij}^{12}$,
    and $\bar{w}_{ij}^{12}$, defined in Eqs.~\eqref{eq:psct_soft_tc_12_weight}
    and~\eqref{eq:psct_soft_tc_so_weight}, on the leading-order
    parton-shower evolution, limited to two emissions.
    We show the differential jet rate $y_{34}$ in the Durham
    algorithm~\cite{Catani:1991hj} as a proxy for the rate change,
    and the angle defined in Eq.~\eqref{eq:soft_helper_funcs}
    as a proxy for the impact on differential distributions.
    The process considered is $e^+e^-\to$hadrons at LEP I energies.
    \label{fig:psw_validation}}
\end{figure}
\begin{figure}[t]
  \subfigure{
    \begin{minipage}{0.31\textwidth}
      \begin{center}
        \includegraphics[scale=0.5]{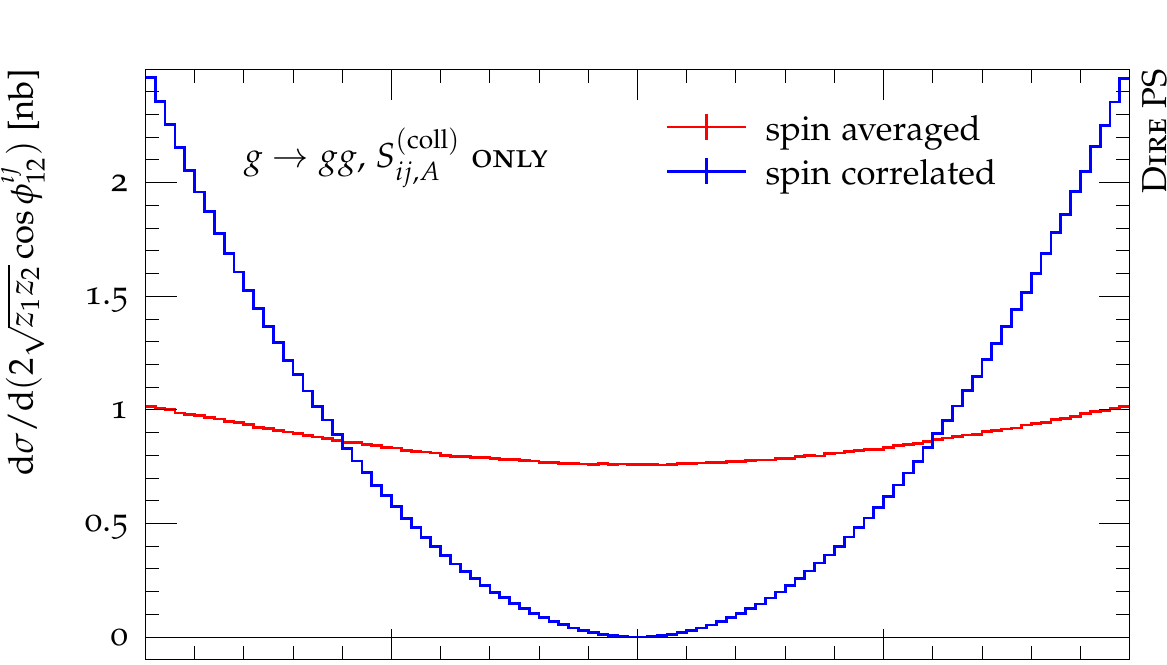}\\[-0.5mm]
        \includegraphics[scale=0.5]{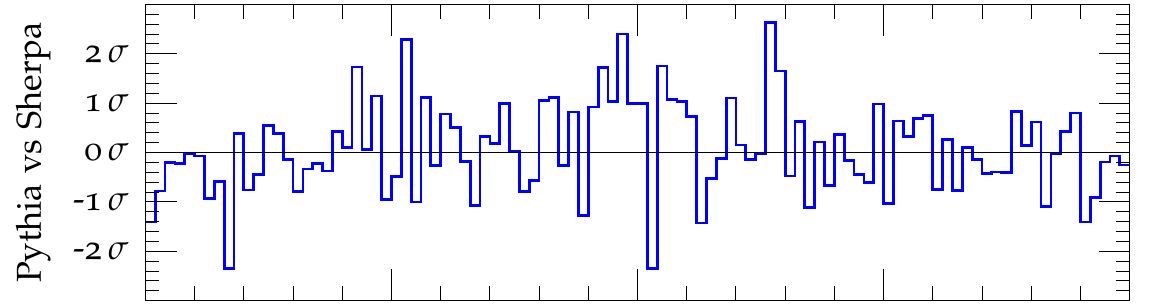}\\[-0.5mm]
        \includegraphics[scale=0.5]{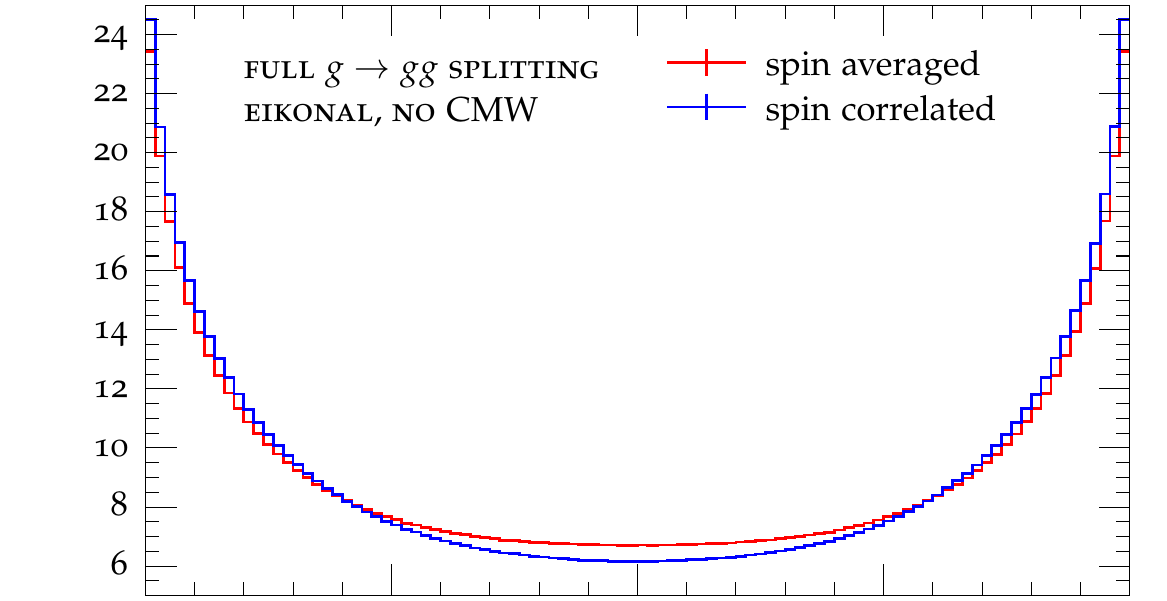}\\[-0.5mm]
        \includegraphics[scale=0.5]{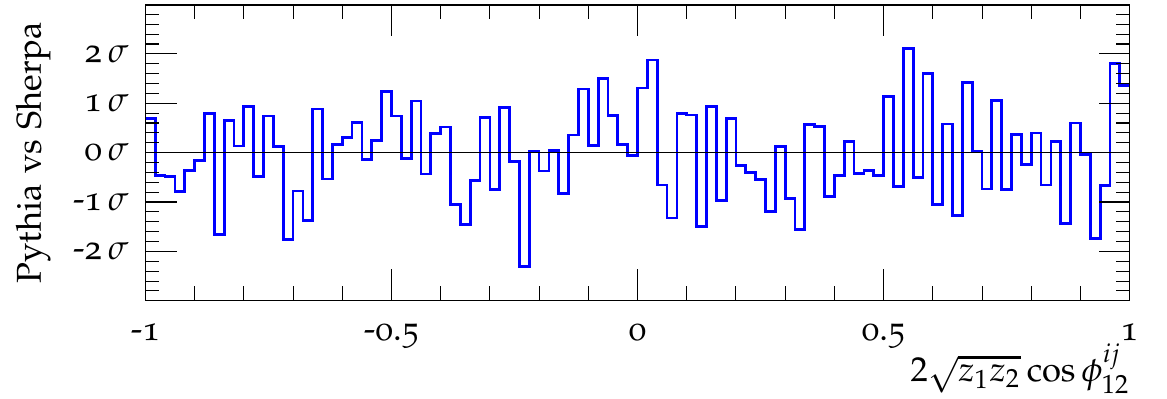}
      \end{center}
    \end{minipage}
    \label{fig:sc_validation_gg}}\hfill
  \subfigure{
    \begin{minipage}{0.31\textwidth}
      \begin{center}
        \includegraphics[scale=0.5]{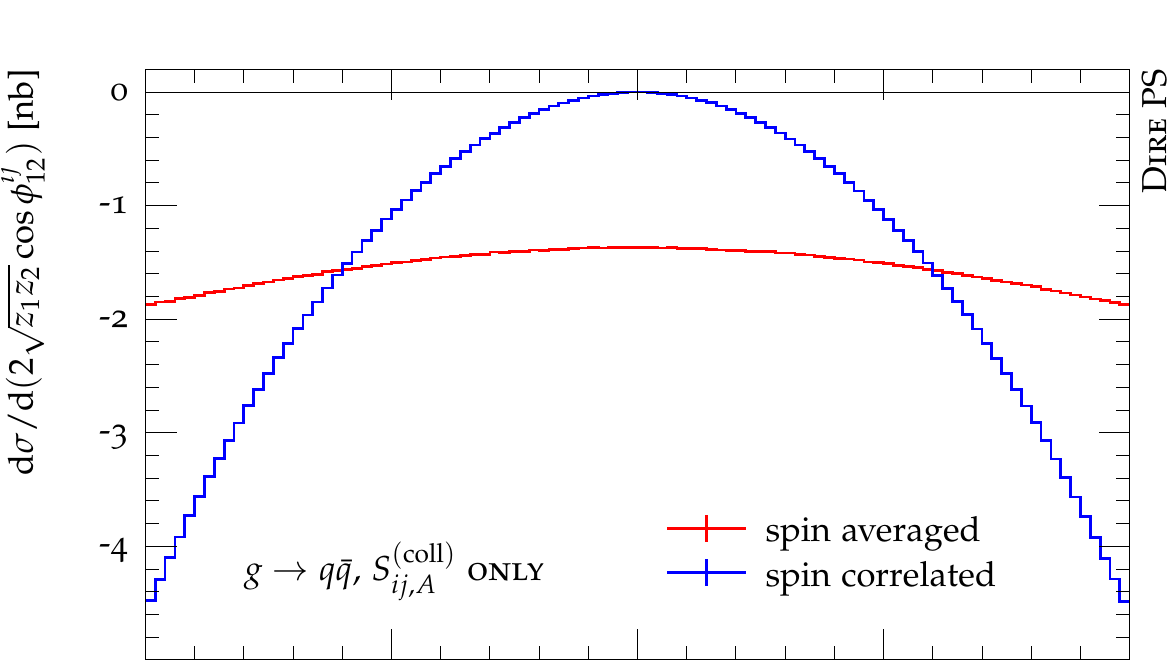}\\[-0.5mm]
        \includegraphics[scale=0.5]{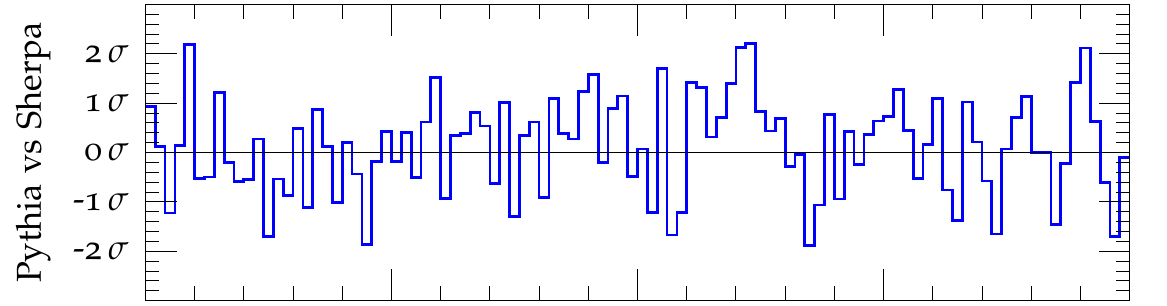}\\[-0.5mm]
        \includegraphics[scale=0.5]{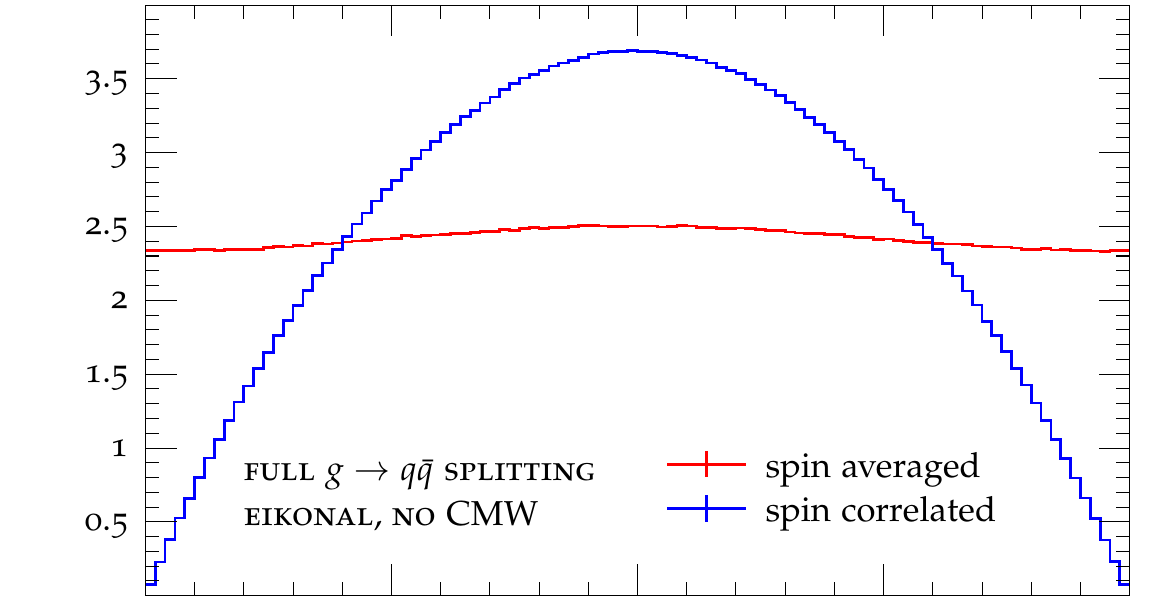}\\[-0.5mm]
        \includegraphics[scale=0.5]{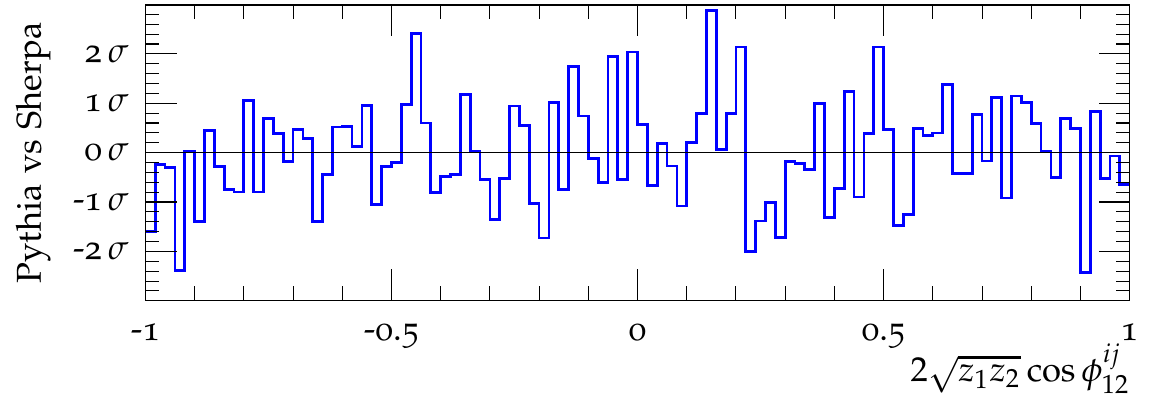}
      \end{center}
    \end{minipage}
    \label{fig:sc_validation_qq}}\hfill
  \subfigure{
    \begin{minipage}{0.31\textwidth}
      \begin{center}
        \includegraphics[scale=0.5]{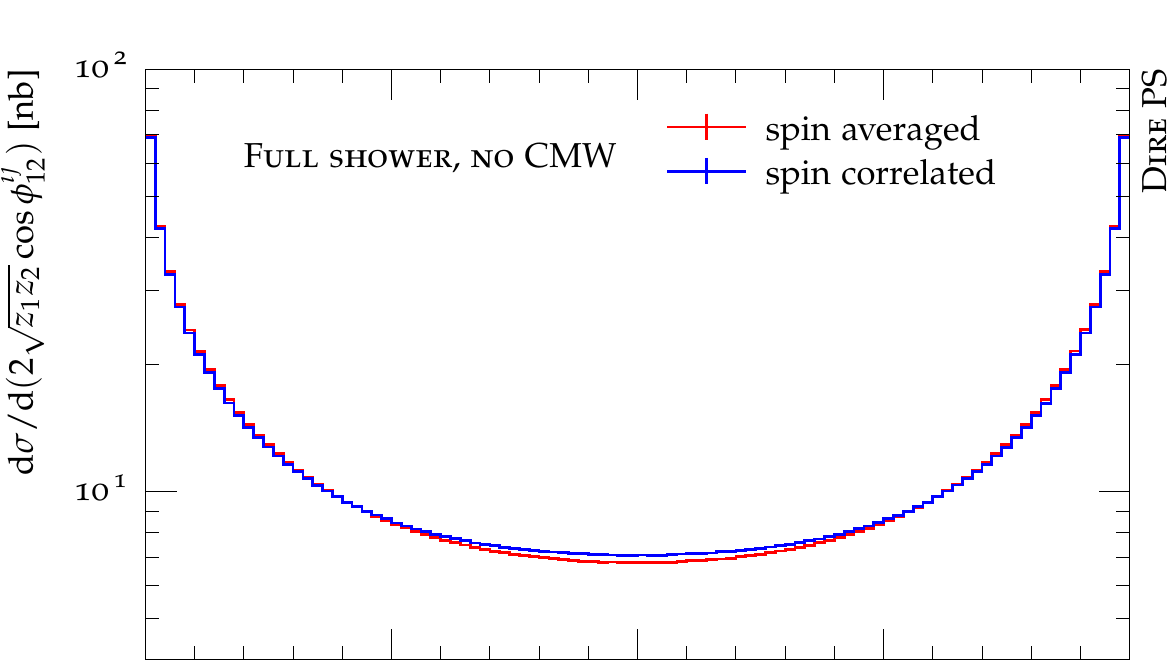}\\[-0.5mm]
        \includegraphics[scale=0.5]{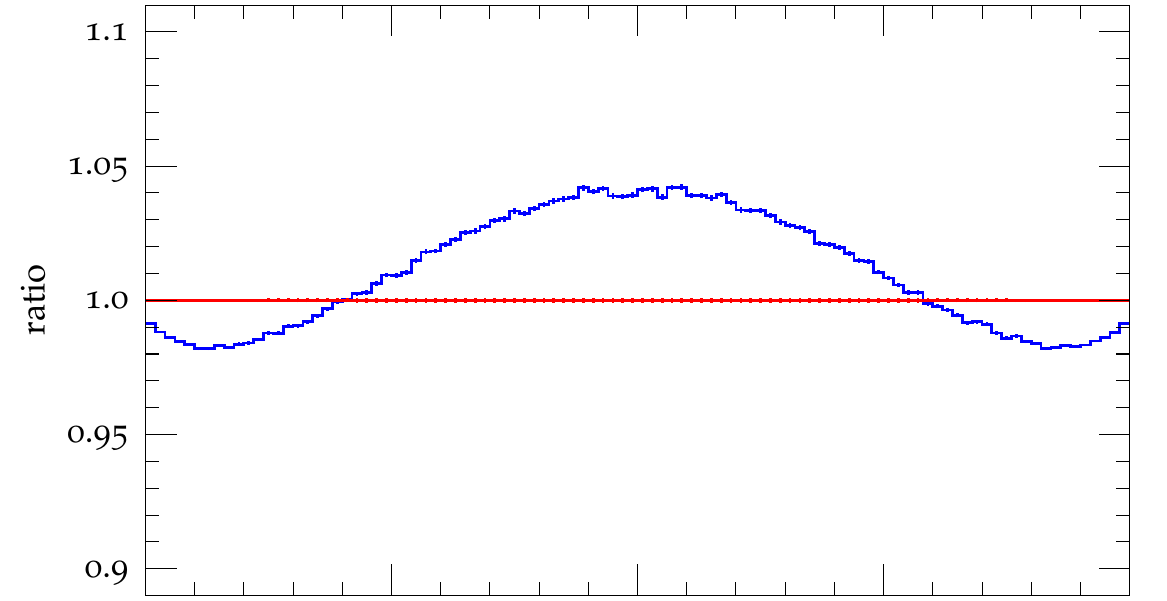}\\[-0.5mm]
        \includegraphics[scale=0.5]{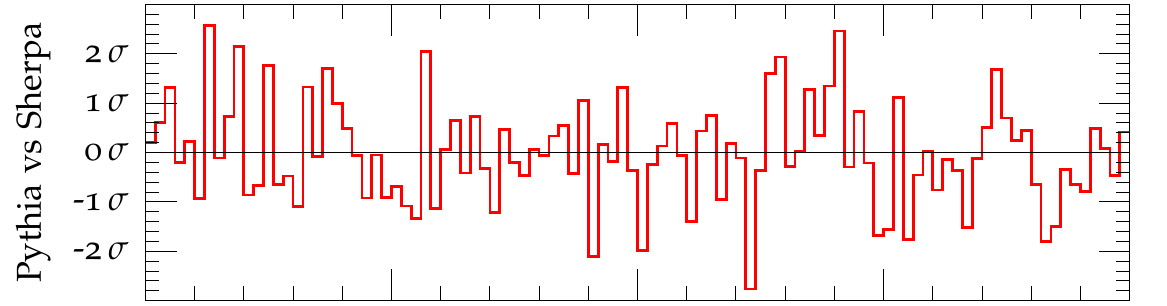}\\[-0.5mm]
        \includegraphics[scale=0.5]{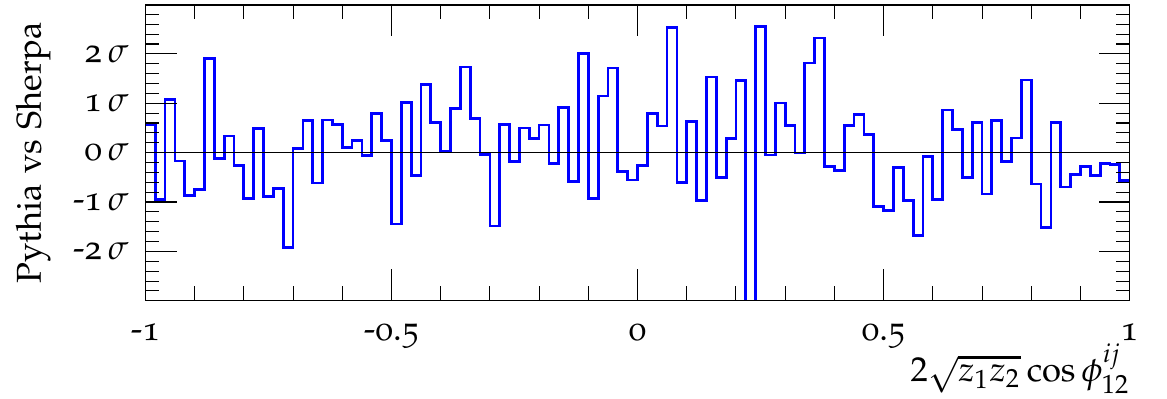}
      \end{center}
    \end{minipage}
    \label{fig:sc_validation_full}}
  \caption{The effect of spin correlations in $\mc{S}_{ij,A}^{\rm(coll)}$ compared
    to uncorrelated parton-shower evolution, including the phase-space suppression
    investigated in Fig.~\ref{fig:psw_validation}. All simulations are limited to two emissions.
    The left and middle panels show the impact on the pure $z(1-z)$ contribution
    (top panels) and on the complete $g\to gg$ and $g\to q\bar{q}$ splitting function
    (bottom panels). In both cases the production of the gluon is described by the
    eikonal part of the $q\to qg$ splitting function only. The right panel shows
    the effect of spin correlations on the complete two-emission pattern. In order
    for the results to be as similar as possible, the weight $w_{ij}^{12}$ from
    Eq.~\eqref{eq:psct_soft_tc_12_weight} is included. The process considered is
    $e^+e^-\to$hadrons at LEP I energies.
    \label{fig:sc_validation}}
\end{figure}
\begin{figure}[t]
  \begin{center}
  \subfigure{
    \begin{minipage}{0.31\textwidth}
      \begin{center}
        \includegraphics[scale=0.5]{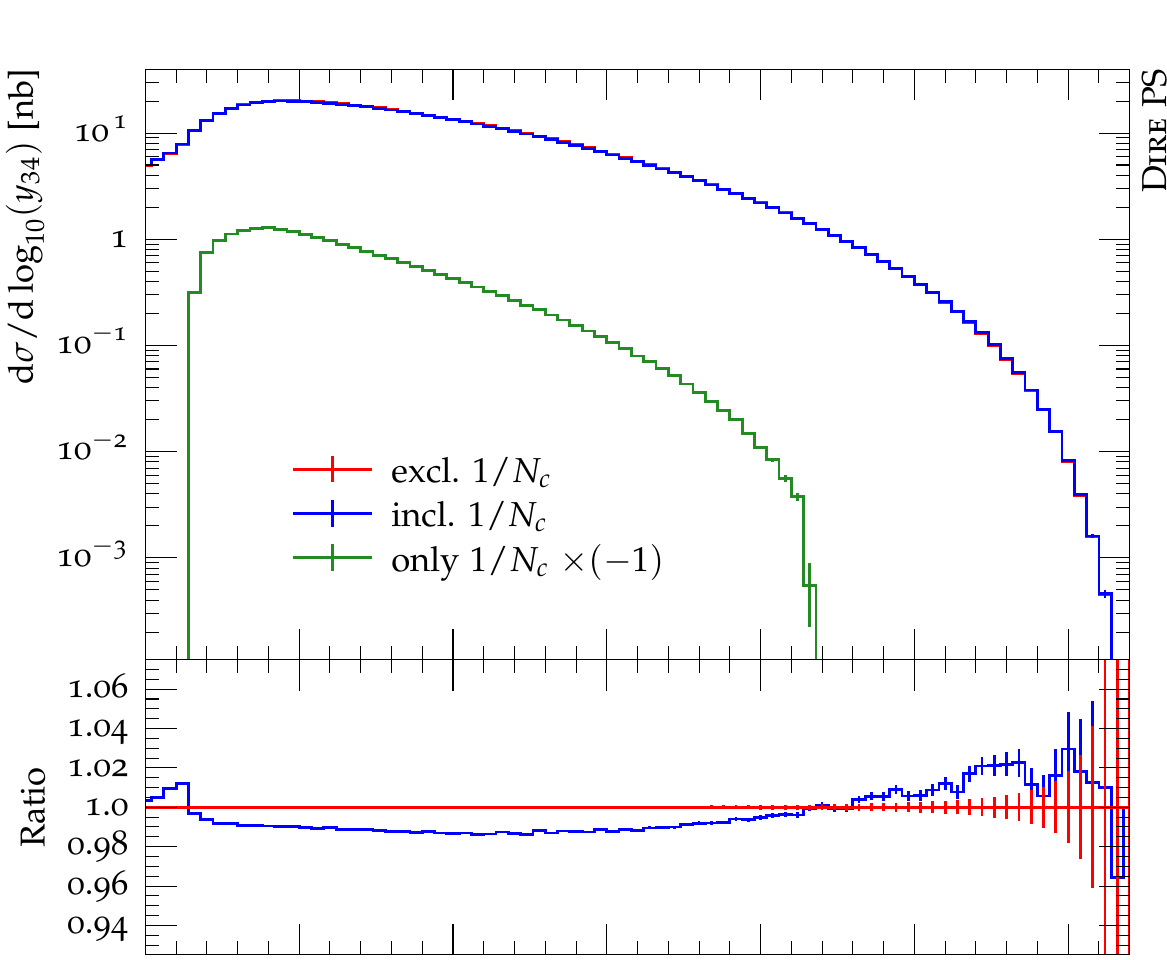}\\[-0.5mm]
        \includegraphics[scale=0.5]{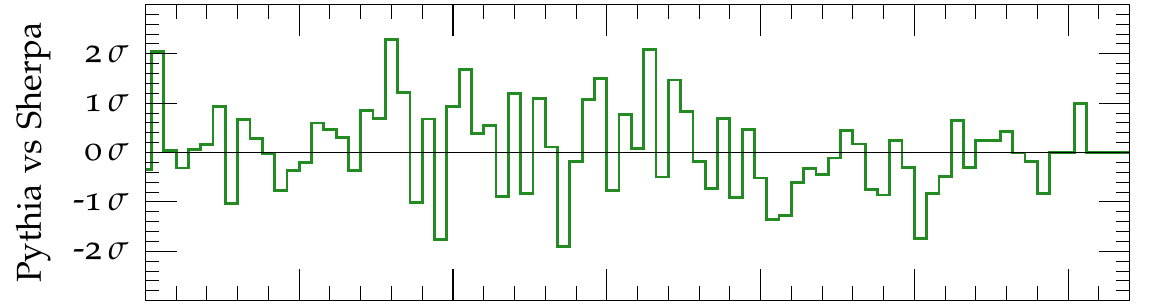}\\[-0.5mm]
        \includegraphics[scale=0.5]{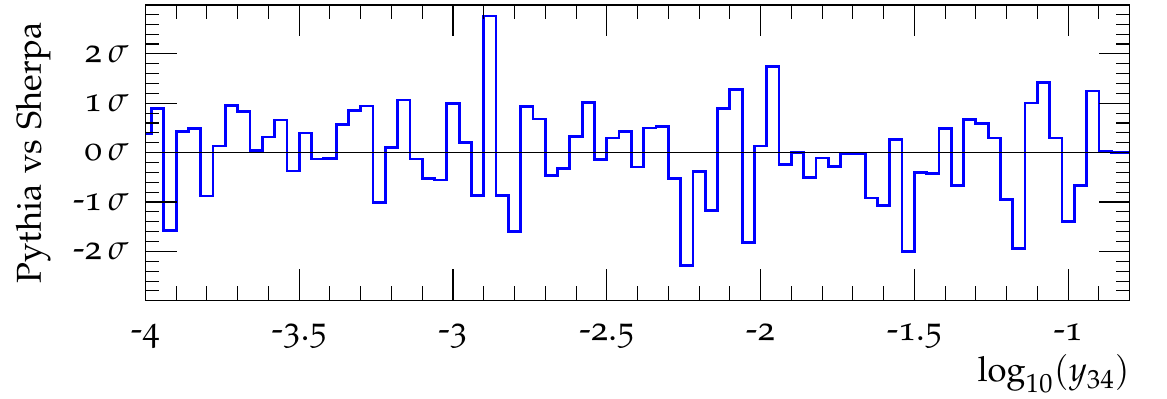}
      \end{center}
    \end{minipage}
    \label{fig:slc_y_34}}\hfill
  \subfigure{
    \begin{minipage}{0.31\textwidth}
      \begin{center}
        \includegraphics[scale=0.5]{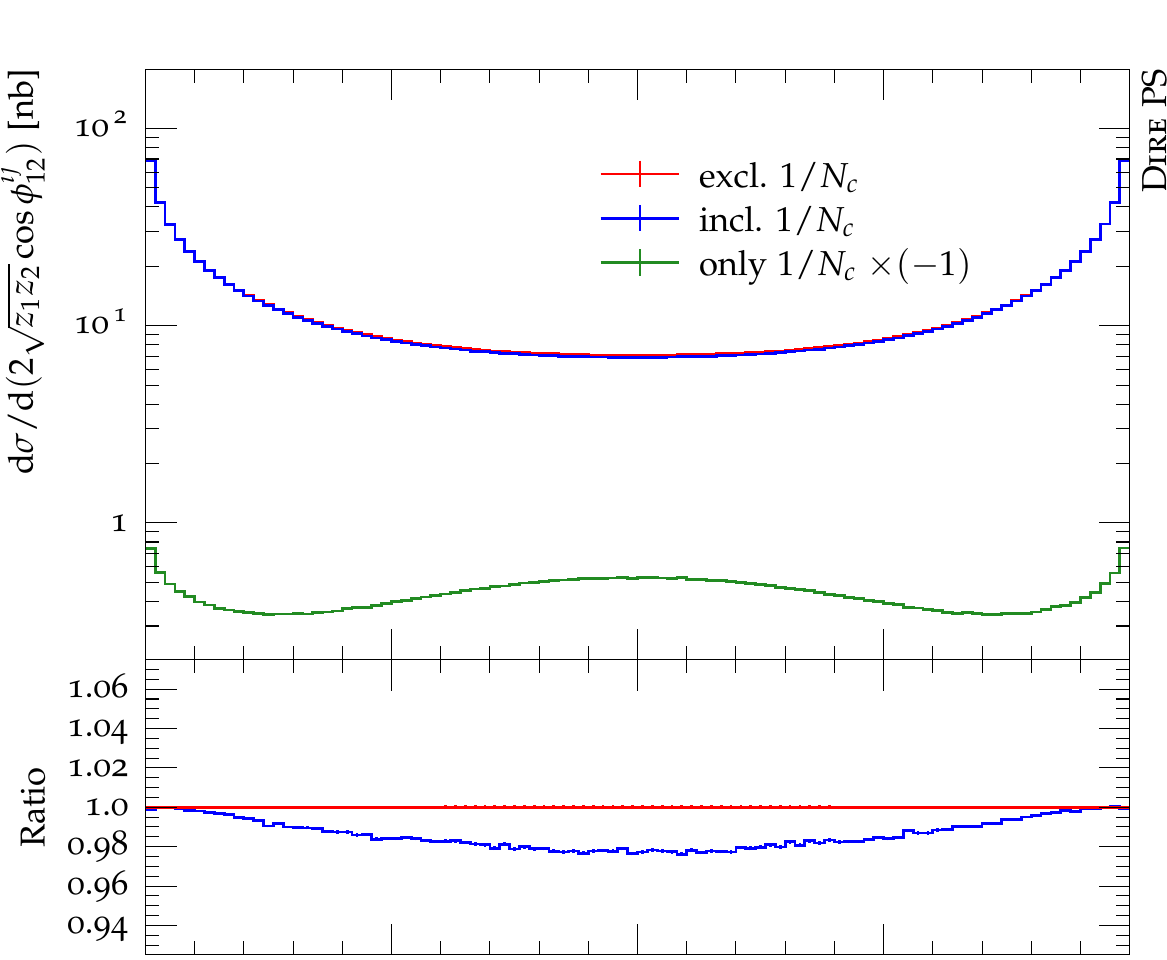}\\[-0.5mm]
        \includegraphics[scale=0.5]{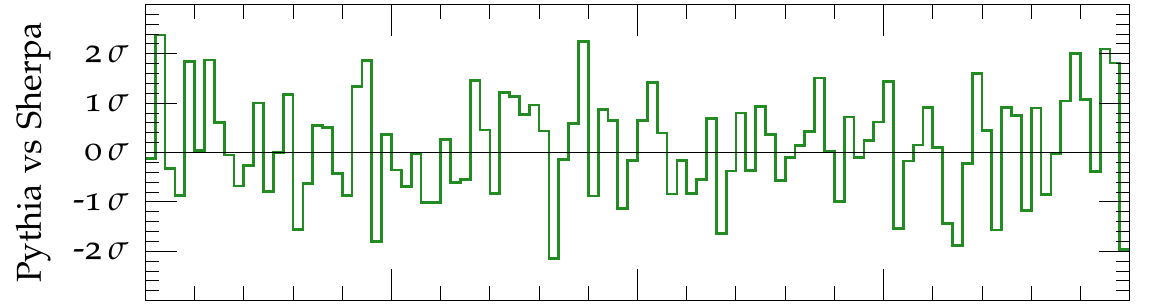}\\[-0.5mm]
        \includegraphics[scale=0.5]{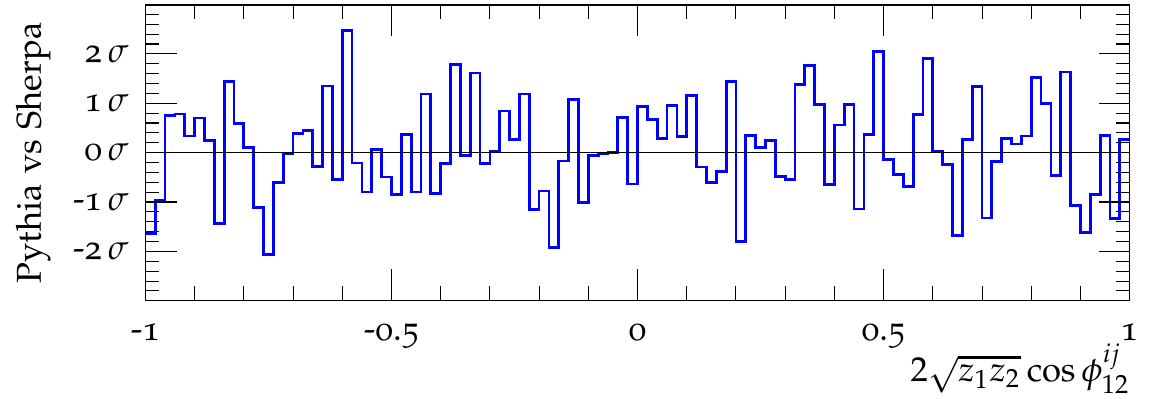}
      \end{center}
    \end{minipage}
    \label{fig:slc_cosphi}}\hfill
  \subfigure{
    \begin{minipage}{0.31\textwidth}
      \begin{center}
        \includegraphics[scale=0.5]{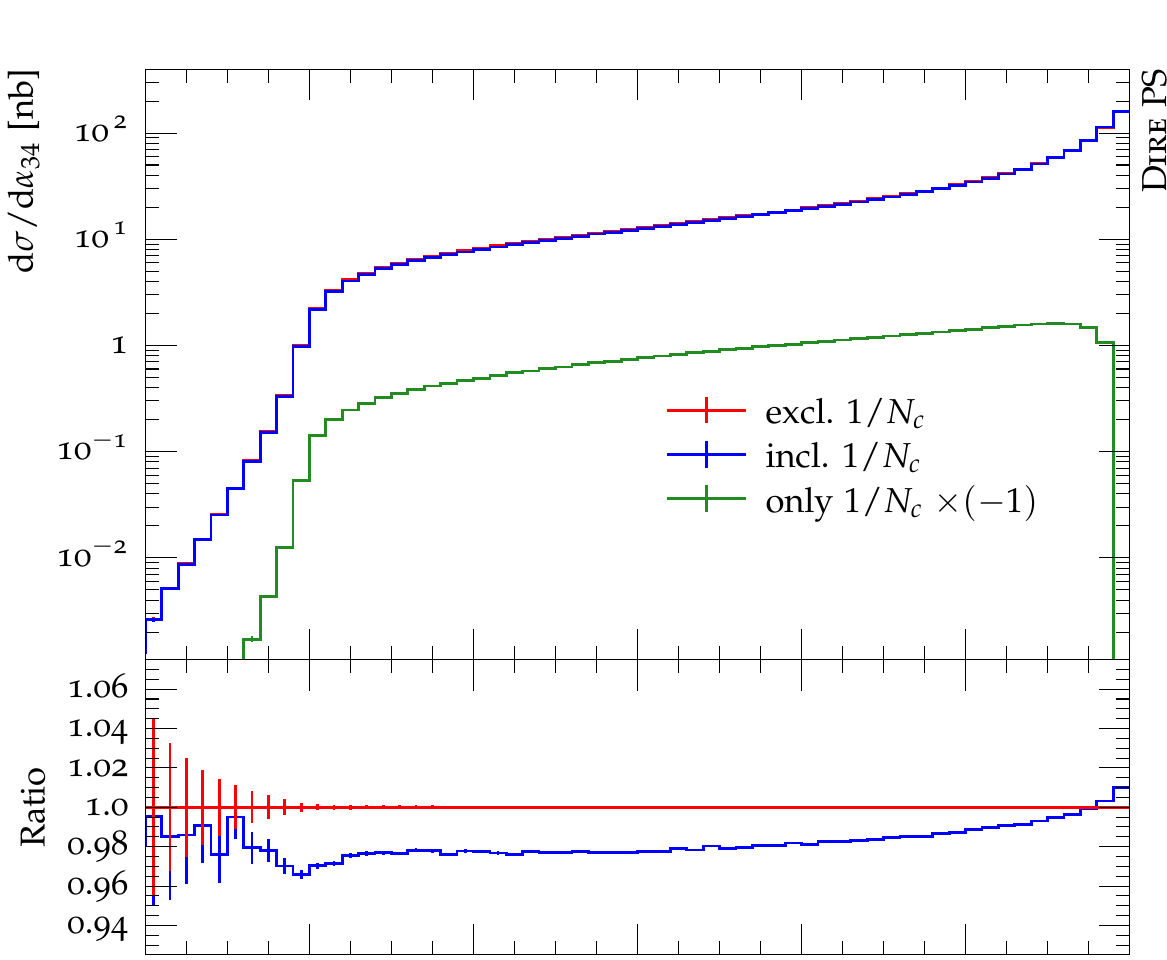}\\[-0.5mm]
        \includegraphics[scale=0.5]{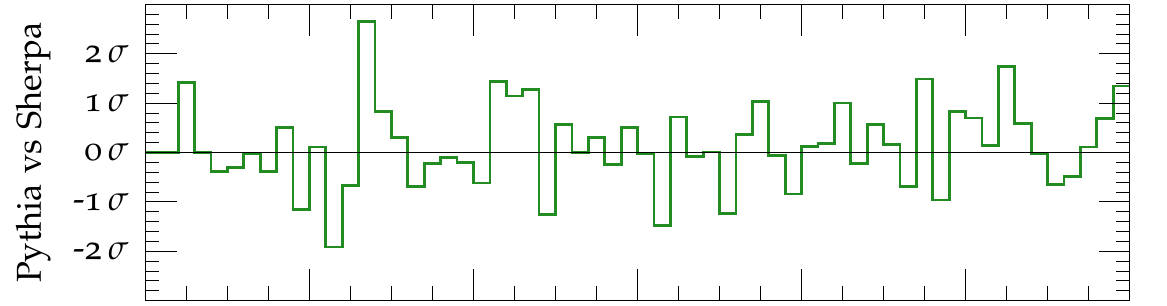}\\[-0.5mm]
        \includegraphics[scale=0.5]{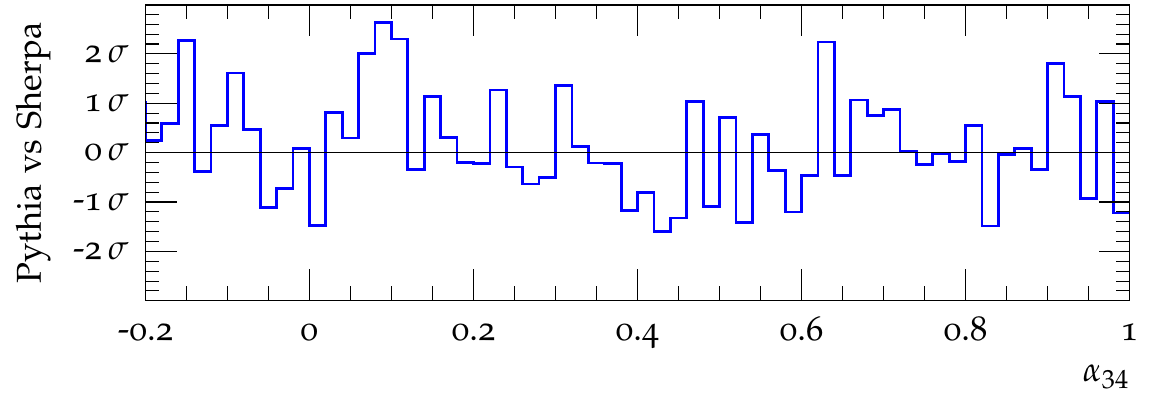}
      \end{center}
    \end{minipage}
    \label{fig:slc_alpha34}}
  \end{center}
  \caption{The impact of the generic sub-leading color corrections,
    Eq.~\eqref{eq:psct_soft_slc}, on the radiation pattern in
    $e^+e^-\to$hadrons at LEP I energies. The reference result (red)
    includes all next-to-leading order effects investigated in
    Figs.~\ref{fig:psw_validation} and~\ref{fig:sc_validation}.
    All simulations are limited to two emissions. Note that the
    simulation results of the sub-leading color corrections alone
    (green) and the baseline (red) do not add up to the full result
    because of the different Sudakov factors.
    \label{fig:slc}}
\end{figure}
\begin{figure}[t]
  \begin{center}
  \subfigure{
    \begin{minipage}{0.31\textwidth}
      \begin{center}
        \includegraphics[scale=0.5]{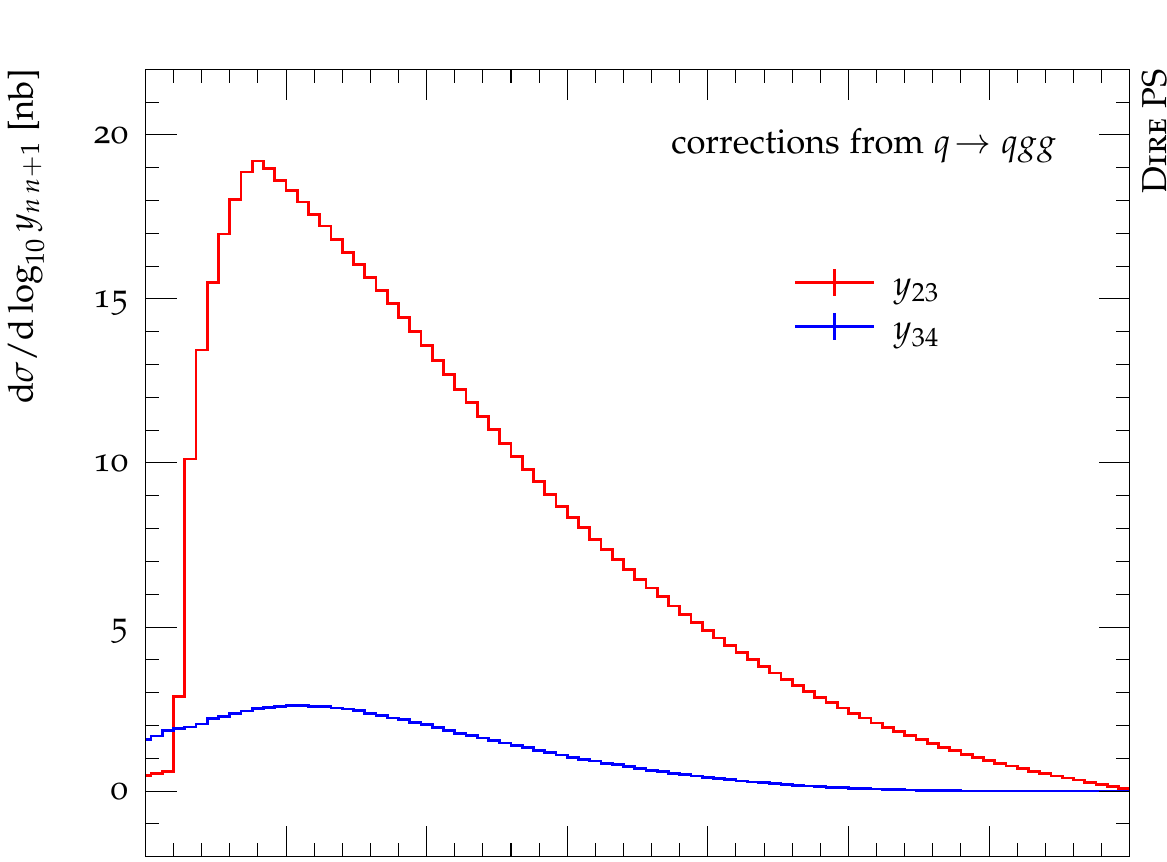}\\[-0.5mm]
        \includegraphics[scale=0.5]{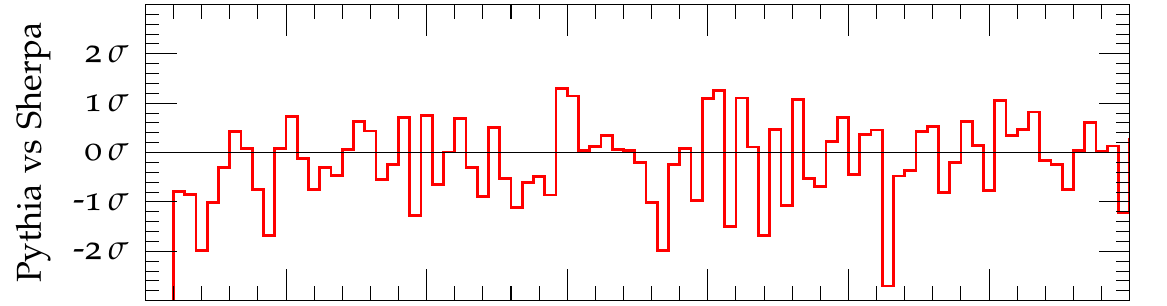}\\[-0.5mm]
        \includegraphics[scale=0.5]{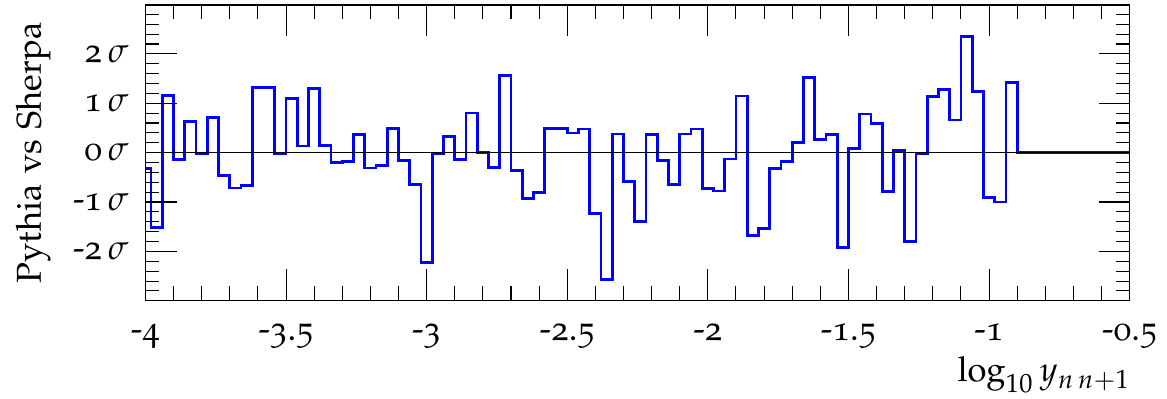}
      \end{center}
    \end{minipage}
    \label{fig:tc_gg}}\hskip 5mm
  \subfigure{
    \begin{minipage}{0.31\textwidth}
      \begin{center}
        \includegraphics[scale=0.5]{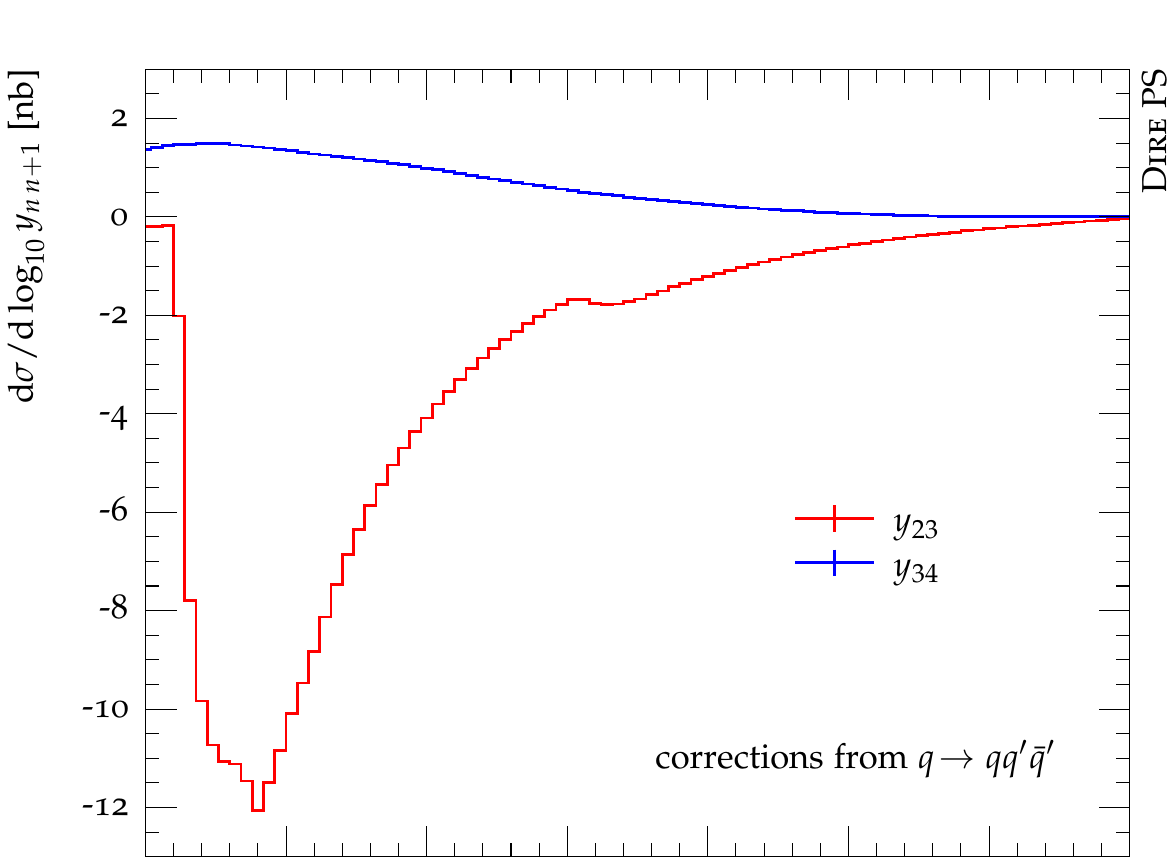}\\[-0.5mm]
        \includegraphics[scale=0.5]{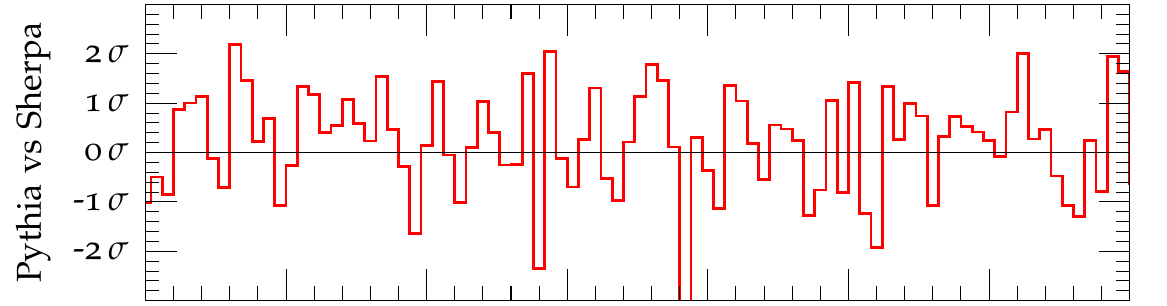}\\[-0.5mm]
        \includegraphics[scale=0.5]{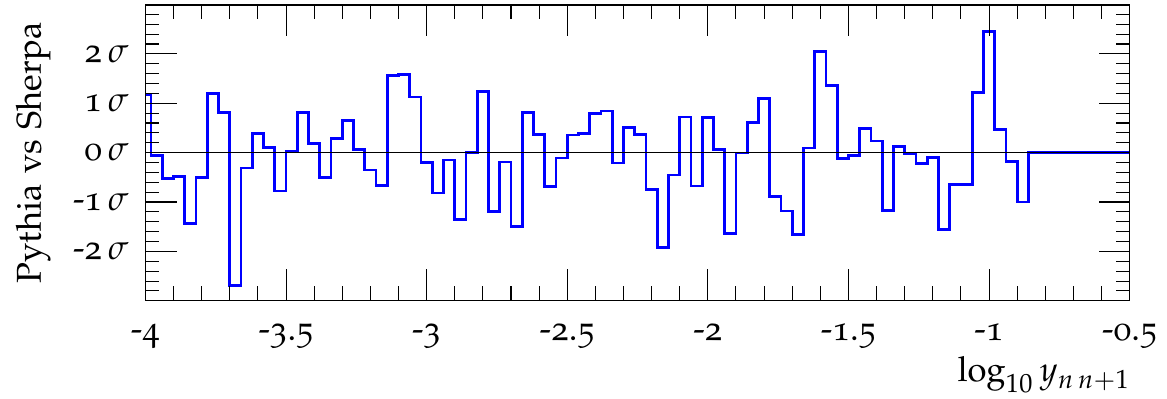}
      \end{center}
    \end{minipage}
    \label{fig:tc_qq}}
  \end{center}
  \caption{The impact of subtracted real-emission corrections,
    Eq.~\eqref{eq:subtracted_real}, and endpoint terms,
    Eq.~\eqref{eq:cusp_diff_endpoints} on the radiation pattern in
    $e^+e^-\to$hadrons at LEP I energies. We show the contributions
    from $q\to qgg$ (left) and $q\to qq'\bar{q}'$ (right) to the
    differential $2\to 3$ (red) and $3\to 4$ (right) jet rates
    in the Durham algorithm.
    \label{fig:tc}}
\end{figure}
\begin{figure}[t]
  \begin{center}
  \subfigure{
    \includegraphics[scale=0.485]{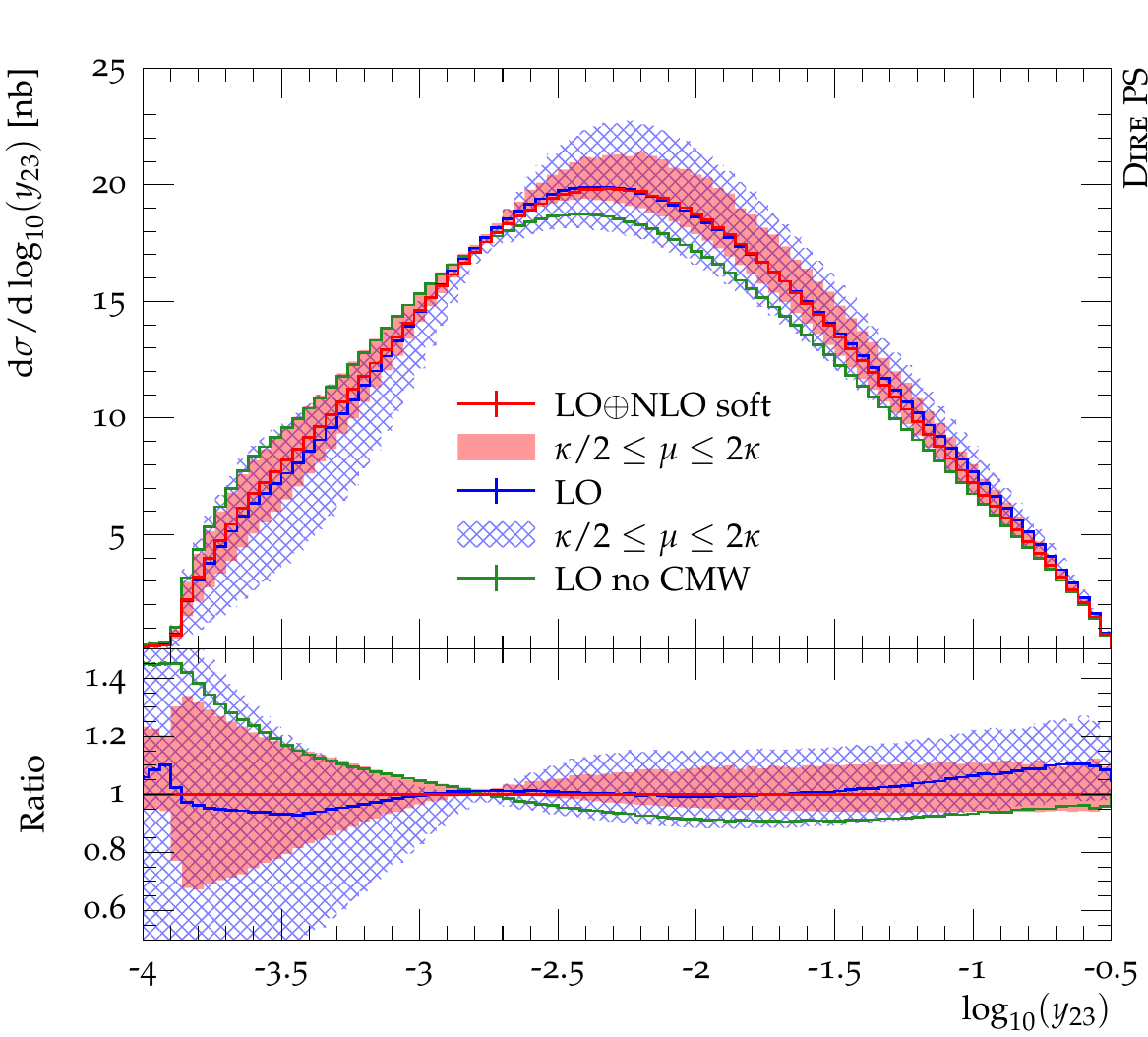}
    \label{fig:sv_y23}}\hfill
  \subfigure{
    \includegraphics[scale=0.485]{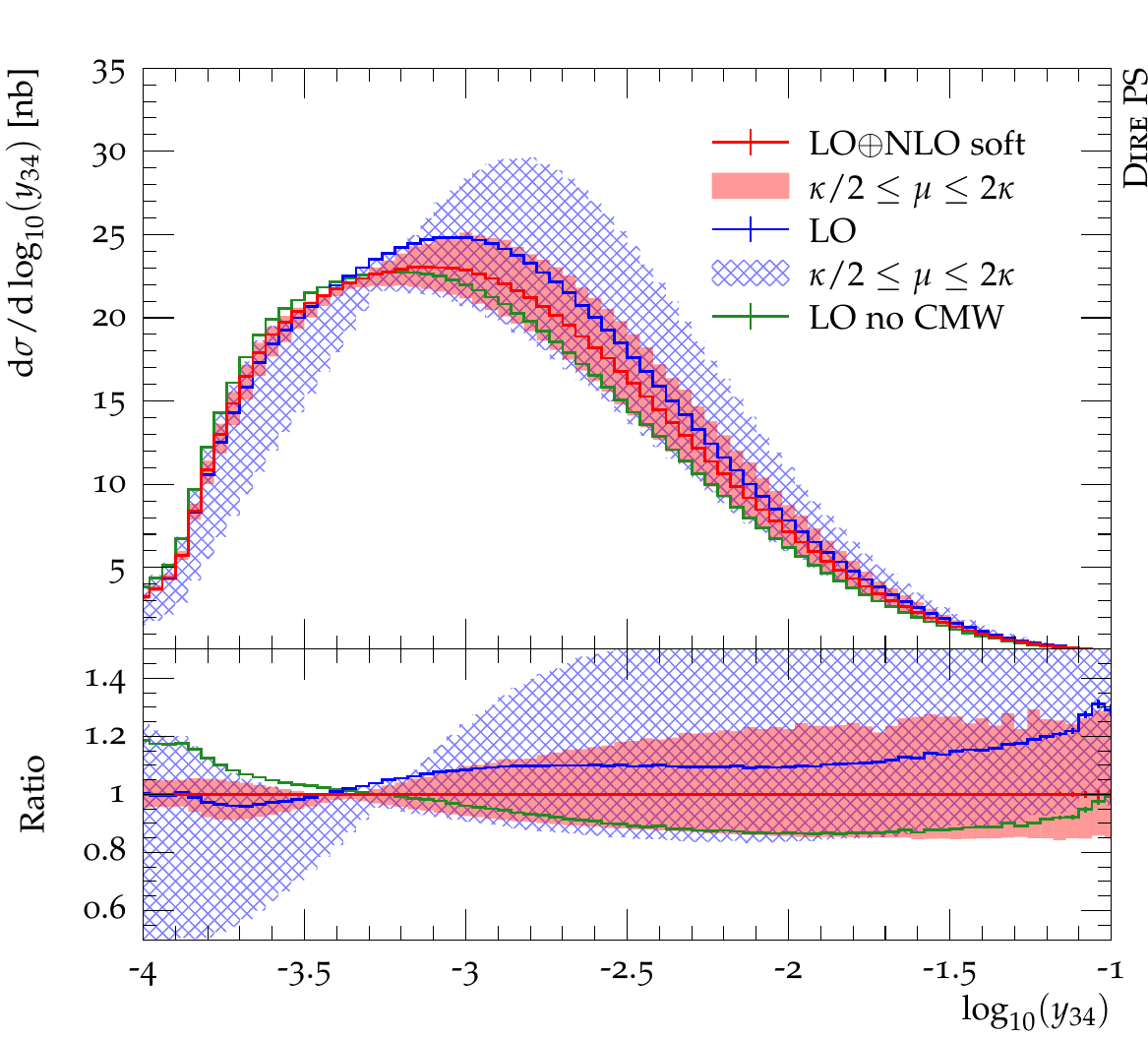}
    \label{fig:sv_y34}}\hfill
  \subfigure{
    \includegraphics[scale=0.485]{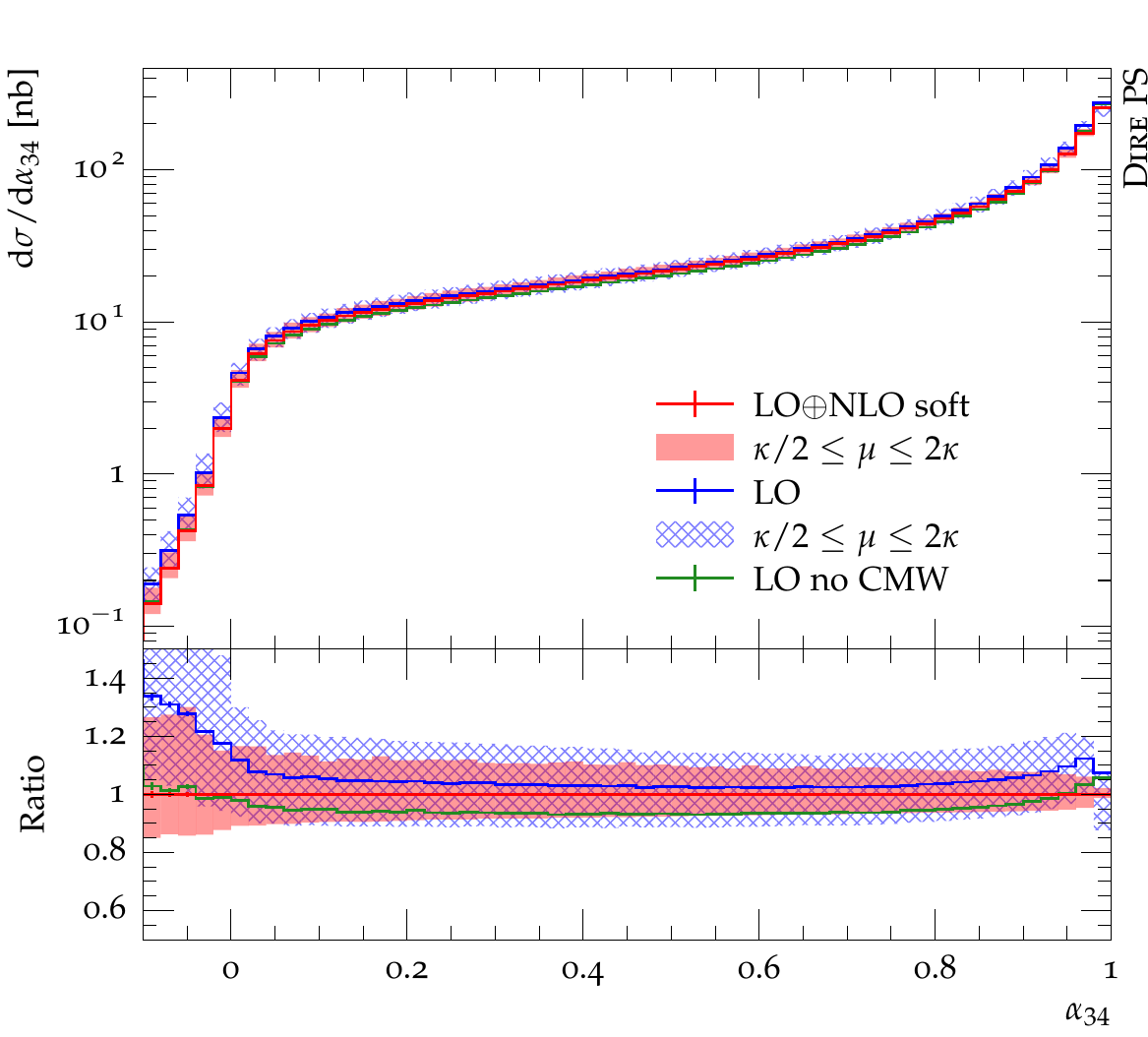}
    \label{fig:sv_a34}}
  \end{center}
  \caption{Scale variations in the leading-order and next-to-leading
    order (soft) parton shower simulation of $e^+e^-\to$hadrons at
    LEP I energies at parton level. We compare to both the plain
    leading-order predictions (green) and the result in the CMW
    scheme (blue).
    \label{fig:sv}}
\end{figure}

Figure~\ref{fig:psw_validation} shows the impact of the phase-space weights, $w_{ij}^{12}$
and $\bar{w}_{ij}^{12}$, defined in Eqs.~\eqref{eq:psct_soft_tc_12_weight}
and~\eqref{eq:psct_soft_tc_so_weight}. These weights generate a strong suppression
of the radiation probability. The effect is eventually compensated by other corrections
(see Fig~\ref{fig:sv}), such that a fairly good agreement with the leading-order
approximation is obtained. The lower panels in Fig.~\ref{fig:psw_validation}
show a comparison between the results from Pythia against those from Sherpa.
The two predictions agree up to statistical fluctuations, providing a strong
cross-check on the consistency of our implementation.
Figure~\ref{fig:sc_validation} shows the impact of the spin correlations
implemented by the $\cos^2\phi_{12}^{\,ij}$ dependence of
Eq.~\eqref{eq:soft_helper_funcs} compared to a spin averaged simulation.
While the related effects are striking when investigating the $z(1-z)$-dependent
parts of the splitting functions in isolation, they are greatly diminished in
the complete calculation.
Figure~\ref{fig:slc} displays the impact of the generic sub-leading color
corrections in Eq.~\eqref{eq:psct_soft_slc}. The effects are generally
smaller than expected based on a naive estimate (i.e.\ $\mc{O}(1/2N_c)$),
because Eq.~\eqref{eq:psct_soft_slc} is suppressed in the collinear
region, cf. the discussion in Sec.~\ref{sec:mc}.
Figure~\ref{fig:tc} shows the impact of the subtracted real-emission
corrections, Eq.~\eqref{eq:subtracted_real}, and the endpoint terms,
Eq.~\eqref{eq:cusp_diff_endpoints}, on the radiation pattern in
$q\to q(gg)$, $q\to q(q'\bar{q}')$ and $q\to q(q\bar{q})$ splittings,
where the particles in parentheses are the soft emissions.
We have verified that exact agreement between our implementations
is obtained also in the case of $g\to g(gg)$ and $g\to g(q\bar{q})$.
The numerical impact of these corrections is similar to the quark-induced case.
Note that the  $3\to 4$ jet rates receive corrections from the
subtracted real emission only, while the $2\to 3$ jet rates are
impacted by both the subtracted real-emission and the endpoint terms.


Figure~\ref{fig:sv} compares the results from a leading-order simulation
according to~\cite{Hoche:2015sya} to our complete next-to-leading order
prediction. In the leading-order case we present the calculation
with and without the CMW scheme~\cite{Catani:1990rr}.
We observe that the CMW prediction matches the rates of the full
next-to-leading order result for the Durham jet rate $y_{23}$ fairly well
in the intermediate-$y$ region, but there are some discrepancies in the
low and high-$y$ region. In addition, there is a considerable rate change
in $y_{34}$. The angular observable $\alpha_{34}$ shows deviations
between the CMW prediction and the full next-to-leading order result
at very small and very large values. In all cases, the scale uncertainty
is greatly reduced at the next-to-leading order,
and the next-to-leading order predictions lie within the leading order
uncertainty bands. Note in particular that our result presents the first genuine
estimate of the perturbative uncertainty in a parton-shower simulation.
Some earlier attempts, despite generating variations of the same order,
treated the problem in an approximate manner~\cite{Bendavid:2018nar}.
Other techniques~\cite{Badger:2016bpw,Hoche:2017hno} assumed the scale variations
on collinear parts of the splitting functions to be identical to the soft parts,
and therefore generate artificially small uncertainty bands, which may not reflect
the true perturbative precision at the order to which the computation is performed.

\section{Conclusions}
\label{sec:conclusions}
We have presented a calculation of the next-to-leading order corrections
to soft-gluon radiation, differentially in the one-emission phase space.
This is a crucial ingredient in the construction of a next-to-leading order parton shower.
We have demonstrated, for the first time, that the soft next-to-leading order
contribution to the evolution of color dipoles can be obtained in a modified
subtraction scheme, such that both one- and two-emission terms are amenable to
Monte-Carlo integration in four dimensions. The two-loop cusp anomalous dimension
emerges naturally in this method. We observe fair agreement between the results
of the fully differential simulation and the approximate treatment using the CMW
scheme, where the two-loop cusp anomalous dimension is included in an
inclusive manner. The similarity of the results is reassuring, because
the individual higher-order contributions have kinematical dependencies
that can differ strongly from the iterated leading-order result.
Our calculation can be seen as a confirmation that the existing leading-order
parton showers developed over the past decades have been amended by the dominant
effects arising from the higher-order soft corrections, but it also confirms
that the higher-order corrections do have an impact beyond a simple $K$-factor.
We are now in place to compute these effects without the need for approximations,
and to include them in phenomenological studies as well as experimental analyses
at the particle level. This allows in particular to obtain meaningful estimates
of the renormalization scale uncertainty.

\begin{acknowledgments}
  \noindent
  We thank Lance Dixon and Thomas Gehrmann for comments on the manuscript.
  This work was supported by the U.S.\ Department of Energy under contracts
  DE--AC02--76SF00515 and DE--AC02--07CH11359.
\end{acknowledgments}

\appendix
\section{Real-emission corrections to soft-gluon radiation}
\label{sec:real_corrections}
This appendix details the computation of the real-emission corrections listed
in Eqs.~\eqref{eq:soft_so_int}-\eqref{eq:soft_rem_coll}. We perform the calculation
separately in the strong ordering approximation, for the soft remainder term,
and for the two collinear contributions in Eqs.~\eqref{eq:soft_helper_funcs}.
\subsection{Strong ordering approximation}
The real corrections in the strong ordering approximation, Eq.~\eqref{eq:soft_so},
lead to one non-trivial integral over transverse momenta, which is given by
\begin{equation}\label{eq:s12_integral_1}
  \begin{split}
    &\Omega(1-2\eps)\int_0^\pi d\phi\,\frac{(\sin^2\phi)^{-\eps}}{s_{12}}
    =\frac{\Omega(1-2\eps)}{Q^2(\alpha_1\beta_2+\alpha_2\beta_1)}
    \int_{0}^{1}d\chi\,\frac{4^{-\eps}(\chi(1-\chi))^{-\tfrac{1}{2}-\eps}}{1+K}\left(1-\frac{2K\,\chi}{1+K}\right)^{-1}\\
    &\qquad=\frac{\Omega(2-2\eps)}{Q^2(\alpha_1\beta_2+\alpha_2\beta_1)}
    \frac{_2F_1\big(1,\tfrac{1}{2}-\eps,1-2\eps;2K/(1+K)\big)}{1+K}\\
    &\qquad=\frac{\Omega(2-2\eps)}{Q^2}
    \left[\frac{(\alpha_1\beta_2)^{2\eps}\Theta(\alpha_1\beta_2-\alpha_2\beta_1)}{
        (\alpha_1\beta_2-\alpha_2\beta_1)^{1+2\eps}}\,
      _2F_1\Big(-\eps,-2\eps,1-\eps,\frac{\alpha_2\beta_1}{\alpha_1\beta_2}\Big)
      +(1\leftrightarrow2)\right]\;.
  \end{split}
\end{equation}
Here we have defined the intermediate variables $\chi=(1+\cos\phi)/2$ and
$K=2\,\sqrt{\alpha_1\beta_1\alpha_2\beta_2}/(\alpha_1\beta_2+\alpha_2\beta_1)$.
The last line was introduced in~\cite{Kramer:1986sr}~\footnote{
  Note that in contrast to~\cite{Kramer:1986sr} the subsequent integrals have a
  divergence at ${\alpha_2\beta_1}\to{\alpha_1\beta_2}$ and vice versa, hence we
  cannot replace $\,_2F_1(-\eps,-2\eps,1-\eps,z)$ by its leading term in $\eps$.}.
It is obtained by the following three transformations of the hypergeometric function
\begin{equation}\label{eq:kl_2f1_trafos}
  \begin{split}
    \,_2F_1\left(2 a,c-\frac{1}{2};2 c-1;\frac{2 z}{z+1}\right)
    =&\;(1+z)^{2 a} \,_2F_1\left(a,a+\frac{1}{2};c;z^2\right)\\
    \,_2F_1\left(\frac{a}{2},\frac{a+1}{2};a-b+1;\frac{4z}{(1+z)^2}\right)
    =&\;(1+z)^a \,_2F_1(a,b;a-b+1;z)\;,
        \quad\text{if}\quad|z|<1\\
    \,_2F_1(a,b;c;z)=&\;(1-z)^{-a-b+c}\,_2F_1(c-a,c-b;c;z)
  \end{split}
\end{equation}
Using Eq.~\eqref{eq:s12_integral_1}, the full integral of the first term
in the strong ordering approximation, Eq.~\eqref{eq:soft_so}, reads
\begin{equation}
  \begin{split}
    S_{ij,A}^{\rm(so)}(q)=&\;
    \int \frac{d^Dp_1}{(2\pi)^{D-1}} \frac{d^Dp_2}{(2\pi)^{D-1}}\,\delta^+(p_1^2)\delta^+(p_2^2)\,
    \mc{F}_{ij}(1,2)\,g_s^4\mu^{4\eps}\,\frac{s_{ij}}{s_{i1}s_{12}s_{2j}}\\
    =&\;\frac{\bar{\alpha}_s^2}{(2\pi)^2}\,\Big(\frac{\kappa}{Q}\Big)^{4\eps}\,
    \int \frac{d\beta_1}{\beta_1^{1+\eps}} \frac{d\alpha_2}{\alpha_2^{1+\eps}}\,
    \left[\frac{(\alpha_1\beta_2)^{2\eps}\Theta(1-\frac{\alpha_2\beta_1}{\alpha_1\beta_2})\,
      _2F_1\left(-\eps,-2\eps,1-\eps,\frac{\alpha_2\beta_1}{\alpha_1\beta_2}\right)}{
        (\alpha_1\beta_2)^{\eps}(\alpha_1\beta_2-\alpha_2\beta_1)^{1+2\eps}}
    +(1\leftrightarrow2)\right]\,.\\
  \end{split}
\end{equation}
Here we have split the integration range in applying the second transformation in
Eq.~\eqref{eq:kl_2f1_trafos} in order to guarantee that that the argument of the hypergeometric function
is between zero and one in the entire range of integration.
Light-cone momentum conservation takes the form $\alpha=\alpha_1+\alpha_2$ and $\beta=\beta_1+\beta_2$.
This suggest the parametrization $\alpha_2=\alpha\,s$ and $\beta_1=\beta\,t$, with $s$ and $t$ ranging
from zero to one. We remap the integration variables to the unit hypercube using the additional
transformations
\begin{equation}\label{eq:st_to_uv_trafo}
  u=\frac{t}{1-s},\quad v=\frac{s}{1-t}
  \qquad\text{and}\qquad
  u=\frac{1-s}{t},\quad v=\frac{1-t}{s}
\end{equation}
in the first and second integral, respectively. We obtain
\begin{equation}
  \begin{split}
    S_{ij,A}^{\rm(so)}(q)
    =&\;\frac{\bar{\alpha}_s^2}{(2\pi)^2}\,\frac{Q^2}{\kappa^2}
    \int \frac{du}{u^{1+\eps}} \frac{dv}{v^{1+\eps}}\,
    \frac{(1+uv)(1-uv)^{2\eps}}{
      ((1-u)(1-v))^{1+2\eps}}\,_2F_1\left(-\eps,-2\eps,1-\eps,uv\right)\;.
  \end{split}
\end{equation}
where we have defined $\kappa^2=Q^2\alpha\beta$. Changing variables to $x=uv$
and $y=(1-v)/(1-uv)$ we can write
\begin{equation}
  \begin{split}
    S_{ij,A}^{\rm(so)}(q)
    =&\;\frac{\bar{\alpha}_s^2}{(2\pi)^2}\,\frac{Q^2}{\kappa^2}
    \int \frac{dx}{x^{1+\eps}} \frac{dy}{y^{1+2\eps}}\,
    \frac{(1+x)(1-y(1-x))^{2\eps}}{
      ((1-x)(1-y))^{1+2\eps}}\,_2F_1\left(-\eps,-2\eps,1-\eps,x\right)\;.
  \end{split}
\end{equation}
We first perform the $y$-integration and obtain
\begin{equation}
  \begin{split}
    S_{ij,A}^{\rm(so)}(q)
    =&\;\frac{\bar{\alpha}_s^2}{(2\pi)^2}\,\frac{Q^2}{\kappa^2}
    \frac{\Gamma(-2\eps)^2}{\Gamma(-4\eps)}\,
    \int \frac{dx}{x^{1+\eps}}\,\frac{1+x}{(1-x)^{1+2\eps}}
    \,_2F_1\left(-2\eps,-2\eps,-4\eps,1-x\right)
    \,_2F_1\left(-\eps,-2\eps,1-\eps,x\right)\;.
  \end{split}
\end{equation}
The last integral can be solved by expanding the two hypergeometric functions
up to $\mc{O}(\eps^3)$ and expanding the result up to the finite term. We use
HypExp~\cite{Huber:2005yg,Huber:2007dx} and obtain
\begin{equation}\label{eq:soft_so_1}
  S_{ij,A}^{\rm(so)}(q)=
  \frac{\bar{\alpha}_s^2}{(2\pi)^2}\,\frac{Q^2}{\kappa^2}
  \left(\frac{3}{\eps^2}-\frac{4}{3}\,\pi^2-22\,\eps\,\zeta_3+\mc{O}(\eps^2)\right)\;.
\end{equation}
The integral of the second term in the strong ordering approximation can be obtained
exploiting the symmetry in $p_1$ and $p_2$. The final term is given by
\begin{equation}
  \begin{split}
    S_{ij,B}^{\rm(so)}(q)=&\;\int\frac{d^Dp_1}{(2\pi)^{D-1}}\frac{d^Dp_2}{(2\pi)^{D-1}}\,
    \delta^+(p_1^2)\delta^+(p_2^2)\,\mc{F}_{ij}(1,2)\,
    g_s^4\mu^{4\eps}\frac{s_{ij}^2}{s_{i1}s_{1j}s_{i2}s_{2j}}\\
    =&\;\frac{\bar{\alpha}_s^2}{(2\pi)^2}\,\frac{Q^2}{\kappa^2}
    \int \frac{d\beta_1}{\beta_1^{1+\eps}} \frac{d\alpha_2}{\alpha_2^{1+\eps}}\,
    \frac{1}{(\alpha_1\beta_2)^{1+\eps}}\;.
  \end{split}
\end{equation}
We use again the parametrization $\alpha_2=\alpha\,s$ and $\beta_1=\beta\,t$,
with $s$ and $t$ ranging from zero to one. The integral can be evaluated
in terms of beta functions and we obtain
\begin{equation}\label{eq:soft_so_3}
  \begin{split}
    S_{ij,B}^{\rm(so)}(q)
    =&\;\frac{\bar{\alpha}_s^2}{(2\pi)^2}\,\frac{Q^2}{\kappa^2}
    \left(\frac{4}{\eps^2}-\frac{4}{3}\,\pi^2-16\,\eps\,\zeta_3
    -\frac{4}{45}\,\eps^2\,\pi^4+\mc{O}(\eps^3)\right)\;.
  \end{split}
\end{equation}
The complete symmetrized result in the strong ordering approximation is
\begin{equation}\label{eq:soft_so_int_1}
  \begin{split}
    S_{ij}^{\rm(so)}(q)
    =&\;\frac{1}{2}\left(2\,S_{ij,A}^{\rm(so)}(1,2)-S_{ij,B}^{\rm(so)}(1,2)\right)
    =\frac{\bar{\alpha}_s^2}{(2\pi)^2}\,\frac{Q^2}{\kappa^2}
    \left(\frac{1}{\eps^2}-\frac{2}{3}\,\pi^2-14\,\eps\,\zeta_3+\mc{O}(\eps^2)\right)\;.
  \end{split}
\end{equation}

\subsection{Soft remainder function}
The full integral of the first term in the remainder function,
Eq.~\eqref{eq:soft_helper_funcs}, reads
\begin{equation}
  \begin{split}
    S_{ij,A}^{\rm(rem)}(q)=&\;
    \int\frac{d^Dp_1}{(2\pi)^{D-1}}\frac{d^Dp_2}{(2\pi)^{D-1}}\,
    \delta^+(p_1^2)\delta^+(p_2^2)\,\mc{F}_{ij}(1,2)\,
    g_s^4\mu^{4\eps}\frac{s_{i1}s_{j2}+s_{i2}s_{j1}}{(s_{i1}+s_{i2})(s_{j1}+s_{j2})}\,
    \frac{s_{ij}}{s_{i1}s_{12}s_{2j}}\;.
  \end{split}
\end{equation}
Again we use the parametrization $\alpha_2=\alpha\,s$ and $\beta_1=\beta\,t$,
with $s$ and $t$ ranging from zero to one. Performing the change of variables
as in the strongly ordered case, Eq.~\eqref{eq:st_to_uv_trafo}, we can again
remap the integration to the unit hypercube and obtain
\begin{equation}
  \begin{split}
    S_{ij,A}^{\rm(rem)}(q)
    =&\;\frac{\bar{\alpha}_s^2}{(2\pi)^2}\,\frac{Q^2}{\kappa^2}
    \int \frac{du}{u^{1+\eps}} \frac{dv}{v^{1+\eps}}\,
    \frac{(1+uv)^2(1-uv)^{-2+2\eps}}{
      ((1-u)(1-v))^{2\eps}}\,_2F_1\left(-\eps,-2\eps,1-\eps,uv\right)\;.
  \end{split}
\end{equation}
Changing variables to $x=uv$ and $y=(1-v)/(1-uv)$ we can write
\begin{equation}
  \begin{split}
    S_{ij,A}^{\rm(rem)}(q)
    =&\;\frac{\bar{\alpha}_s^2}{(2\pi)^2}\,\frac{Q^2}{\kappa^2}
    \int \frac{dx}{x^{1+\eps}} \frac{dy}{y^{2\eps}}\,
    \frac{(1+x)^2(1-y(1-x))^{-1+2\eps}}{
      (1-x)^{1+2\eps}(1-y)^{2\eps}}\,_2F_1\left(-\eps,-2\eps,1-\eps,x\right)\;.
  \end{split}
\end{equation}
We first perform the $y$-integration and obtain
\begin{equation}
  \begin{split}
    S_{ij,A}^{\rm(rem)}(q)
    =&\;\frac{\bar{\alpha}_s^2}{(2\pi)^2}\,\frac{Q^2}{\kappa^2}
    \frac{\Gamma(1-2\eps)^2}{\Gamma(2-4\eps)}\,
    \int \frac{dx}{x^{1+\eps}}\,\frac{(1+x)^2}{(1-x)^{1+2\eps}}\\
    &\qquad\times\,_2F_1\left(1-2\eps,1-2\eps,2-4\eps,1-x\right)
    \,_2F_1\left(-\eps,-2\eps,1-\eps,x\right)\;.
  \end{split}
\end{equation}
The last integral can be solved by expanding the two hypergeometric functions
up to $\mc{O}(\eps^3)$ and expanding the result up to the finite term. We use
HypExp~\cite{Huber:2005yg,Huber:2007dx} and obtain
\begin{equation}\label{eq:soft_rem_1}
  S_{ij,A}^{\rm(rem)}(q)=
  \frac{\bar{\alpha}_s^2}{(2\pi)^2}\,\frac{Q^2}{\kappa^2}
  \left(\frac{1}{\eps^2}-\frac{2}{\eps}-4-\frac{2}{3}\,\pi^2
  +\eps\left(\frac{2}{3}\,\pi^2-8-14\,\zeta_3\right)+\mc{O}(\eps^2)\right)\;.
\end{equation}
The integral of the second remainder term can be obtained by symmetry.
The final term is given by
\begin{equation}
  \begin{split}
    S_{ij,B}^{\rm(rem)}(q)=&\;
    \int\frac{d^Dp_1}{(2\pi)^{D-1}}\frac{d^Dp_2}{(2\pi)^{D-1}}\,
    \delta^+(p_1^2)\delta^+(p_2^2)\,\mc{F}_{ij}(1,2)\,
    g_s^4\mu^{4\eps}\frac{s_{i1}s_{j2}+s_{i2}s_{j1}}{(s_{i1}+s_{i2})(s_{j1}+s_{j2})}\,
    \frac{s_{ij}^2}{s_{i1}s_{1j}s_{i2}s_{2j}}\\
    =&\;\frac{\bar{\alpha}_s^2}{(2\pi)^2}\,\frac{Q^2}{\kappa^2}
    \int \frac{d\beta_1}{\beta_1^{1+\eps}} \frac{d\alpha_2}{\alpha_2^{1+\eps}}\,
    \frac{\alpha_2\beta_1+\alpha_1\beta_2}{\alpha\beta\,(\alpha_1\beta_2)^{1+\eps}}\;.
  \end{split}
\end{equation}
The integral can again be evaluated in terms of beta functions and we obtain
\begin{equation}\label{eq:soft_rem_3}
  \begin{split}
    S_{ij,B}^{\rm(rem)}(q)
    =&\;\frac{\bar{\alpha}_s^2}{(2\pi)^2}\,\frac{Q^2}{\kappa^2}
    \left(\frac{2}{\eps^2}-\frac{2}{3}\,\pi^2-8\,\eps\,\zeta_3
    -\frac{2}{45}\,\eps^2\,\pi^4+\mc{O}(\eps^3)\right)\;.
  \end{split}
\end{equation}
The complete symmetrized soft remainder is
\begin{equation}\label{eq:soft_rem}
  \begin{split}
    S_{ij}^{\rm(rem)}(q)
    =&\;\frac{1}{2}\left(2\,S_{ij,A}^{\rm(rem)}(1,2)-S_{ij,B}^{\rm(rem)}(1,2)\right)\\
    =&\;\frac{\bar{\alpha}_s^2}{(2\pi)^2}\,\frac{Q^2}{\kappa^2}
    \left(-\frac{2}{\eps}-4-\frac{\pi^2}{3}
    +\eps\left(\frac{2}{3}\,\pi^2-8-10\,\zeta_3\right)+\mc{O}(\eps^2)\right)\;.
  \end{split}
\end{equation}

\subsection{Collinear terms}
The integral of the first collinear term, Eq.~\eqref{eq:soft_helper_funcs}, reads
\begin{equation}
  \begin{split}
    S_{ij,A}^{\rm(coll)}(q)=&\;
    \int\frac{d^Dp_1}{(2\pi)^{D-1}}\frac{d^Dp_2}{(2\pi)^{D-1}}\,
    \delta^+(p_1^2)\delta^+(p_2^2)\,\mc{F}_{ij}(1,2)\,
    g_s^4\mu^{4\eps}\frac{1}{s_{12}^2}\,
    \frac{(s_{i1}s_{j2}-s_{i2}s_{j1})^2}{(s_{i1}+s_{i2})^2(s_{j1}+s_{j2})^2}\;.
  \end{split}
\end{equation}
This term contains a new type of integral over transverse momenta, which gives
\begin{equation}\label{eq:s122_integral}
  \begin{split}
    &\Omega(1-2\eps)\int_0^\pi d\phi\,\frac{(\sin^2\phi)^{-\eps}}{s_{12}^2}
    =\frac{\Omega(1-2\eps)}{Q^4(\alpha_1\beta_2+\alpha_2\beta_1)^2}
    \int_{0}^{1}d\chi\,\frac{4^{-\eps}(\chi(1-\chi))^{-\tfrac{1}{2}-\eps}}{(1+K)^2}
    \left(1-\frac{2K\,\chi}{1+K}\right)^{-2}\\
    &\qquad=\frac{\Omega(2-2\eps)}{Q^4(\alpha_1\beta_2+\alpha_2\beta_1)^2}
    \left(\frac{1+2\eps}{1-K^2}\frac{_2F_1\big(1,\tfrac{1}{2}-\eps,1-2\eps;2K/(1+K)\big)}{1+K}-\frac{2\eps}{1-K^2}\right)\\
    &\qquad=-2\eps\,\frac{\Omega(2-2\eps)}{Q^4(\alpha_1\beta_2-\alpha_2\beta_1)^2}
    +(1+2\eps)\,\frac{\Omega(2-2\eps)}{Q^4}(\alpha_1\beta_2+\alpha_2\beta_1)\\
    &\qquad\qquad\times\bigg[\frac{(\alpha_1\beta_2)^{2\eps}\Theta(\alpha_1\beta_2-\alpha_2\beta_1)}{
        (\alpha_1\beta_2-\alpha_2\beta_1)^{3+2\eps}}\,
      _2F_1\Big(-\eps,-2\eps,1-\eps,\frac{\alpha_2\beta_1}{\alpha_1\beta_2}\Big)
      +(1\leftrightarrow2)\bigg]\;,
  \end{split}
\end{equation}
where we have again used the transformations in Eq.~\eqref{eq:kl_2f1_trafos}.
To perform the remaining integrations we use again the parametrization
$\alpha_2=\alpha\,s$ and $\beta_1=\beta\,t$, with $s$ and $t$ ranging
from zero to one. The first term in Eq.~\eqref{eq:s122_integral} is evaluated
in terms of beta functions, and we obtain
\begin{equation}
  \begin{split}
    S_{ij,A}^{\rm(coll)}(q)
    =&\;-\frac{\bar{\alpha}_s^2}{(2\pi)^2}\,\frac{Q^2}{\kappa^2}
    \frac{2\eps\,\Gamma(1-\eps)^4}{\Gamma(2-2\eps)^2}+
    \tilde{S}_{ij,A}^{\rm(coll)}(1,2)\;.
  \end{split}
\end{equation}
The second term, $\tilde{S}_{ij,A}^{\rm(coll)}(1,2)$, is evaluated by
performing the same change of variables as in the strongly ordered case,
Eq.~\eqref{eq:st_to_uv_trafo}. The two contributions are identical
and can be combined. We obtain
\begin{equation}
  \begin{split}
    \tilde{S}_{ij,A}^{\rm(coll)}(q)
    =&\;\frac{\bar{\alpha}_s^2}{(2\pi)^2}\,\frac{Q^2}{\kappa^2}
    \int \frac{du}{u^{\eps}} \frac{dv}{v^{\eps}}\,
    \frac{(2+4\eps)(1+uv)(1-uv)^{-4+2\eps}}{
      ((1-u)(1-v))^{-1+2\eps}}\,_2F_1\left(-\eps,-2\eps,1-\eps,uv\right)\;.
  \end{split}
\end{equation}
Changing variables to $x=uv$ and $y=(1-v)/(1-uv)$ we can write
\begin{equation}
  \begin{split}
    \tilde{S}_{ij,A}^{\rm(coll)}(q)
    =&\;\frac{\bar{\alpha}_s^2}{(2\pi)^2}\,\frac{Q^2}{\kappa^2}
    \int \frac{dx}{x^{\eps}} \frac{dy}{y^{-1+2\eps}}\,
    \frac{(2+4\eps)(1+x)(1-y(1-x))^{-2+2\eps}}{
      (1-x)^{1+2\eps}(1-y)^{-1+2\eps}}\,_2F_1\left(-\eps,-2\eps,1-\eps,x\right)\;.
  \end{split}
\end{equation}
We first perform the $y$-integration and obtain
\begin{equation}
  \begin{split}
    \tilde{S}_{ij,A}^{\rm(coll)}(q)
    =&\;\frac{\bar{\alpha}_s^2}{(2\pi)^2}\,\frac{Q^2}{\kappa^2}
    \frac{\Gamma(2-2\eps)^2}{\Gamma(4-4\eps)}\,
    \int \frac{dx}{x^\eps}\,\frac{(2+4\eps)(1+x)}{(1-x)^{1+2\eps}}\\
    &\qquad\times\,_2F_1\left(2-2\eps,2-2\eps,4-4\eps,1-x\right)
    \,_2F_1\left(-\eps,-2\eps,1-\eps,x\right)\;.
  \end{split}
\end{equation}
The last integral can be solved by expanding the two hypergeometric functions
up to $\mc{O}(\eps^2)$ and expanding the result up to the finite term.
The integrals are evaluated to the required accuracy with the help of
HypExp~\cite{Huber:2005yg,Huber:2007dx}. The final result is
\begin{equation}\label{eq:coll_1}
  \begin{split}
    S_{ij,A}^{\rm(coll)}(q)
    =&\;\frac{\bar{\alpha}_s^2}{(2\pi)^2}\,\frac{Q^2}{\kappa^2}
    \left(-\frac{1}{3\eps}-\frac{8}{9}
    -\eps\left(\frac{52}{27}-\frac{\pi^2}{9}\right)+\mc{O}(\eps^2)\right)\\
  \end{split}
\end{equation}
The integral of the second collinear term is given by
\begin{equation}
  \begin{split}
    S_{ij,B}^{\rm(coll)}(q)=&\;
    \int\frac{d^Dp_1}{(2\pi)^{D-1}}\frac{d^Dp_2}{(2\pi)^{D-1}}\,
    \delta^+(p_1^2)\delta^+(p_2^2)\,\mc{F}_{ij}(1,2)\,
    g_s^4\mu^{4\eps}\frac{1}{s_{12}}\,\frac{s_{ij}}{(s_{i1}+s_{i2})(s_{j1}+s_{j2})}\;.
  \end{split}
\end{equation}
Again we use the parametrization $\alpha_2=\alpha\,s$ and $\beta_1=\beta\,t$,
with $s$ and $t$ ranging from zero to one to obtain
\begin{equation}
  \begin{split}
    S_{ij,B}^{\rm(coll)}(q)=
    &\;\frac{\bar{\alpha}_s^2}{(2\pi)^2}\,\frac{Q^2}{\kappa^2}
    \int \frac{ds}{s^\eps} \frac{dt}{t^\eps}\,
    \frac{((1-s)(1-t))^{\eps}}{(1-s-t)^{1+2\eps}}\,\\
    &\qquad\times\Theta(1-s-t)
      \,_2F_1\Big(-\eps,-2\eps,1-\eps,\frac{s\,t}{(1-s)(1-t)}\Big)
      +\Big(\begin{array}{c}s\leftrightarrow 1-s\\t\leftrightarrow 1-t\end{array}\Big)\;.
  \end{split}
\end{equation}
The two terms are identical and can be combined. To perform the remaining
integrations we use again the change of variables as in the strongly
ordered case, Eq.~\eqref{eq:st_to_uv_trafo}. We obtain
\begin{equation}
  \begin{split}
    S_{ij,B}^{\rm(coll)}(q)
    =&\;\frac{\bar{\alpha}_s^2}{(2\pi)^2}\,\frac{Q^2}{\kappa^2}
    \int \frac{du}{u^{\eps}} \frac{dv}{v^{\eps}}\,
    \frac{2\,(1-uv)^{-2+2\eps}}{
      ((1-u)(1-v))^{2\eps}}\,_2F_1\left(-\eps,-2\eps,1-\eps,uv\right)\;.
  \end{split}
\end{equation}
Changing variables to $x=uv$ and $y=(1-v)/(1-uv)$ we can write
\begin{equation}
  \begin{split}
    S_{ij,B}^{\rm(coll)}(q)
    =&\;\frac{\bar{\alpha}_s^2}{(2\pi)^2}\,\frac{Q^2}{\kappa^2}
    \int \frac{dx}{x^{\eps}} \frac{dy}{y^{2\eps}}\,
    \frac{2\,(1-y(1-x))^{-1+2\eps}}{
      (1-x)^{1+2\eps}(1-y)^{2\eps}}\,_2F_1\left(-\eps,-2\eps,1-\eps,x\right)\;.
  \end{split}
\end{equation}
We first perform the $y$-integration and obtain
\begin{equation}
  \begin{split}
    S_{ij,B}^{\rm(coll)}(q)
    =&\;\frac{\bar{\alpha}_s^2}{(2\pi)^2}\,\frac{Q^2}{\kappa^2}
    \frac{\Gamma(1-2\eps)^2}{\Gamma(2-4\eps)}\,
    \int \frac{dx}{x^\eps}\,\frac{4}{(1-x)^{1+2\eps}}\\
    &\qquad\times\,_2F_1\left(1-2\eps,1-2\eps,2-4\eps,1-x\right)
    \,_2F_1\left(-\eps,-2\eps,1-\eps,x\right)\;.
  \end{split}
\end{equation}
The last integral can be solved by expanding the two hypergeometric functions
up to $\mc{O}(\eps^2)$ and expanding the result up to the finite term.
The integrals are evaluated to the required accuracy with the help of
HypExp~\cite{Huber:2005yg,Huber:2007dx}. The final result is
\begin{equation}\label{eq:coll_2}
  \begin{split}
    S_{ij,B}^{\rm(coll)}(q)
    =&\;\frac{\bar{\alpha}_s^2}{(2\pi)^2}\,\frac{Q^2}{\kappa^2}
    \left(-\frac{1}{\eps}-2
    +\eps\left(\frac{\pi^2}{3}-4\right)+\mc{O}(\eps^2)\right)\\
  \end{split}
\end{equation}
The complete symmetrized collinear term in the two soft gluon final state is
\begin{equation}\label{eq:coll_gg}
  \begin{split}
    S_{ij,gg}^{\rm(coll)}(q)
    =&\;\frac{1}{2}\left((1-\eps)S_{ij,A}^{\rm(coll)}(1,2)-2S_{ij,B}^{\rm(coll)}(1,2)\right)
    =\frac{\bar{\alpha}_s^2}{(2\pi)^2}\,\frac{Q^2}{\kappa^2}
    \left(\frac{5}{6\eps}+\frac{31}{18}
    +\eps\left(\frac{94}{27}-\frac{5}{18}\pi^2\right)+\mc{O}(\eps^2)\right)\;.
  \end{split}
\end{equation}
The complete collinear term in the soft quark-antiquark final state is
\begin{equation}\label{eq:coll_qq}
  \begin{split}
    S_{ij,q\bar{q}}^{\rm(coll)}(q)
    =&\;\left(S_{ij,A}^{\rm(coll)}(1,2)-S_{ij,B}^{\rm(coll)}(1,2)\right)
    =\frac{\bar{\alpha}_s^2}{(2\pi)^2}\,\frac{Q^2}{\kappa^2}
    \left(\frac{2}{3\eps}+\frac{10}{9}
    +\eps\left(\frac{56}{27}-\frac{2}{9}\pi^2\right)+\mc{O}(\eps^2)\right)\;.
  \end{split}
\end{equation}

\section{Relation to soft-gluon resummation}
\label{sec:resummation}
This section will establish the connection between our parton-shower simulation
and soft gluon resummation formalism developed in~\cite{Korchemsky:1992xv,Korchemsky:1993uz}.
According to the resummation formalism we can express the soft part of the factorized
Drell-Yan cross section as
\begin{equation}\label{eq:dy_bspace_soft}
  \frac{1}{\sigma_{DY}^{(0)}}\frac{d\sigma_{DY}(z,Q^2)}{d\ln Q^2}
  =\big|\mc{H}_{DY}(Q^2)\big|^2\widetilde{W}_{DY}(z)\;,
\end{equation}
where $\mc{H}_{DY}(Q^2)$ is the hard matrix element and $\widetilde{W}_{DY}$ is the Fourier transform of the vacuum expectation value
of the Wilson loop, which accounts for the eikonal emission of soft gluons,
\begin{equation}\label{eq:wilson_loop}
  \widetilde{W}_{DY}(z)=\int_{-\infty}^\infty\frac{dy_0}{2\pi}\,e^{iy_0Q(1-z)}\,
  \langle0|\mc{P}\exp\left(ig\int_{C_{DY}}dx_\mu A^\mu\right)|0\rangle\;,
\end{equation}
and $Q(1-z)$ is the total soft gluon energy. The Wilson loop, Eq.~\eqref{eq:wilson_loop}
is related to Eqs.~\eqref{eq:lo_soft} and~\eqref{eq:nlo} in momentum space
by the Fourier transform at $b_\perp=0$, as derived in~\cite{Belitsky:1998tc,Li:2011zp}
\begin{equation}\label{eq:bspace_trafo}
  \begin{split}
    \int_{-\infty}^{\;\infty}\frac{{\rm d}^{D}p}{(2\pi)^{D-1}}\,e^{-ibp}\,
    \delta^{(2)}(b_\perp)\left(\frac{p_+p_-}{\mu^2}\right)^{-1-k\eps}
    =\;&\frac{1}{8\pi^2}\frac{(4\pi)^\eps}{\Gamma(1-\eps)}
    \frac{\Gamma(1-k\eps)^2}{\eps^2\,k^2\,e^{2k\eps\gamma_E}}\,L^{k\eps}\;,
  \end{split}
\end{equation}
where we have defined $L=-b_+b_-/(4b_0^2)$ and $b_0=e^{-\gamma_E}/\mu$. 
Applying this transformation to Eqs.~\eqref{eq:lo_soft} and~\eqref{eq:nlo},
we derive the unrenormalized one- and two-loop soft functions
\begin{equation}\label{eq:lo_b}
  \begin{split}
    S_{ij}^{\rm(1)}(b_+b_-)=
    \frac{\alpha_s(\mu)}{2\pi}\,\frac{Q^2b_+b_-}{4}
    \frac{\Gamma(1-\eps)}{\eps^2\,e^{\eps\gamma_E}}\,L^\eps
    =\frac{\alpha_s(\mu)}{2\pi}\,\frac{Q^2b_+b_-}{4}
    \left[\,\frac{1}{\eps^2}+\frac{1}{\eps}\ln L
      +\frac{1}{2}\ln^2 L+\frac{\pi^2}{12}+\mc{O}(\eps)\,\right]\;,
  \end{split}
\end{equation}
and
\begin{equation}\label{eq:nlo_b}
  \begin{split}
    S_{ij}^{\rm(2)}(b_+b_-)=-
    \frac{\alpha_s^2(\mu)}{(2\pi)^2}\,\frac{Q^2b_+b_-}{4}
    \Bigg[\,&\frac{\beta_0}{6}\ln^3 L
      +\frac{\Gamma_{\rm cusp}^{(2)}}{2}\ln^2 L
      +\ln L\left(-\frac{\beta_0}{2\eps^2}+\frac{\Gamma_{\rm cusp}^{(2)}}{2\eps}
        +\Gamma_{\rm soft}^{(2)}+\frac{\pi^2}{12}\beta_0\right)+\,\ldots\,\Bigg]\;,
  \end{split}
\end{equation}
where the dots stand for non-logarithmic contributions.
After renormalization, Eq.~\eqref{eq:nlo_b} yields the coefficients
$w_3^{(2)}=\beta_0/6$, $w_2^{(2)}=\Gamma_{\rm cusp}^{(2)}/2$
and $w_1^{(2)}=\Gamma_{\rm soft}^{(2)}+(\pi^2/12)\,\beta_0$ computed
in~\cite{Belitsky:1998tc}. Upon implementing the NLO corrections
computed in Sec.~\ref{sec:analytic} in the parton shower,
we could in principle claim the same formal accuracy in the resummation
of soft-gluon effects. However, we need to take into account that,
in contrast to Eq.~\eqref{eq:bspace_trafo} the phase-space volume
in parton-shower simulations is not infinite, but limited by
the hadronic center-of-mass energy of the collider.

In Ref.~\cite{Dasgupta:2018nvj}, an improved framework to assess the accuracy of
parton showers at fixed number of branchings was proposed. As shown above, our approach
satisfies these criteria up to the second emission. In addition, the weight defined
in Eq.~\eqref{eq:psct_soft_tc_so_weight} generalizes to higher particle multiplicity
and can be used to correct for the kinematical mismatches in the single
strong ordering region identified in Sec.~3.3 of~\cite{Dasgupta:2018nvj}.
Nevertheless, the color structure beyond two gluon emissions will not be
accurately reflected. Therefore, the problems identified in Sec.~3.2
of~\cite{Dasgupta:2018nvj} remain at higher particle multiplicity. They can
be solved by employing full color parton shower algorithms that are valid
for an arbitrary number of emissions, such as those proposed and implemented
in~\cite{Isaacson:2018zdi,Martinez:2018ffw}.

\section{Overlap between double soft and triple collinear splitting functions}
In this section we compute the overlap between the double soft splitting functions,
Eqs.~\eqref{eq:real_soft_2}, and the triple-collinear splitting functions
of~\cite{Catani:1999ss}. The triple-collinear splitting functions can be included
in the parton shower using the techniques described in~\cite{Hoche:2017iem}.
The results presented in this appendix will allow, in a future publication,
to remove double counted contributions and construct splitting functions 
that are valid in the full parton shower phase space. In terms of the
collinear variables $z_k=s_{kj}/(s_{j1}+s_{j2}+s_{ij})$~\cite{Catani:1999ss},
the $(12)$ double soft enhanced $q\bar{q}$ and $gg$ emission parts
of a $(i12)$ triple collinear splitting function can be written as
\begin{equation}\label{eq:ps_tc_doublesoft_coll}
  \begin{split}
  P_{\bar{q}_1q_2a_i}^{\rm(ds)}=&-C_a\frac{T_R}{2}\,
  \frac{s_{i12}^2}{(s_{i12}-s_{12})^2}
  \left[\left(\frac{t_{12,i}}{s_{12}}-
    \frac{z_1-z_2}{z_1+z_2}\right)^2+\frac{4z_i}{z_1+z_2}
    \left(1-\frac{s_{i12}}{s_{12}}\right)\right]\;,\\
  P_{\bar{g}_1g_2a_i}^{\rm(ds)}=&\;
  P_{\bar{g}_1g_2a_i}^{\rm(ps)}+C_a\frac{C_A}{2}\,
  \frac{s_{i12}^2}{(s_{i12}-s_{12})^2}
  \left[\,(1-\eps)\left(\frac{t_{12,i}}{s_{12}}-
    \frac{z_1-z_2}{z_1+z_2}\right)^2+\frac{8z_i}{z_1+z_2}
    \left(1-\frac{s_{i12}}{s_{12}}\right)\right]\;,
  \end{split}
\end{equation}
where the pure soft contribution is given by
\begin{equation}\label{eq:ps_tc_doublesoft_soft}
  \begin{split}
  P_{\bar{g}_1g_2a_i}^{\rm(ps)}=&\;C_a\frac{C_A}{2}
  \frac{s_{i12}}{(s_{i12}-s_{12})}
  \left[\,\frac{s_{i12}}{s_{i1}}\left(\frac{s_{i2}}{s_{12}}-\frac{1}{z_1}\right)
    \frac{1-z_i+z_2}{z_2(1-z_i)}
    +\frac{s_{i12}}{s_{12}}\left(\frac{1}{z_1}+\frac{1}{z_2}-\frac{1}{z_1+z_2}\right)\right]
  +\Big(1\leftrightarrow 2\Big)\;,
  \end{split}
\end{equation}
and where~\cite{Catani:1999ss}
\begin{equation}\label{eq:t123}
  t_{12,i}=2\,\frac{z_1s_{i2}-z_2s_{i1}}{z_1+z_2}
  +\frac{z_1-z_2}{z_1+z_2}\,s_{12}\;.
\end{equation}
The above terms may occur multiple times in each triple-collinear
splitting function, as required by the symmetry of the final state.

\bibliography{journal}

\begin{thebibliography}{59}%
\makeatletter
\providecommand \@ifxundefined [1]{%
 \@ifx{#1\undefined}
}%
\providecommand \@ifnum [1]{%
 \ifnum #1\expandafter \@firstoftwo
 \else \expandafter \@secondoftwo
 \fi
}%
\providecommand \@ifx [1]{%
 \ifx #1\expandafter \@firstoftwo
 \else \expandafter \@secondoftwo
 \fi
}%
\providecommand \natexlab [1]{#1}%
\providecommand \enquote  [1]{``#1''}%
\providecommand \bibnamefont  [1]{#1}%
\providecommand \bibfnamefont [1]{#1}%
\providecommand \citenamefont [1]{#1}%
\providecommand \href@noop [0]{\@secondoftwo}%
\providecommand \href [0]{\begingroup \@sanitize@url \@href}%
\providecommand \@href[1]{\@@startlink{#1}\@@href}%
\providecommand \@@href[1]{\endgroup#1\@@endlink}%
\providecommand \@sanitize@url [0]{\catcode `\\12\catcode `\$12\catcode
  `\&12\catcode `\#12\catcode `\^12\catcode `\_12\catcode `\%12\relax}%
\providecommand \@@startlink[1]{}%
\providecommand \@@endlink[0]{}%
\providecommand \url  [0]{\begingroup\@sanitize@url \@url }%
\providecommand \@url [1]{\endgroup\@href {#1}{\urlprefix }}%
\providecommand \urlprefix  [0]{URL }%
\providecommand \Eprint [0]{\href }%
\providecommand \doibase [0]{http://dx.doi.org/}%
\providecommand \selectlanguage [0]{\@gobble}%
\providecommand \bibinfo  [0]{\@secondoftwo}%
\providecommand \bibfield  [0]{\@secondoftwo}%
\providecommand \translation [1]{[#1]}%
\providecommand \BibitemOpen [0]{}%
\providecommand \bibitemStop [0]{}%
\providecommand \bibitemNoStop [0]{.\EOS\space}%
\providecommand \EOS [0]{\spacefactor3000\relax}%
\providecommand \BibitemShut  [1]{\csname bibitem#1\endcsname}%
\let\auto@bib@innerbib\@empty
\bibitem [{\citenamefont {Webber}(1986)}]{Webber:1986mc}%
  \BibitemOpen
  \bibfield  {author} {\bibinfo {author} {\bibfnamefont {B.}~\bibnamefont
  {Webber}},\ }\href {http://inspirebeta.net/record/18125} {\bibfield
  {journal} {\bibinfo  {journal} {Ann. Rev. Nucl. Part. Sci.}\ }\textbf
  {\bibinfo {volume} {36}},\ \bibinfo {pages} {253} (\bibinfo {year}
  {1986})}\BibitemShut {NoStop}%
\bibitem [{\citenamefont {Buckley}\ \emph {et~al.}(2011)\citenamefont {Buckley}
  \emph {et~al.}}]{Buckley:2011ms}%
  \BibitemOpen
  \bibfield  {author} {\bibinfo {author} {\bibfnamefont {A.}~\bibnamefont
  {Buckley}} \emph {et~al.},\ }\href {\doibase
  http://dx.doi.org/10.1016/j.physrep.2011.03.005} {\bibfield  {journal}
  {\bibinfo  {journal} {Phys. Rept.}\ }\textbf {\bibinfo {volume} {504}},\
  \bibinfo {pages} {145} (\bibinfo {year} {2011})},\ \Eprint
  {http://arxiv.org/abs/1101.2599} {arXiv:1101.2599 [hep-ph]} \BibitemShut
  {NoStop}%
\bibitem [{\citenamefont {Alwall}\ \emph {et~al.}(2008)\citenamefont {Alwall}
  \emph {et~al.}}]{Alwall:2007fs}%
  \BibitemOpen
  \bibfield  {author} {\bibinfo {author} {\bibfnamefont {J.}~\bibnamefont
  {Alwall}} \emph {et~al.},\ }\href {http://inspirehep.net/record/753397}
  {\bibfield  {journal} {\bibinfo  {journal} {Eur. Phys. J.}\ }\textbf
  {\bibinfo {volume} {C53}},\ \bibinfo {pages} {473} (\bibinfo {year}
  {2008})},\ \Eprint {http://arxiv.org/abs/0706.2569} {arXiv:0706.2569
  [hep-ph]} \BibitemShut {NoStop}%
\bibitem [{\citenamefont {Nason}\ and\ \citenamefont
  {Webber}(2012)}]{Nason:2012pr}%
  \BibitemOpen
  \bibfield  {author} {\bibinfo {author} {\bibfnamefont {P.}~\bibnamefont
  {Nason}}\ and\ \bibinfo {author} {\bibfnamefont {B.}~\bibnamefont {Webber}},\
  }\href {\doibase 10.1146/annurev-nucl-102711-094928} {\bibfield  {journal}
  {\bibinfo  {journal} {Ann.Rev.Nucl.Part.Sci.}\ }\textbf {\bibinfo {volume}
  {62}},\ \bibinfo {pages} {187} (\bibinfo {year} {2012})},\ \Eprint
  {http://arxiv.org/abs/1202.1251} {arXiv:1202.1251 [hep-ph]} \BibitemShut
  {NoStop}%
\bibitem [{\citenamefont {Kato}\ and\ \citenamefont
  {Munehisa}(1987)}]{Kato:1986sg}%
  \BibitemOpen
  \bibfield  {author} {\bibinfo {author} {\bibfnamefont {K.}~\bibnamefont
  {Kato}}\ and\ \bibinfo {author} {\bibfnamefont {T.}~\bibnamefont
  {Munehisa}},\ }\href {\doibase 10.1103/PhysRevD.36.61} {\bibfield  {journal}
  {\bibinfo  {journal} {Phys. Rev.}\ }\textbf {\bibinfo {volume} {D36}},\
  \bibinfo {pages} {61} (\bibinfo {year} {1987})}\BibitemShut {NoStop}%
\bibitem [{\citenamefont {Kato}\ and\ \citenamefont
  {Munehisa}(1989)}]{Kato:1988ii}%
  \BibitemOpen
  \bibfield  {author} {\bibinfo {author} {\bibfnamefont {K.}~\bibnamefont
  {Kato}}\ and\ \bibinfo {author} {\bibfnamefont {T.}~\bibnamefont
  {Munehisa}},\ }\href {\doibase 10.1103/PhysRevD.39.156} {\bibfield  {journal}
  {\bibinfo  {journal} {Phys. Rev.}\ }\textbf {\bibinfo {volume} {D39}},\
  \bibinfo {pages} {156} (\bibinfo {year} {1989})}\BibitemShut {NoStop}%
\bibitem [{\citenamefont {Kato}\ and\ \citenamefont
  {Munehisa}(1991)}]{Kato:1990as}%
  \BibitemOpen
  \bibfield  {author} {\bibinfo {author} {\bibfnamefont {K.}~\bibnamefont
  {Kato}}\ and\ \bibinfo {author} {\bibfnamefont {T.}~\bibnamefont
  {Munehisa}},\ }\href {\doibase 10.1016/0010-4655(91)90051-L} {\bibfield
  {journal} {\bibinfo  {journal} {Comput. Phys. Commun.}\ }\textbf {\bibinfo
  {volume} {64}},\ \bibinfo {pages} {67} (\bibinfo {year} {1991})}\BibitemShut
  {NoStop}%
\bibitem [{\citenamefont {Kato}\ \emph {et~al.}(1992)\citenamefont {Kato},
  \citenamefont {Munehisa},\ and\ \citenamefont {Tanaka}}]{Kato:1991fs}%
  \BibitemOpen
  \bibfield  {author} {\bibinfo {author} {\bibfnamefont {K.}~\bibnamefont
  {Kato}}, \bibinfo {author} {\bibfnamefont {T.}~\bibnamefont {Munehisa}}, \
  and\ \bibinfo {author} {\bibfnamefont {H.}~\bibnamefont {Tanaka}},\ }\href
  {\doibase 10.1007/BF01559457} {\bibfield  {journal} {\bibinfo  {journal} {Z.
  Phys.}\ }\textbf {\bibinfo {volume} {C54}},\ \bibinfo {pages} {397} (\bibinfo
  {year} {1992})}\BibitemShut {NoStop}%
\bibitem [{\citenamefont {Giele}\ \emph {et~al.}(2011)\citenamefont {Giele},
  \citenamefont {Kosower},\ and\ \citenamefont {Skands}}]{Giele:2011cb}%
  \BibitemOpen
  \bibfield  {author} {\bibinfo {author} {\bibfnamefont {W.~T.}\ \bibnamefont
  {Giele}}, \bibinfo {author} {\bibfnamefont {D.~A.}\ \bibnamefont {Kosower}},
  \ and\ \bibinfo {author} {\bibfnamefont {P.~Z.}\ \bibnamefont {Skands}},\
  }\href {\doibase 10.1103/PhysRevD.84.054003} {\bibfield  {journal} {\bibinfo
  {journal} {Phys. Rev.}\ }\textbf {\bibinfo {volume} {D84}},\ \bibinfo {pages}
  {054003} (\bibinfo {year} {2011})},\ \Eprint {http://arxiv.org/abs/1102.2126}
  {arXiv:1102.2126 [hep-ph]} \BibitemShut {NoStop}%
\bibitem [{\citenamefont {Hartgring}\ \emph {et~al.}(2013)\citenamefont
  {Hartgring}, \citenamefont {Laenen},\ and\ \citenamefont
  {Skands}}]{Hartgring:2013jma}%
  \BibitemOpen
  \bibfield  {author} {\bibinfo {author} {\bibfnamefont {L.}~\bibnamefont
  {Hartgring}}, \bibinfo {author} {\bibfnamefont {E.}~\bibnamefont {Laenen}}, \
  and\ \bibinfo {author} {\bibfnamefont {P.}~\bibnamefont {Skands}},\ }\href
  {\doibase 10.1007/JHEP10(2013)127} {\bibfield  {journal} {\bibinfo  {journal}
  {JHEP}\ }\textbf {\bibinfo {volume} {10}},\ \bibinfo {pages} {127} (\bibinfo
  {year} {2013})},\ \Eprint {http://arxiv.org/abs/1303.4974} {arXiv:1303.4974
  [hep-ph]} \BibitemShut {NoStop}%
\bibitem [{\citenamefont {Li}\ and\ \citenamefont {Skands}(2017)}]{Li:2016yez}%
  \BibitemOpen
  \bibfield  {author} {\bibinfo {author} {\bibfnamefont {H.~T.}\ \bibnamefont
  {Li}}\ and\ \bibinfo {author} {\bibfnamefont {P.}~\bibnamefont {Skands}},\
  }\href {\doibase 10.1016/j.physletb.2017.05.011} {\bibfield  {journal}
  {\bibinfo  {journal} {Phys. Lett.}\ }\textbf {\bibinfo {volume} {B771}},\
  \bibinfo {pages} {59} (\bibinfo {year} {2017})},\ \Eprint
  {http://arxiv.org/abs/1611.00013} {arXiv:1611.00013 [hep-ph]} \BibitemShut
  {NoStop}%
\bibitem [{\citenamefont {Nagy}\ and\ \citenamefont
  {Soper}(2018)}]{Nagy:2017ggp}%
  \BibitemOpen
  \bibfield  {author} {\bibinfo {author} {\bibfnamefont {Z.}~\bibnamefont
  {Nagy}}\ and\ \bibinfo {author} {\bibfnamefont {D.~E.}\ \bibnamefont
  {Soper}},\ }\href {\doibase 10.1103/PhysRevD.98.014034} {\bibfield  {journal}
  {\bibinfo  {journal} {Phys. Rev.}\ }\textbf {\bibinfo {volume} {D98}},\
  \bibinfo {pages} {014034} (\bibinfo {year} {2018})},\ \Eprint
  {http://arxiv.org/abs/1705.08093} {arXiv:1705.08093 [hep-ph]} \BibitemShut
  {NoStop}%
\bibitem [{\citenamefont {H{\"o}che}\ and\ \citenamefont
  {Prestel}(2017)}]{Hoche:2017iem}%
  \BibitemOpen
  \bibfield  {author} {\bibinfo {author} {\bibfnamefont {S.}~\bibnamefont
  {H{\"o}che}}\ and\ \bibinfo {author} {\bibfnamefont {S.}~\bibnamefont
  {Prestel}},\ }\href {\doibase 10.1103/PhysRevD.96.074017} {\bibfield
  {journal} {\bibinfo  {journal} {Phys. Rev.}\ }\textbf {\bibinfo {volume}
  {D96}},\ \bibinfo {pages} {074017} (\bibinfo {year} {2017})},\ \Eprint
  {http://arxiv.org/abs/1705.00742} {arXiv:1705.00742 [hep-ph]} \BibitemShut
  {NoStop}%
\bibitem [{\citenamefont {H{\"o}che}\ \emph {et~al.}(2017)\citenamefont
  {H{\"o}che}, \citenamefont {Krauss},\ and\ \citenamefont
  {Prestel}}]{Hoche:2017hno}%
  \BibitemOpen
  \bibfield  {author} {\bibinfo {author} {\bibfnamefont {S.}~\bibnamefont
  {H{\"o}che}}, \bibinfo {author} {\bibfnamefont {F.}~\bibnamefont {Krauss}}, \
  and\ \bibinfo {author} {\bibfnamefont {S.}~\bibnamefont {Prestel}},\ }\href
  {\doibase 10.1007/JHEP10(2017)093} {\bibfield  {journal} {\bibinfo  {journal}
  {JHEP}\ }\textbf {\bibinfo {volume} {10}},\ \bibinfo {pages} {093} (\bibinfo
  {year} {2017})},\ \Eprint {http://arxiv.org/abs/1705.00982} {arXiv:1705.00982
  [hep-ph]} \BibitemShut {NoStop}%
\bibitem [{\citenamefont {H{\"o}che}\ \emph {et~al.}(2018)\citenamefont
  {H{\"o}che}, \citenamefont {Reichelt},\ and\ \citenamefont
  {Siegert}}]{Hoeche:2017jsi}%
  \BibitemOpen
  \bibfield  {author} {\bibinfo {author} {\bibfnamefont {S.}~\bibnamefont
  {H{\"o}che}}, \bibinfo {author} {\bibfnamefont {D.}~\bibnamefont {Reichelt}},
  \ and\ \bibinfo {author} {\bibfnamefont {F.}~\bibnamefont {Siegert}},\ }\href
  {\doibase 10.1007/JHEP01(2018)118} {\bibfield  {journal} {\bibinfo  {journal}
  {JHEP}\ }\textbf {\bibinfo {volume} {01}},\ \bibinfo {pages} {118} (\bibinfo
  {year} {2018})},\ \Eprint {http://arxiv.org/abs/1711.03497} {arXiv:1711.03497
  [hep-ph]} \BibitemShut {NoStop}%
\bibitem [{\citenamefont {Kodaira}\ and\ \citenamefont
  {Trentadue}(1982)}]{Kodaira:1981nh}%
  \BibitemOpen
  \bibfield  {author} {\bibinfo {author} {\bibfnamefont {J.}~\bibnamefont
  {Kodaira}}\ and\ \bibinfo {author} {\bibfnamefont {L.}~\bibnamefont
  {Trentadue}},\ }\href {\doibase 10.1016/0370-2693(82)90907-8} {\bibfield
  {journal} {\bibinfo  {journal} {Phys.Lett.}\ }\textbf {\bibinfo {volume}
  {B112}},\ \bibinfo {pages} {66} (\bibinfo {year} {1982})}\BibitemShut
  {NoStop}%
\bibitem [{\citenamefont {Davies}\ and\ \citenamefont
  {Stirling}(1984)}]{Davies:1984hs}%
  \BibitemOpen
  \bibfield  {author} {\bibinfo {author} {\bibfnamefont {C.}~\bibnamefont
  {Davies}}\ and\ \bibinfo {author} {\bibfnamefont {W.~J.}\ \bibnamefont
  {Stirling}},\ }\href {\doibase 10.1016/0550-3213(84)90316-X} {\bibfield
  {journal} {\bibinfo  {journal} {Nucl.Phys.}\ }\textbf {\bibinfo {volume}
  {B244}},\ \bibinfo {pages} {337} (\bibinfo {year} {1984})}\BibitemShut
  {NoStop}%
\bibitem [{\citenamefont {Davies}\ \emph {et~al.}(1985)\citenamefont {Davies},
  \citenamefont {Webber},\ and\ \citenamefont {Stirling}}]{Davies:1984sp}%
  \BibitemOpen
  \bibfield  {author} {\bibinfo {author} {\bibfnamefont {C.}~\bibnamefont
  {Davies}}, \bibinfo {author} {\bibfnamefont {B.}~\bibnamefont {Webber}}, \
  and\ \bibinfo {author} {\bibfnamefont {W.}~\bibnamefont {Stirling}},\ }\href
  {\doibase 10.1016/0550-3213(85)90402-X} {\bibfield  {journal} {\bibinfo
  {journal} {Nucl. Phys.}\ }\textbf {\bibinfo {volume} {B256}},\ \bibinfo
  {pages} {413} (\bibinfo {year} {1985})}\BibitemShut {NoStop}%
\bibitem [{\citenamefont {Catani}\ \emph {et~al.}(1988)\citenamefont {Catani},
  \citenamefont {D'Emilio},\ and\ \citenamefont {Trentadue}}]{Catani:1988vd}%
  \BibitemOpen
  \bibfield  {author} {\bibinfo {author} {\bibfnamefont {S.}~\bibnamefont
  {Catani}}, \bibinfo {author} {\bibfnamefont {E.}~\bibnamefont {D'Emilio}}, \
  and\ \bibinfo {author} {\bibfnamefont {L.}~\bibnamefont {Trentadue}},\ }\href
  {\doibase 10.1016/0370-2693(88)90912-4} {\bibfield  {journal} {\bibinfo
  {journal} {Phys.Lett.}\ }\textbf {\bibinfo {volume} {B211}},\ \bibinfo
  {pages} {335} (\bibinfo {year} {1988})}\BibitemShut {NoStop}%
\bibitem [{\citenamefont {Catani}\ \emph
  {et~al.}(1991{\natexlab{a}})\citenamefont {Catani}, \citenamefont {Webber},\
  and\ \citenamefont {Marchesini}}]{Catani:1990rr}%
  \BibitemOpen
  \bibfield  {author} {\bibinfo {author} {\bibfnamefont {S.}~\bibnamefont
  {Catani}}, \bibinfo {author} {\bibfnamefont {B.~R.}\ \bibnamefont {Webber}},
  \ and\ \bibinfo {author} {\bibfnamefont {G.}~\bibnamefont {Marchesini}},\
  }\href {\doibase 10.1016/0550-3213(91)90390-J} {\bibfield  {journal}
  {\bibinfo  {journal} {Nucl. Phys.}\ }\textbf {\bibinfo {volume} {B349}},\
  \bibinfo {pages} {635} (\bibinfo {year} {1991}{\natexlab{a}})}\BibitemShut
  {NoStop}%
\bibitem [{\citenamefont {H{\"o}che}\ and\ \citenamefont
  {Prestel}(2015)}]{Hoche:2015sya}%
  \BibitemOpen
  \bibfield  {author} {\bibinfo {author} {\bibfnamefont {S.}~\bibnamefont
  {H{\"o}che}}\ and\ \bibinfo {author} {\bibfnamefont {S.}~\bibnamefont
  {Prestel}},\ }\href {\doibase 10.1140/epjc/s10052-015-3684-2} {\bibfield
  {journal} {\bibinfo  {journal} {Eur. Phys. J.}\ }\textbf {\bibinfo {volume}
  {C75}},\ \bibinfo {pages} {461} (\bibinfo {year} {2015})},\ \Eprint
  {http://arxiv.org/abs/1506.05057} {arXiv:1506.05057 [hep-ph]} \BibitemShut
  {NoStop}%
\bibitem [{\citenamefont {Sterman}(1987)}]{Sterman:1986aj}%
  \BibitemOpen
  \bibfield  {author} {\bibinfo {author} {\bibfnamefont {G.~F.}\ \bibnamefont
  {Sterman}},\ }\href {\doibase 10.1016/0550-3213(87)90258-6} {\bibfield
  {journal} {\bibinfo  {journal} {Nucl. Phys.}\ }\textbf {\bibinfo {volume}
  {B281}},\ \bibinfo {pages} {310} (\bibinfo {year} {1987})}\BibitemShut
  {NoStop}%
\bibitem [{\citenamefont {Collins}\ \emph
  {et~al.}(1988{\natexlab{a}})\citenamefont {Collins}, \citenamefont {Soper},\
  and\ \citenamefont {Sterman}}]{Collins:1988ig}%
  \BibitemOpen
  \bibfield  {author} {\bibinfo {author} {\bibfnamefont {J.~C.}\ \bibnamefont
  {Collins}}, \bibinfo {author} {\bibfnamefont {D.~E.}\ \bibnamefont {Soper}},
  \ and\ \bibinfo {author} {\bibfnamefont {G.}~\bibnamefont {Sterman}},\ }\href
  {http://inspirehep.net/search?j=NUPHA,B308,833} {\bibfield  {journal}
  {\bibinfo  {journal} {Nucl. Phys.}\ }\textbf {\bibinfo {volume} {B308}},\
  \bibinfo {pages} {833} (\bibinfo {year} {1988}{\natexlab{a}})}\BibitemShut
  {NoStop}%
\bibitem [{\citenamefont {Collins}\ \emph
  {et~al.}(1988{\natexlab{b}})\citenamefont {Collins}, \citenamefont {Soper},\
  and\ \citenamefont {Sterman}}]{Collins:1989gx}%
  \BibitemOpen
  \bibfield  {author} {\bibinfo {author} {\bibfnamefont {J.~C.}\ \bibnamefont
  {Collins}}, \bibinfo {author} {\bibfnamefont {D.~E.}\ \bibnamefont {Soper}},
  \ and\ \bibinfo {author} {\bibfnamefont {G.}~\bibnamefont {Sterman}},\ }\href
  {http://inspirehep.net/search?p=hep-ph/0409313} {\bibfield  {journal}
  {\bibinfo  {journal} {Adv. Ser. Direct. High Energy Phys.}\ }\textbf
  {\bibinfo {volume} {5}},\ \bibinfo {pages} {1} (\bibinfo {year}
  {1988}{\natexlab{b}})},\ \Eprint {http://arxiv.org/abs/hep-ph/0409313}
  {hep-ph/0409313} \BibitemShut {NoStop}%
\bibitem [{\citenamefont {Monni}\ \emph {et~al.}(2011)\citenamefont {Monni},
  \citenamefont {Gehrmann},\ and\ \citenamefont {Luisoni}}]{Monni:2011gb}%
  \BibitemOpen
  \bibfield  {author} {\bibinfo {author} {\bibfnamefont {P.~F.}\ \bibnamefont
  {Monni}}, \bibinfo {author} {\bibfnamefont {T.}~\bibnamefont {Gehrmann}}, \
  and\ \bibinfo {author} {\bibfnamefont {G.}~\bibnamefont {Luisoni}},\ }\href
  {\doibase 10.1007/JHEP08(2011)010} {\bibfield  {journal} {\bibinfo  {journal}
  {JHEP}\ }\textbf {\bibinfo {volume} {08}},\ \bibinfo {pages} {010} (\bibinfo
  {year} {2011})},\ \Eprint {http://arxiv.org/abs/1105.4560} {arXiv:1105.4560
  [hep-ph]} \BibitemShut {NoStop}%
\bibitem [{\citenamefont {Catani}\ and\ \citenamefont
  {Seymour}(1997)}]{Catani:1996vz}%
  \BibitemOpen
  \bibfield  {author} {\bibinfo {author} {\bibfnamefont {S.}~\bibnamefont
  {Catani}}\ and\ \bibinfo {author} {\bibfnamefont {M.~H.}\ \bibnamefont
  {Seymour}},\ }\href {http://inspirehep.net/search?p=hep-ph/9605323}
  {\bibfield  {journal} {\bibinfo  {journal} {Nucl. Phys.}\ }\textbf {\bibinfo
  {volume} {B485}},\ \bibinfo {pages} {291} (\bibinfo {year} {1997})},\ \Eprint
  {http://arxiv.org/abs/hep-ph/9605323} {hep-ph/9605323} \BibitemShut {NoStop}%
\bibitem [{\citenamefont {Cornwall}\ and\ \citenamefont
  {Tiktopoulos}(1976)}]{Cornwall:1975ty}%
  \BibitemOpen
  \bibfield  {author} {\bibinfo {author} {\bibfnamefont {J.~M.}\ \bibnamefont
  {Cornwall}}\ and\ \bibinfo {author} {\bibfnamefont {G.}~\bibnamefont
  {Tiktopoulos}},\ }\href {http://inspirehep.net/search?j=PHRVA,D13,3370}
  {\bibfield  {journal} {\bibinfo  {journal} {Phys. Rev.}\ }\textbf {\bibinfo
  {volume} {D13}},\ \bibinfo {pages} {3370} (\bibinfo {year}
  {1976})}\BibitemShut {NoStop}%
\bibitem [{\citenamefont {Frenkel}\ and\ \citenamefont
  {Taylor}(1976)}]{Frenkel:1976bj}%
  \BibitemOpen
  \bibfield  {author} {\bibinfo {author} {\bibfnamefont {J.}~\bibnamefont
  {Frenkel}}\ and\ \bibinfo {author} {\bibfnamefont {J.~C.}\ \bibnamefont
  {Taylor}},\ }\href {http://inspirehep.net/search?j=NUPHA,B116,185} {\bibfield
   {journal} {\bibinfo  {journal} {Nucl. Phys.}\ }\textbf {\bibinfo {volume}
  {B116}},\ \bibinfo {pages} {185} (\bibinfo {year} {1976})}\BibitemShut
  {NoStop}%
\bibitem [{\citenamefont {Bern}\ \emph {et~al.}(1999)\citenamefont {Bern},
  \citenamefont {Del~Duca}, \citenamefont {Kilgore},\ and\ \citenamefont
  {Schmidt}}]{Bern:1999ry}%
  \BibitemOpen
  \bibfield  {author} {\bibinfo {author} {\bibfnamefont {Z.}~\bibnamefont
  {Bern}}, \bibinfo {author} {\bibfnamefont {V.}~\bibnamefont {Del~Duca}},
  \bibinfo {author} {\bibfnamefont {W.~B.}\ \bibnamefont {Kilgore}}, \ and\
  \bibinfo {author} {\bibfnamefont {C.~R.}\ \bibnamefont {Schmidt}},\ }\href
  {\doibase 10.1103/PhysRevD.60.116001} {\bibfield  {journal} {\bibinfo
  {journal} {Phys. Rev.}\ }\textbf {\bibinfo {volume} {D60}},\ \bibinfo {pages}
  {116001} (\bibinfo {year} {1999})},\ \Eprint
  {http://arxiv.org/abs/hep-ph/9903516} {hep-ph/9903516} \BibitemShut {NoStop}%
\bibitem [{\citenamefont {Catani}\ and\ \citenamefont
  {Grazzini}(2000{\natexlab{a}})}]{Catani:2000pi}%
  \BibitemOpen
  \bibfield  {author} {\bibinfo {author} {\bibfnamefont {S.}~\bibnamefont
  {Catani}}\ and\ \bibinfo {author} {\bibfnamefont {M.}~\bibnamefont
  {Grazzini}},\ }\href {\doibase 10.1016/S0550-3213(00)00572-1} {\bibfield
  {journal} {\bibinfo  {journal} {Nucl. Phys.}\ }\textbf {\bibinfo {volume}
  {B591}},\ \bibinfo {pages} {435} (\bibinfo {year} {2000}{\natexlab{a}})},\
  \Eprint {http://arxiv.org/abs/hep-ph/0007142} {hep-ph/0007142} \BibitemShut
  {NoStop}%
\bibitem [{\citenamefont {Catani}\ and\ \citenamefont
  {Grazzini}(2000{\natexlab{b}})}]{Catani:1999ss}%
  \BibitemOpen
  \bibfield  {author} {\bibinfo {author} {\bibfnamefont {S.}~\bibnamefont
  {Catani}}\ and\ \bibinfo {author} {\bibfnamefont {M.}~\bibnamefont
  {Grazzini}},\ }\href {\doibase 10.1016/S0550-3213(99)00778-6} {\bibfield
  {journal} {\bibinfo  {journal} {Nucl. Phys.}\ }\textbf {\bibinfo {volume}
  {B570}},\ \bibinfo {pages} {287} (\bibinfo {year} {2000}{\natexlab{b}})},\
  \Eprint {http://arxiv.org/abs/hep-ph/9908523} {hep-ph/9908523} \BibitemShut
  {NoStop}%
\bibitem [{\citenamefont {Gardi}\ and\ \citenamefont
  {Magnea}(2009)}]{Gardi:2009qi}%
  \BibitemOpen
  \bibfield  {author} {\bibinfo {author} {\bibfnamefont {E.}~\bibnamefont
  {Gardi}}\ and\ \bibinfo {author} {\bibfnamefont {L.}~\bibnamefont {Magnea}},\
  }\href {\doibase 10.1088/1126-6708/2009/03/079} {\bibfield  {journal}
  {\bibinfo  {journal} {JHEP}\ }\textbf {\bibinfo {volume} {03}},\ \bibinfo
  {pages} {079} (\bibinfo {year} {2009})},\ \Eprint
  {http://arxiv.org/abs/0901.1091} {arXiv:0901.1091 [hep-ph]} \BibitemShut
  {NoStop}%
\bibitem [{\citenamefont {Becher}\ and\ \citenamefont
  {Neubert}(2009)}]{Becher:2009cu}%
  \BibitemOpen
  \bibfield  {author} {\bibinfo {author} {\bibfnamefont {T.}~\bibnamefont
  {Becher}}\ and\ \bibinfo {author} {\bibfnamefont {M.}~\bibnamefont
  {Neubert}},\ }\href {\doibase 10.1103/PhysRevLett.102.162001} {\bibfield
  {journal} {\bibinfo  {journal} {Phys. Rev. Lett.}\ }\textbf {\bibinfo
  {volume} {102}},\ \bibinfo {pages} {162001} (\bibinfo {year} {2009})},\
  \Eprint {http://arxiv.org/abs/0901.0722} {arXiv:0901.0722 [hep-ph]}
  \BibitemShut {NoStop}%
\bibitem [{\citenamefont {Belitsky}(1998)}]{Belitsky:1998tc}%
  \BibitemOpen
  \bibfield  {author} {\bibinfo {author} {\bibfnamefont {A.~V.}\ \bibnamefont
  {Belitsky}},\ }\href {\doibase 10.1016/S0370-2693(98)01249-0} {\bibfield
  {journal} {\bibinfo  {journal} {Phys. Lett.}\ }\textbf {\bibinfo {volume}
  {B442}},\ \bibinfo {pages} {307} (\bibinfo {year} {1998})},\ \Eprint
  {http://arxiv.org/abs/hep-ph/9808389} {hep-ph/9808389} \BibitemShut {NoStop}%
\bibitem [{\citenamefont {Li}\ \emph {et~al.}(2011)\citenamefont {Li},
  \citenamefont {Mantry},\ and\ \citenamefont {Petriello}}]{Li:2011zp}%
  \BibitemOpen
  \bibfield  {author} {\bibinfo {author} {\bibfnamefont {Y.}~\bibnamefont
  {Li}}, \bibinfo {author} {\bibfnamefont {S.}~\bibnamefont {Mantry}}, \ and\
  \bibinfo {author} {\bibfnamefont {F.}~\bibnamefont {Petriello}},\ }\href
  {\doibase 10.1103/PhysRevD.84.094014} {\bibfield  {journal} {\bibinfo
  {journal} {Phys. Rev.}\ }\textbf {\bibinfo {volume} {D84}},\ \bibinfo {pages}
  {094014} (\bibinfo {year} {2011})},\ \Eprint {http://arxiv.org/abs/1105.5171}
  {arXiv:1105.5171 [hep-ph]} \BibitemShut {NoStop}%
\bibitem [{\citenamefont {Amati}\ \emph {et~al.}(1980)\citenamefont {Amati},
  \citenamefont {Bassetto}, \citenamefont {Ciafaloni}, \citenamefont
  {Marchesini},\ and\ \citenamefont {Veneziano}}]{Amati:1980ch}%
  \BibitemOpen
  \bibfield  {author} {\bibinfo {author} {\bibfnamefont {D.}~\bibnamefont
  {Amati}}, \bibinfo {author} {\bibfnamefont {A.}~\bibnamefont {Bassetto}},
  \bibinfo {author} {\bibfnamefont {M.}~\bibnamefont {Ciafaloni}}, \bibinfo
  {author} {\bibfnamefont {G.}~\bibnamefont {Marchesini}}, \ and\ \bibinfo
  {author} {\bibfnamefont {G.}~\bibnamefont {Veneziano}},\ }\href {\doibase
  10.1016/0550-3213(80)90012-7} {\bibfield  {journal} {\bibinfo  {journal}
  {Nucl. Phys.}\ }\textbf {\bibinfo {volume} {B173}},\ \bibinfo {pages} {429}
  (\bibinfo {year} {1980})}\BibitemShut {NoStop}%
\bibitem [{\citenamefont {Frixione}\ and\ \citenamefont
  {Webber}(2002)}]{Frixione:2002ik}%
  \BibitemOpen
  \bibfield  {author} {\bibinfo {author} {\bibfnamefont {S.}~\bibnamefont
  {Frixione}}\ and\ \bibinfo {author} {\bibfnamefont {B.~R.}\ \bibnamefont
  {Webber}},\ }\href {http://inspirehep.net/search?p=hep-ph/0204244} {\bibfield
   {journal} {\bibinfo  {journal} {JHEP}\ }\textbf {\bibinfo {volume} {06}},\
  \bibinfo {pages} {029} (\bibinfo {year} {2002})},\ \Eprint
  {http://arxiv.org/abs/hep-ph/0204244} {hep-ph/0204244} \BibitemShut {NoStop}%
\bibitem [{\citenamefont {Catani}\ \emph {et~al.}(2002)\citenamefont {Catani},
  \citenamefont {Dittmaier}, \citenamefont {Seymour},\ and\ \citenamefont
  {Trocsanyi}}]{Catani:2002hc}%
  \BibitemOpen
  \bibfield  {author} {\bibinfo {author} {\bibfnamefont {S.}~\bibnamefont
  {Catani}}, \bibinfo {author} {\bibfnamefont {S.}~\bibnamefont {Dittmaier}},
  \bibinfo {author} {\bibfnamefont {M.~H.}\ \bibnamefont {Seymour}}, \ and\
  \bibinfo {author} {\bibfnamefont {Z.}~\bibnamefont {Trocsanyi}},\ }\href
  {http://inspirehep.net/search?p=hep-ph/0201036} {\bibfield  {journal}
  {\bibinfo  {journal} {Nucl. Phys.}\ }\textbf {\bibinfo {volume} {B627}},\
  \bibinfo {pages} {189} (\bibinfo {year} {2002})},\ \Eprint
  {http://arxiv.org/abs/hep-ph/0201036} {hep-ph/0201036} \BibitemShut {NoStop}%
\bibitem [{\citenamefont {Fischer}\ and\ \citenamefont
  {Prestel}(2017)}]{Fischer:2017yja}%
  \BibitemOpen
  \bibfield  {author} {\bibinfo {author} {\bibfnamefont {N.}~\bibnamefont
  {Fischer}}\ and\ \bibinfo {author} {\bibfnamefont {S.}~\bibnamefont
  {Prestel}},\ }\href {\doibase 10.1140/epjc/s10052-017-5160-7} {\bibfield
  {journal} {\bibinfo  {journal} {Eur. Phys. J.}\ }\textbf {\bibinfo {volume}
  {C77}},\ \bibinfo {pages} {601} (\bibinfo {year} {2017})},\ \Eprint
  {http://arxiv.org/abs/1706.06218} {arXiv:1706.06218 [hep-ph]} \BibitemShut
  {NoStop}%
\bibitem [{\citenamefont {Curci}\ \emph {et~al.}(1980)\citenamefont {Curci},
  \citenamefont {Furmanski},\ and\ \citenamefont {Petronzio}}]{Curci:1980uw}%
  \BibitemOpen
  \bibfield  {author} {\bibinfo {author} {\bibfnamefont {G.}~\bibnamefont
  {Curci}}, \bibinfo {author} {\bibfnamefont {W.}~\bibnamefont {Furmanski}}, \
  and\ \bibinfo {author} {\bibfnamefont {R.}~\bibnamefont {Petronzio}},\ }\href
  {\doibase 10.1016/0550-3213(80)90003-6} {\bibfield  {journal} {\bibinfo
  {journal} {Nucl. Phys.}\ }\textbf {\bibinfo {volume} {B175}},\ \bibinfo
  {pages} {27} (\bibinfo {year} {1980})}\BibitemShut {NoStop}%
\bibitem [{\citenamefont {Chiu}\ \emph {et~al.}(2012)\citenamefont {Chiu},
  \citenamefont {Jain}, \citenamefont {Neill},\ and\ \citenamefont
  {Rothstein}}]{Chiu:2011qc}%
  \BibitemOpen
  \bibfield  {author} {\bibinfo {author} {\bibfnamefont {J.-y.}\ \bibnamefont
  {Chiu}}, \bibinfo {author} {\bibfnamefont {A.}~\bibnamefont {Jain}}, \bibinfo
  {author} {\bibfnamefont {D.}~\bibnamefont {Neill}}, \ and\ \bibinfo {author}
  {\bibfnamefont {I.~Z.}\ \bibnamefont {Rothstein}},\ }\href {\doibase
  10.1103/PhysRevLett.108.151601} {\bibfield  {journal} {\bibinfo  {journal}
  {Phys. Rev. Lett.}\ }\textbf {\bibinfo {volume} {108}},\ \bibinfo {pages}
  {151601} (\bibinfo {year} {2012})},\ \Eprint {http://arxiv.org/abs/1104.0881}
  {arXiv:1104.0881 [hep-ph]} \BibitemShut {NoStop}%
\bibitem [{\citenamefont {Li}\ \emph {et~al.}(2016)\citenamefont {Li},
  \citenamefont {Neill},\ and\ \citenamefont {Zhu}}]{Li:2016axz}%
  \BibitemOpen
  \bibfield  {author} {\bibinfo {author} {\bibfnamefont {Y.}~\bibnamefont
  {Li}}, \bibinfo {author} {\bibfnamefont {D.}~\bibnamefont {Neill}}, \ and\
  \bibinfo {author} {\bibfnamefont {H.~X.}\ \bibnamefont {Zhu}},\ }\href@noop
  {} {\  (\bibinfo {year} {2016})},\ \Eprint {http://arxiv.org/abs/1604.00392}
  {arXiv:1604.00392 [hep-ph]} \BibitemShut {NoStop}%
\bibitem [{\citenamefont {Sj{\"o}strand}(1985)}]{Sjostrand:1985xi}%
  \BibitemOpen
  \bibfield  {author} {\bibinfo {author} {\bibfnamefont {T.}~\bibnamefont
  {Sj{\"o}strand}},\ }\href {http://inspirehep.net/search?j=PHLTA,B157,321}
  {\bibfield  {journal} {\bibinfo  {journal} {Phys. Lett.}\ }\textbf {\bibinfo
  {volume} {B157}},\ \bibinfo {pages} {321} (\bibinfo {year}
  {1985})}\BibitemShut {NoStop}%
\bibitem [{\citenamefont {Sj{\"o}strand}\ \emph {et~al.}(2015)\citenamefont
  {Sj{\"o}strand}, \citenamefont {Ask}, \citenamefont {Christiansen},
  \citenamefont {Corke}, \citenamefont {Desai}, \citenamefont {Ilten},
  \citenamefont {Mrenna}, \citenamefont {Prestel}, \citenamefont {Rasmussen},\
  and\ \citenamefont {Skands}}]{Sjostrand:2014zea}%
  \BibitemOpen
  \bibfield  {author} {\bibinfo {author} {\bibfnamefont {T.}~\bibnamefont
  {Sj{\"o}strand}}, \bibinfo {author} {\bibfnamefont {S.}~\bibnamefont {Ask}},
  \bibinfo {author} {\bibfnamefont {J.~R.}\ \bibnamefont {Christiansen}},
  \bibinfo {author} {\bibfnamefont {R.}~\bibnamefont {Corke}}, \bibinfo
  {author} {\bibfnamefont {N.}~\bibnamefont {Desai}}, \bibinfo {author}
  {\bibfnamefont {P.}~\bibnamefont {Ilten}}, \bibinfo {author} {\bibfnamefont
  {S.}~\bibnamefont {Mrenna}}, \bibinfo {author} {\bibfnamefont
  {S.}~\bibnamefont {Prestel}}, \bibinfo {author} {\bibfnamefont {C.~O.}\
  \bibnamefont {Rasmussen}}, \ and\ \bibinfo {author} {\bibfnamefont {P.~Z.}\
  \bibnamefont {Skands}},\ }\href {\doibase 10.1016/j.cpc.2015.01.024}
  {\bibfield  {journal} {\bibinfo  {journal} {Comput. Phys. Commun.}\ }\textbf
  {\bibinfo {volume} {191}},\ \bibinfo {pages} {159} (\bibinfo {year}
  {2015})},\ \Eprint {http://arxiv.org/abs/1410.3012} {arXiv:1410.3012
  [hep-ph]} \BibitemShut {NoStop}%
\bibitem [{\citenamefont {Gleisberg}\ \emph {et~al.}(2004)\citenamefont
  {Gleisberg}, \citenamefont {H{\"o}che}, \citenamefont {Krauss}, \citenamefont
  {Sch{\"a}licke}, \citenamefont {Schumann},\ and\ \citenamefont
  {Winter}}]{Gleisberg:2003xi}%
  \BibitemOpen
  \bibfield  {author} {\bibinfo {author} {\bibfnamefont {T.}~\bibnamefont
  {Gleisberg}}, \bibinfo {author} {\bibfnamefont {S.}~\bibnamefont
  {H{\"o}che}}, \bibinfo {author} {\bibfnamefont {F.}~\bibnamefont {Krauss}},
  \bibinfo {author} {\bibfnamefont {A.}~\bibnamefont {Sch{\"a}licke}}, \bibinfo
  {author} {\bibfnamefont {S.}~\bibnamefont {Schumann}}, \ and\ \bibinfo
  {author} {\bibfnamefont {J.}~\bibnamefont {Winter}},\ }\href
  {http://inspirehep.net/search?irn=5730570} {\bibfield  {journal} {\bibinfo
  {journal} {JHEP}\ }\textbf {\bibinfo {volume} {02}},\ \bibinfo {pages} {056}
  (\bibinfo {year} {2004})},\ \Eprint {http://arxiv.org/abs/hep-ph/0311263}
  {hep-ph/0311263} \BibitemShut {NoStop}%
\bibitem [{\citenamefont {Gleisberg}\ \emph {et~al.}(2009)\citenamefont
  {Gleisberg}, \citenamefont {H{\"o}che}, \citenamefont {Krauss}, \citenamefont
  {Sch\"{o}nherr}, \citenamefont {Schumann}, \citenamefont {Siegert},\ and\
  \citenamefont {Winter}}]{Gleisberg:2008ta}%
  \BibitemOpen
  \bibfield  {author} {\bibinfo {author} {\bibfnamefont {T.}~\bibnamefont
  {Gleisberg}}, \bibinfo {author} {\bibfnamefont {S.}~\bibnamefont
  {H{\"o}che}}, \bibinfo {author} {\bibfnamefont {F.}~\bibnamefont {Krauss}},
  \bibinfo {author} {\bibfnamefont {M.}~\bibnamefont {Sch\"{o}nherr}}, \bibinfo
  {author} {\bibfnamefont {S.}~\bibnamefont {Schumann}}, \bibinfo {author}
  {\bibfnamefont {F.}~\bibnamefont {Siegert}}, \ and\ \bibinfo {author}
  {\bibfnamefont {J.}~\bibnamefont {Winter}},\ }\href {\doibase
  10.1088/1126-6708/2009/02/007} {\bibfield  {journal} {\bibinfo  {journal}
  {JHEP}\ }\textbf {\bibinfo {volume} {02}},\ \bibinfo {pages} {007} (\bibinfo
  {year} {2009})},\ \Eprint {http://arxiv.org/abs/0811.4622} {arXiv:0811.4622
  [hep-ph]} \BibitemShut {NoStop}%
\bibitem [{\citenamefont {Lai}\ \emph {et~al.}(2010)\citenamefont {Lai},
  \citenamefont {Guzzi}, \citenamefont {Huston}, \citenamefont {Li},
  \citenamefont {Nadolsky}, \citenamefont {Pumplin},\ and\ \citenamefont
  {Yuan}}]{Lai:2010vv}%
  \BibitemOpen
  \bibfield  {author} {\bibinfo {author} {\bibfnamefont {H.-L.}\ \bibnamefont
  {Lai}}, \bibinfo {author} {\bibfnamefont {M.}~\bibnamefont {Guzzi}}, \bibinfo
  {author} {\bibfnamefont {J.}~\bibnamefont {Huston}}, \bibinfo {author}
  {\bibfnamefont {Z.}~\bibnamefont {Li}}, \bibinfo {author} {\bibfnamefont
  {P.~M.}\ \bibnamefont {Nadolsky}}, \bibinfo {author} {\bibfnamefont
  {J.}~\bibnamefont {Pumplin}}, \ and\ \bibinfo {author} {\bibfnamefont
  {C.-P.}\ \bibnamefont {Yuan}},\ }\href {\doibase 10.1103/PhysRevD.82.074024}
  {\bibfield  {journal} {\bibinfo  {journal} {Phys.Rev.}\ }\textbf {\bibinfo
  {volume} {D82}},\ \bibinfo {pages} {074024} (\bibinfo {year} {2010})},\
  \Eprint {http://arxiv.org/abs/1007.2241} {arXiv:1007.2241 [hep-ph]}
  \BibitemShut {NoStop}%
\bibitem [{\citenamefont {Catani}\ \emph
  {et~al.}(1991{\natexlab{b}})\citenamefont {Catani}, \citenamefont
  {Dokshitzer}, \citenamefont {Olsson}, \citenamefont {Turnock},\ and\
  \citenamefont {Webber}}]{Catani:1991hj}%
  \BibitemOpen
  \bibfield  {author} {\bibinfo {author} {\bibfnamefont {S.}~\bibnamefont
  {Catani}}, \bibinfo {author} {\bibfnamefont {Y.~L.}\ \bibnamefont
  {Dokshitzer}}, \bibinfo {author} {\bibfnamefont {M.}~\bibnamefont {Olsson}},
  \bibinfo {author} {\bibfnamefont {G.}~\bibnamefont {Turnock}}, \ and\
  \bibinfo {author} {\bibfnamefont {B.~R.}\ \bibnamefont {Webber}},\ }\href
  {http://inspirehep.net/search?j=PHLTA,B269,432} {\bibfield  {journal}
  {\bibinfo  {journal} {Phys. Lett.}\ }\textbf {\bibinfo {volume} {B269}},\
  \bibinfo {pages} {432} (\bibinfo {year} {1991}{\natexlab{b}})}\BibitemShut
  {NoStop}%
\bibitem [{\citenamefont {Abreu}\ \emph {et~al.}(1991)\citenamefont {Abreu}
  \emph {et~al.}}]{Abreu:1990ce}%
  \BibitemOpen
  \bibfield  {author} {\bibinfo {author} {\bibfnamefont {P.}~\bibnamefont
  {Abreu}} \emph {et~al.} (\bibinfo {collaboration} {DELPHI}),\ }\href
  {\doibase 10.1016/0370-2693(91)90796-S} {\bibfield  {journal} {\bibinfo
  {journal} {Phys. Lett.}\ }\textbf {\bibinfo {volume} {B255}},\ \bibinfo
  {pages} {466} (\bibinfo {year} {1991})}\BibitemShut {NoStop}%
\bibitem [{\citenamefont {Andersen}\ \emph {et~al.}(2018)\citenamefont
  {Andersen} \emph {et~al.}}]{Bendavid:2018nar}%
  \BibitemOpen
  \bibfield  {author} {\bibinfo {author} {\bibfnamefont {J.~R.}\ \bibnamefont
  {Andersen}} \emph {et~al.},\ }\href
  {http://inspirehep.net/record/1663483/files/1803.07977.pdf} {\  (\bibinfo
  {year} {2018})},\ \Eprint {http://arxiv.org/abs/1803.07977} {arXiv:1803.07977
  [hep-ph]} \BibitemShut {NoStop}%
\bibitem [{\citenamefont {Andersen}\ \emph {et~al.}(2016)\citenamefont
  {Andersen} \emph {et~al.}}]{Badger:2016bpw}%
  \BibitemOpen
  \bibfield  {author} {\bibinfo {author} {\bibfnamefont {J.~R.}\ \bibnamefont
  {Andersen}} \emph {et~al.},\ }\href
  {http://lss.fnal.gov/archive/2016/conf/fermilab-conf-16-175-ppd-t.pdf} {\
  (\bibinfo {year} {2016})},\ \Eprint {http://arxiv.org/abs/1605.04692}
  {arXiv:1605.04692 [hep-ph]} \BibitemShut {NoStop}%
\bibitem [{\citenamefont {Kramer}\ and\ \citenamefont
  {Lampe}(1987)}]{Kramer:1986sr}%
  \BibitemOpen
  \bibfield  {author} {\bibinfo {author} {\bibfnamefont {G.}~\bibnamefont
  {Kramer}}\ and\ \bibinfo {author} {\bibfnamefont {B.}~\bibnamefont {Lampe}},\
  }\href {\doibase 10.1063/1.527586} {\bibfield  {journal} {\bibinfo  {journal}
  {J.Math.Phys.}\ }\textbf {\bibinfo {volume} {28}},\ \bibinfo {pages} {945}
  (\bibinfo {year} {1987})}\BibitemShut {NoStop}%
\bibitem [{\citenamefont {Huber}\ and\ \citenamefont
  {Ma{\^i}tre}(2006)}]{Huber:2005yg}%
  \BibitemOpen
  \bibfield  {author} {\bibinfo {author} {\bibfnamefont {T.}~\bibnamefont
  {Huber}}\ and\ \bibinfo {author} {\bibfnamefont {D.}~\bibnamefont
  {Ma{\^i}tre}},\ }\href {\doibase 10.1016/j.cpc.2006.01.007} {\bibfield
  {journal} {\bibinfo  {journal} {Comput. Phys. Commun.}\ }\textbf {\bibinfo
  {volume} {175}},\ \bibinfo {pages} {122} (\bibinfo {year} {2006})},\ \Eprint
  {http://arxiv.org/abs/hep-ph/0507094} {hep-ph/0507094} \BibitemShut {NoStop}%
\bibitem [{\citenamefont {Huber}\ and\ \citenamefont
  {Ma{\^i}tre}(2008)}]{Huber:2007dx}%
  \BibitemOpen
  \bibfield  {author} {\bibinfo {author} {\bibfnamefont {T.}~\bibnamefont
  {Huber}}\ and\ \bibinfo {author} {\bibfnamefont {D.}~\bibnamefont
  {Ma{\^i}tre}},\ }\href {\doibase 10.1016/j.cpc.2007.12.008} {\bibfield
  {journal} {\bibinfo  {journal} {Comput. Phys. Commun.}\ }\textbf {\bibinfo
  {volume} {178}},\ \bibinfo {pages} {755} (\bibinfo {year} {2008})},\ \Eprint
  {http://arxiv.org/abs/0708.2443} {arXiv:0708.2443 [hep-ph]} \BibitemShut
  {NoStop}%
\bibitem [{\citenamefont {Korchemsky}\ and\ \citenamefont
  {Marchesini}(1993{\natexlab{a}})}]{Korchemsky:1992xv}%
  \BibitemOpen
  \bibfield  {author} {\bibinfo {author} {\bibfnamefont {G.~P.}\ \bibnamefont
  {Korchemsky}}\ and\ \bibinfo {author} {\bibfnamefont {G.}~\bibnamefont
  {Marchesini}},\ }\href {\doibase 10.1016/0550-3213(93)90167-N} {\bibfield
  {journal} {\bibinfo  {journal} {Nucl. Phys.}\ }\textbf {\bibinfo {volume}
  {B406}},\ \bibinfo {pages} {225} (\bibinfo {year} {1993}{\natexlab{a}})},\
  \Eprint {http://arxiv.org/abs/hep-ph/9210281} {hep-ph/9210281} \BibitemShut
  {NoStop}%
\bibitem [{\citenamefont {Korchemsky}\ and\ \citenamefont
  {Marchesini}(1993{\natexlab{b}})}]{Korchemsky:1993uz}%
  \BibitemOpen
  \bibfield  {author} {\bibinfo {author} {\bibfnamefont {G.~P.}\ \bibnamefont
  {Korchemsky}}\ and\ \bibinfo {author} {\bibfnamefont {G.}~\bibnamefont
  {Marchesini}},\ }\href {\doibase 10.1016/0370-2693(93)90015-A} {\bibfield
  {journal} {\bibinfo  {journal} {Phys. Lett.}\ }\textbf {\bibinfo {volume}
  {B313}},\ \bibinfo {pages} {433} (\bibinfo {year}
  {1993}{\natexlab{b}})}\BibitemShut {NoStop}%
\bibitem [{\citenamefont {Dasgupta}\ \emph {et~al.}(2018)\citenamefont
  {Dasgupta}, \citenamefont {Dreyer}, \citenamefont {Hamilton}, \citenamefont
  {Monni},\ and\ \citenamefont {Salam}}]{Dasgupta:2018nvj}%
  \BibitemOpen
  \bibfield  {author} {\bibinfo {author} {\bibfnamefont {M.}~\bibnamefont
  {Dasgupta}}, \bibinfo {author} {\bibfnamefont {F.~A.}\ \bibnamefont
  {Dreyer}}, \bibinfo {author} {\bibfnamefont {K.}~\bibnamefont {Hamilton}},
  \bibinfo {author} {\bibfnamefont {P.~F.}\ \bibnamefont {Monni}}, \ and\
  \bibinfo {author} {\bibfnamefont {G.~P.}\ \bibnamefont {Salam}},\ }\href
  {\doibase 10.1007/JHEP09(2018)033} {\bibfield  {journal} {\bibinfo  {journal}
  {JHEP}\ }\textbf {\bibinfo {volume} {09}},\ \bibinfo {pages} {033} (\bibinfo
  {year} {2018})},\ \Eprint {http://arxiv.org/abs/1805.09327} {arXiv:1805.09327
  [hep-ph]} \BibitemShut {NoStop}%
\bibitem [{\citenamefont {Isaacson}\ and\ \citenamefont
  {Prestel}(2018)}]{Isaacson:2018zdi}%
  \BibitemOpen
  \bibfield  {author} {\bibinfo {author} {\bibfnamefont {J.}~\bibnamefont
  {Isaacson}}\ and\ \bibinfo {author} {\bibfnamefont {S.}~\bibnamefont
  {Prestel}},\ }\href@noop {} {\  (\bibinfo {year} {2018})},\ \Eprint
  {http://arxiv.org/abs/1806.10102} {arXiv:1806.10102 [hep-ph]} \BibitemShut
  {NoStop}%
\bibitem [{\citenamefont {{\'A}ngeles~Mart{\'i}nez}\ \emph
  {et~al.}(2018)\citenamefont {{\'A}ngeles~Mart{\'i}nez}, \citenamefont
  {De~Angelis}, \citenamefont {Forshaw}, \citenamefont {Pl{\"a}tzer},\ and\
  \citenamefont {Seymour}}]{Martinez:2018ffw}%
  \BibitemOpen
  \bibfield  {author} {\bibinfo {author} {\bibfnamefont {R.}~\bibnamefont
  {{\'A}ngeles~Mart{\'i}nez}}, \bibinfo {author} {\bibfnamefont
  {M.}~\bibnamefont {De~Angelis}}, \bibinfo {author} {\bibfnamefont {J.~R.}\
  \bibnamefont {Forshaw}}, \bibinfo {author} {\bibfnamefont {S.}~\bibnamefont
  {Pl{\"a}tzer}}, \ and\ \bibinfo {author} {\bibfnamefont {M.~H.}\ \bibnamefont
  {Seymour}},\ }\href {\doibase 10.1007/JHEP05(2018)044} {\bibfield  {journal}
  {\bibinfo  {journal} {JHEP}\ }\textbf {\bibinfo {volume} {05}},\ \bibinfo
  {pages} {044} (\bibinfo {year} {2018})},\ \Eprint
  {http://arxiv.org/abs/1802.08531} {arXiv:1802.08531 [hep-ph]} \BibitemShut
  {NoStop}%
\end{thebibliography}%
\end{document}